\documentclass{ar-1col}

\usepackage[comma]{natbib}
\usepackage{url}
\usepackage{epic,eepic,units,booktabs}
\usepackage{soul}
\usepackage{psfrag}
\usepackage{latexsym}
\usepackage{longtable}
\usepackage{mathrsfs}
\usepackage{bigstrut}
\usepackage{pifont}
\usepackage{amsmath}
\usepackage{setspace}
\usepackage{subfigure}
\usepackage{mathtools}
\usepackage{mathcomp}
\usepackage{mathdots}
\usepackage{rotating}
\usepackage{array}
\usepackage{multirow}
\usepackage{multicol}
\usepackage{color}
\usepackage{algorithm}
\usepackage{algorithmic}
\usepackage{graphicx}
\usepackage[dvipsnames]{xcolor}
\usepackage{euscript,yfonts,dsfont,bbm,wasysym,pdfsync,arydshln,framed}
\usepackage{amstext}
\usepackage{amssymb}

\usepackage{tikz}
\usetikzlibrary{arrows,shapes,trees,calc,positioning}
\usetikzlibrary{matrix}

\newcommand{\bbZ}{\mathbb{Z}}

\newcommand{\inprod}[2]{\left< #1 , #2 \right>}

\newcommand{\non}{\nonumber}
\newcommand{\ds}{\displaystyle}

\newcommand{\mrd}{\mathrm{d}}
\newcommand{\mre}{\mathrm{e}}
\newcommand{\mri}{\mathrm{i}}

\newcommand{\bd}{{\bf d}}

\newcommand{\bu}{{\bf u}}

\newcommand{\bphi}{\mbox{\boldmath$\phi$}}

\newcommand{\bpsi}{\mbox{\boldmath$\psi$}}
\newcommand{\bxi}{\mbox{\boldmath$\xi$}}

\newcommand{\btau}{\mbox{\boldmath$\tau$}}

\newcommand{\py}{\partial_y}

\newcommand{\toep}{\mbox{toep}}

\newcommand{\tc}{\textcolor}

\newcommand{\DefinedAs}[0]{\mathrel{\mathop:}=}

\DeclareMathOperator*{\logdet}{log\,det}

\DeclareMathOperator*{\minimize}{minimize}

\DeclareMathOperator*{\subject}{subject~to}
\DeclareMathOperator*{\rank}{rank}

\definecolor{bgblue}{rgb}{0.04,0.19,0.53}
\definecolor{dblue1}{rgb}{0,0.3,0.7}
\definecolor{dred}{rgb}{0.4,0.2,0}

\newcommand{\enma}[1]   {\ensuremath{#1}}

\newcommand{\beq}{\begin{equation}}
\newcommand{\eeq}{\end{equation}}
\newcommand{\bseq}{\begin{subequations}}
\newcommand{\eseq}{\end{subequations}}
\newcommand{\beqn}{\begin{eqnarray}}
\newcommand{\eeqn}{\end{eqnarray}}
\newcommand{\ba}{\begin{array}}
\newcommand{\ea}{\end{array}}
\newcommand{\bct}{\begin{center}}
\newcommand{\ect}{\end{center}}
\newcommand{\btmz}{\begin{itemize}}
\newcommand{\etmz}{\end{itemize}}
\newcommand{\benum}{\begin{enumerate}}
\newcommand{\eenum}{\end{enumerate}}

\newcommand{\norm}[1]{\| #1 \|}                 %does not make large \|

\newcommand{\diag}      {\enma{\mathrm{diag}}}

\newcommand{\trace}     {\enma{\mathrm{trace}}}

\newcommand{\col}       {\enma{\mathrm{col}}}

\newcommand{\inner}[2]{\left\langle #1,#2 \right\rangle}

\newcommand{\bv}{{\bf v}}

\newcommand{\matbegin}{
        \left[
}
\newcommand{\matend}{
        \right]
}

\newcommand{\tbo}[2]{
  \matbegin \begin{array}{c}
       #1 \\ #2
       \end{array} \matend }
\newcommand{\thbo}[3]{
  \matbegin \begin{array}{c}
       #1 \\ #2 \\ #3
       \end{array} \matend }
\newcommand{\obt}[2]{
  \matbegin \begin{array}{cc}
       #1 & #2
       \end{array} \matend }

\newcommand{\tbt}[4]{
  \matbegin \begin{array}{cc}
       #1 & #2 \\ #3 & #4
       \end{array} \matend }
\newcommand{\thbt}[6]{
  \matbegin \begin{array}{cc}
       #1 & #2 \\ #3 & #4 \\ #5 & #6
       \end{array} \matend }
\newcommand{\tbth}[6]{
  \matbegin \begin{array}{ccc}
       #1 & #2 & #3\\ #4 & #5 & #6
       \end{array} \matend }

\newcommand{\be}{\begin{equation}}
\newcommand{\ee}{\end{equation}}

\newcommand{\cplxs}{ C\kern -.35em \rule{0.03 em}{.7 ex}~   }

\def\complex{\hbox{C\kern -.45em \rule{0.03 em}{1.5 ex}}~}

\newcommand{\bi}{\begin{itemize}}
\newcommand{\ei}{\end{itemize}}

\newcommand{\btab}{\begin{tabular}}
\newcommand{\etab}{\end{tabular}}
\newcommand{\We}{W\!e}
\newcommand{\bk}{{\bf{k}}}
\newcommand{\bw}{{\bf{w}}}
\newcommand{\bE}{{\bf E}}

\newcommand{\Aos}{A_{\mathrm{os}}}
\newcommand{\Asq}{A_{\mathrm{sq}}}

\newcommand{\Acpn}{A_{\mathrm{cp1}}}
\newcommand{\Acpv}{A_{\mathrm{cp2}}}
\newcommand{\Ak}{A_\bk}
\newcommand{\Bk}{B_\bk}
\newcommand{\Ck}{C_\bk}
\newcommand{\Ek}{E_\bk}
\newcommand{\Gk}{G_\bk}
\newcommand{\Hk}{H_\bk}
\newcommand{\Kk}{K_\bk}
\newcommand{\Pk}{P_\bk}

\newcommand{\Tk}{T_\bk}
\newcommand{\Xk}{X_\bk}
\newcommand{\Zk}{Z_\bk}
\newcommand{\Vk}{V_\bk}
\newcommand{\Wk}{W_\bk}
\newcommand{\cAk}{{\cal A}_\bk}
\newcommand{\cBk}{{\cal B}_\bk}
\newcommand{\cCk}{{\cal C}_\bk}
\newcommand{\cXk}{{\cal X}_\bk}
\newcommand{\cWk}{{\cal W}_\bk}
\newcommand{\Sin}{\Omega_{\bk} (\mri \omega)}
\newcommand{\Sout}{S_{\bk} (\mri \omega)}
\newcommand{\Piout}{\Pi_\bk (\omega)}
\newcommand{\Eout}{E_{\bk}^{\mathrm{out}}}
\newcommand{\Ein}{E_{\bk}^{\mathrm{in}}}
\newcommand{\tcr}{\textcolor{red}}
\newcommand{\tcb}{\textcolor{blue}}
\newcommand{\tcg}{\textcolor{ForestGreen}}

\definecolor{bgblue}{rgb}{0.04,0.39,0.53}
\definecolor{dred}{cmyk}{0,1.0,1.0,0.30}
\jname{Xxxx. Xxx. Xxx. Xxx.}
\jvol{AA}
\jyear{2021}
\doi{10.1146/((please add article doi))}

\begin{document}

% Page header
\markboth{M.\ R.\ Jovanovi\'c}{Input-output viewpoint of transitional and turbulent flows}

% Title
 \title{\mbox{From bypass transition to} \mbox{flow control and data-driven} turbulence modeling: An input-output viewpoint}

%Authors, affiliations address.
\author{Mihailo R. Jovanovi\'c
\affil{Ming Hsieh Department of Electrical and Computer Engineering, University of Southern California, Los Angeles, California 90089, USA; email: mihailo@usc.edu}}

%Abstract
\begin{abstract}
	\vspace*{-2ex}
Transient growth and resolvent analyses are routinely used to assess non-asymptotic properties of fluid flows. In particular, resolvent analysis can be interpreted as a special case of viewing flow dynamics as an open system in which free-stream turbulence, surface roughness, and other irregularities provide sources of input forcing. We offer a comprehensive summary of the tools that can be employed to probe the dynamics of fluctuations around a laminar or turbulent base flow in the presence of such stochastic or deterministic input forcing and describe how input-output techniques enhance resolvent analysis. Specifically, physical insights that may remain hidden in the resolvent analysis are gained by detailed examination of input-output responses between spatially-localized body forces and selected linear combinations of state variables. This differentiating feature plays a key role in quantifying the importance of different mechanisms for bypass transition in wall-bounded shear flows and in explaining how turbulent jets generate noise. We highlight the utility of a stochastic framework, with white or colored inputs, in addressing a variety of open challenges including transition in complex fluids, flow control, and physics-aware data-driven turbulence modeling. Applications with time- or spatially-periodic base flows are discussed and future research directions are outlined.
	\end{abstract}
		
%Keywords, etc.
\begin{keywords}
% keywords, separated by comma, no full stop, lowercase
	\vspace*{-1ex}
input-output analysis, flow modeling and control, physics-aware data-driven modeling, stochastic dynamics, frequency responses, transition to turbulence, turbulent flows, convex optimization
\end{keywords}
\maketitle

% Table of Contents
 \tableofcontents

 \newpage

% Heading 1
\section{INTRODUCTION}

Hydrodynamic stability theory focuses on spectral analysis of the dynamical generator in the linearized Navier-Stokes (NS) equations while seeking the critical Reynolds number at which exponentially growing modes emerge~\citep{schhen01}. Although in many flows predictions agree well with experiments, in wall-bounded shear flows both the critical Reynolds number and the spatial structure of the least-stable or unstable modes are at odds with experimental observations. A broader viewpoint, based on nonmodal analysis of the linearized NS equations, provides reconciliation with experiments and identifies mechanisms for the early stages of subcritical transition~\citep{sch07}.

In the words of~\citet{treemb05}, the eigenvalue decomposition {\em gives a square matrix, or an operator, a personality}. However, this ``personality test'' is conclusive only for normal (i.e., unitarily diagonalizable) operators. For non-normal operators, it is the singular value decomposition (SVD) that offers a robust predictor of  ``personality''~\citep{treemb05}. In wall-bounded shear flows, non-normality of the linearized dynamical operator introduces coupling of exponentially decaying modes which explains high sensitivity of the laminar flow~\citep{sch07}. The high sensitivity degrades the accuracy of analytical and computational predictions that do not explicitly account for modeling imperfections. These are typically difficult to model and may arise from a variety of sources, including surface roughness, thermal fluctuations, and irregularities in the incoming stream.

% The study of dynamical systems with input forcing has a rich history in several branches of electrical engineering including circuit theory, communications, signal processing, and control. In this, dynamical systems are decomposed into essential pieces and viewed as interconnections of input-output ``blocks'' that are brought together via cascade, parallel, and feedback arrangements. This input-output viewpoint facilitates analysis, optimization, and design of complex systems which are composed of sub-systems that are easier to characterize. It also allows to quantify the influence of modeling imperfections (e.g., background noise or experimental uncertainty that is unavoidable in physical \mbox{systems) on quantities of~interest.}

The study of dynamical systems with input forcing has a rich history in several branches of electrical engineering including circuit theory, communications, signal processing, and control. In this, dynamical systems are decomposed into essential pieces and represented as interconnections of input-output ``blocks''. This input-output viewpoint facilitates the analysis, design, and optimization of complex systems, since they can be viewed as simpler sub-systems placed in cascade, parallel, and feedback arrangements with one another. It also allows us to quantify the influence of modeling imperfections (e.g., background noise or experimental uncertainty that is unavoidable in physical systems) on quantities of~interest.

In fluid mechanics, input-output analysis addresses the influence of deterministic as well as stochastic inputs on transient and asymptotic properties of fluid flows. It offers a complementary viewpoint to transient growth~\citep{butfar92} and resolvent~\citep{tretrereddri93} analyses and brings in an appealing robustness interpretation. Specifically, additional insight about the dynamics is gained by carrying out SVD of the operator that maps excitation sources (i.e., inputs such as body forcing fluctuations) to the quantities of interest (i.e., outputs such as velocity fluctuations). In contrast to the resolvent, this operator is not necessarily a square object; it captures the effect of different inputs to particular physical quantities and thereby reveals finer physical aspects~\citep{jovbamJFM05}. In wall-bounded shear flows, input-output analysis exposes large amplification of disturbances and high sensitivity of the laminar flow to uncertainty in the geometry or base velocity~\citep{tretrereddri93,farioa93,bamdah01,mj-phd04}, and provides insights into structural features of turbulent flows~\citep{mcksha10,hwacosJFM10a,hwacosJFM10b}. Additional successful applications range from discovering mechanisms for transition to elastic turbulence in viscoelastic fluids~\citep{hodjovkumJFM08,hodjovkumJFM09,jovkumJNNFM11}, to revealing how turbulent jets generate noise~\citep{jeunicjovPOF16}, and explaining the origin of reattachment streaks in hypersonic flows~\citep{dwisidniccanjovJFM19}.

This review highlights the merits, effectiveness, and versatility of the input-output framework for modeling, analysis, and control of fluid flows. We offer a comprehensive summary of the tools that can be used to probe the dynamics of infinitesimal fluctuations around a given laminar or turbulent base flow, and explain how the framework augments resolvent analysis~\citep{tretrereddri93}. We illustrate how the componentwise input-output approach~\citep{jovbamJFM05} identifies key mechanisms for bypass transition in channel flows of Newtonian and viscoelastic fluids. We then describe how periodic base flow modifications, induced by streamwise traveling waves and spanwise wall-oscillations, can be designed to, respectively, control the onset of turbulence~\citep{moajovJFM10} and identify the optimal period of oscillation for turbulent drag reduction~\citep{moajovJFM12}. The input-output framework is also well-suited for data-driven turbulence modeling; in contrast to physics-agnostic machine learning techniques, the tools from control theory and convex optimization allow for strategic use of data in order to capture second-order statistics of turbulent flows via first-principle models~\citep{zarjovgeoJFM17,zargeojovARC20}.

%Heading 1
	\vspace*{-3ex}
\section{INPUT-OUTPUT VIEWPOINT: BEYOND RESOLVENT~ANALYSIS}
	\label{sec.io}

\begin{figure}[bth]
    \centering
    \vspace*{-0.35cm}
    {
    \begin{tabular}{ccc}
    \hspace*{-5cm}
    \begin{tabular}{c}
   \subfigure[]
   {\includegraphics[width=0.32\textwidth,height=0.08\textheight]{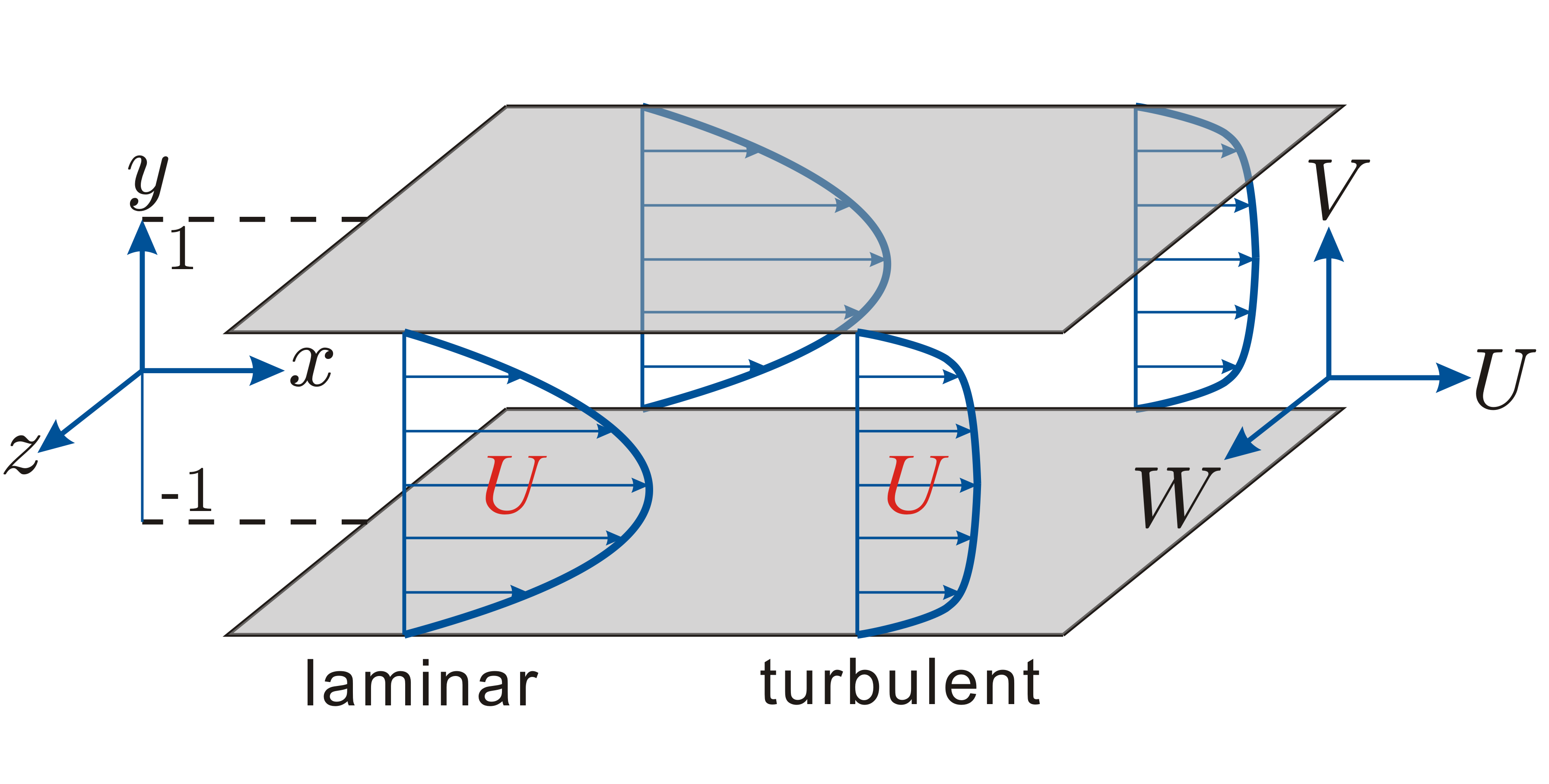}
           \label{fig.channel-U}}
           \end{tabular}   
           &    
   \hspace*{-9.85cm}
   \begin{tabular}{c}
    \subfigure[]
    {\includegraphics[width=0.3\textwidth,,height=0.08\textheight]{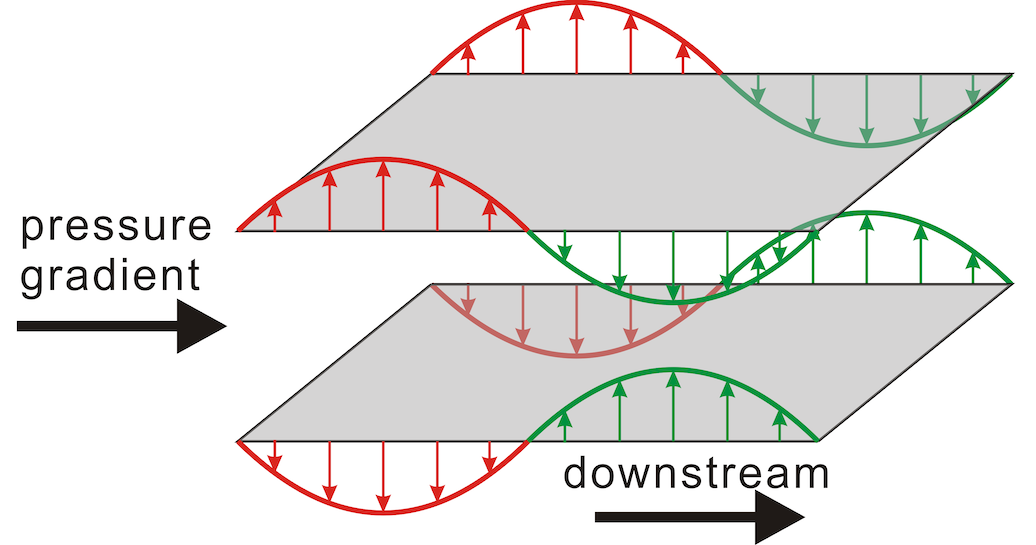}
           \label{fig.channel-stw}}
           \end{tabular}
           &    
   \hspace*{-9.85cm}
   \begin{tabular}{c}
    \subfigure[]
    {\includegraphics[width=0.26\textwidth,,height=0.08\textheight]{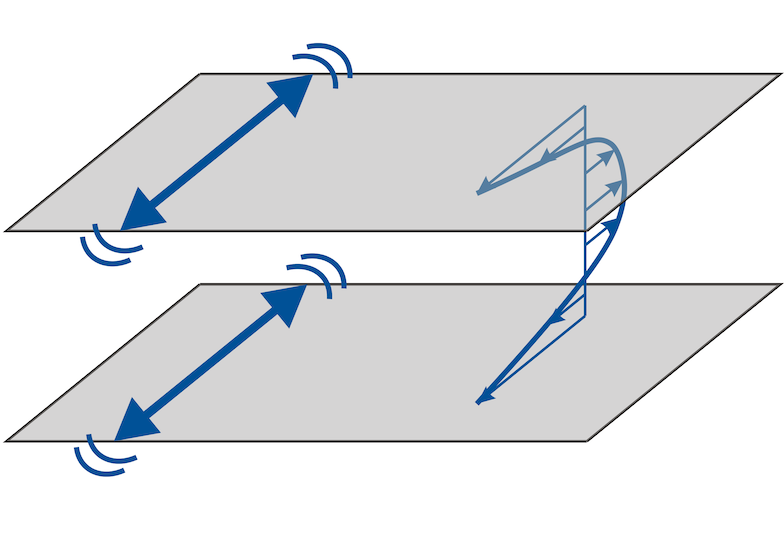}
           \label{fig.channel-swo}}
           \end{tabular}
	\end{tabular}
    }
    \caption{A pressure-driven channel flow (a) between two parallel infinite walls with base flow ($U(y),0,0$), inhomogeneous wall-normal ($y$), and homogeneous streamwise and spanwise ($x,z$) directions; (b) subject to blowing and suction along the walls; and (c) subject to spanwise wall-oscillations.}
    \label{fig.channel}
    \end{figure}

We first review the tools that can be used to probe the dynamics of infinitesimal fluctuations around a given base flow. While this framework can be utilized in a variety of flow regimes and geometries, we resort to a channel flow with homogeneous wall-parallel directions to illustrate the key concepts; see {\bf Figure~\ref{fig.channel-U}}. Even in this simple setup a variety of non-trivial fundamental questions can be addressed by employing an input-output viewpoint, including transition in complex fluids, flow control, and data-driven turbulence modeling.  

	% Margin Note - basic definitions
\begin{marginnote}[]
\entry{Time}{$t$}
\entry{Complex number}{$s = \sigma + \mri \omega$}
\entry{Frequency}{$\omega$}
\entry{Wavenumbers}{$\bk$}
\entry{Dynamical generator}{$\Ak$}		
\entry{State-transition operator}{$\mre^{\Ak t}$}	
\entry{Resolvent operator}{$(sI \, - \, \Ak)^{-1}$}
\entry{Input operator}{$\Bk$}
\entry{Output operator}{$\Ck$}	
\entry{Impulse response}{$\Ck \;\! \mre^{\Ak t} \Bk$}	
\entry{Transfer function}{$\Ck \left( s I - \Ak \right)^{-1} \! \Bk$}
\entry{Frequency response}{$\Ck \left( \mri \omega I - \Ak \right)^{-1} \! \Bk$}
\end{marginnote}

	%=========
	% Figure 1  %
	%=========

% Heading 2
	\vspace*{-6ex}
\subsection{From evolution model to input-output representation}
	\label{sec.evolution}

The linearized NS equations govern the dynamics of infinitesimal fluctuations around a given base flow. Fluctuations can arise from a variety of sources, including surface roughness, imperfections in the incoming stream, acoustics, vibrations, particles, and impurities. In turbulent flows, nonlinear interactions between different length scales can also provide forcing that sustains fluctuations. The linearized NS equations, with an input forcing $\bd_{\bk} (t)$ and an output of interest $\bxi_\bk (t)$, can be brought to an evolution~form,
	\begin{align}
	\label{eq.lnse1}
        \ba{rcl}
             % \dot{\bpsi}_\bk (t)
             \dfrac{\mrd \bpsi_\bk (t)}{\mrd t}
            & \; = \;\, &
            \Ak
            \,
            \bpsi_\bk (t)
            \; + \;
            \Bk
            \,
            \bd_{\bk} (t),
            \\[0.15cm]
            \bxi_\bk (t)
            & \; = \;\, &
            \Ck
            \,
            \bpsi_\bk (t),
        \ea
\end{align}
where $\bpsi_\bk (t)$ is the state and $\bk$ is the vector of wavenumbers. The operator $\Ak$ characterizes dynamical interactions between the states, $\Bk$ specifies the way the input $\bd_\bk (t)$ enters into the dynamics, and $\Ck$ maps the state $\bpsi_\bk (t)$ to the output $\bxi_\bk (t)$. Equation~\ref{eq.lnse1} is a standard state-space model in the controls literature, and it provides a convenient starting point for modal and nonmodal analysis, system identification, turbulence \mbox{modeling, and flow control.}

	%% Margin Note
\begin{marginnote}[]
\entry{Evolution Model~\ref{eq.lnse1}}{A $1$st order (in time) differential equation governs the evolution of the state $\bpsi_{\bk} (t)$ and a static-in-time output equation relates $\bpsi_{\bk} (t)$ to the output 
	$
	\bxi_{\bk} (t) 
	= 
         \Ck \bpsi_\bk (t).
	$
Apart from the boundary conditions, no additional constraints are imposed on $\bpsi_k (t)$.}
\end{marginnote}

For a pressure-driven channel flow of an incompressible Newtonian fluid, the base flow $\bar{\bu}$ is either given by the laminar parabolic profile (Poiseuille flow) or the turbulent mean velocity. In both cases, the flow is fully-developed and $\bar{\bu}$ only depends on the wall-normal distance $y$, $\bar{\bu} = (U(y),0,0)$. Thus, the linearized NS equations are translationally-invariant in wall-parallel directions and in time and fluctuations can be decomposed in terms of the normal modes in $x$ and $z$ as 
	$
	\bpsi (x,y,z,t)
	= 
	\bpsi_{\bk} (y,t) \, \mre^{\mri (k_x x + k_z z)}.
	$
Here, $\bk \DefinedAs (k_x,k_z)$ denotes the vector of wall-parallel wavenumbers and $\Ak$ is the Orr-Sommerfeld/Squire operator~\citep{schhen01}. In addition to $\bk$, System~\ref{eq.lnse1} is parameterized by the base flow $\bar{\bu}$ and the Reynolds number $Re$. For any ($\bk,t$), the state $\bpsi_\bk (t)$, input $\bd_\bk (t)$, and output $\bxi_\bk (t)$ are functions of $y$ but, for notational convenience, we suppress this dependence.

	\vspace*{-2ex}
\subsubsection*{\tc{dred}{Derivation of Equation~\ref{eq.lnse1}}.} The linearized model is obtained by expressing the flow as the sum of the base and fluctuation components and by neglecting the quadratic fluctuation terms. In incompressible flows of Newtonian fluids, the velocity obeys a continuity equation and a Poisson equation for the pressure $p$ is obtained by applying the divergence operator to the linearized NS equations. The Orr-Sommerfeld equation is obtained by acting with the Laplacian $\Delta$ on the wall-normal velocity equation and using the expression for $\Delta p$ to eliminate $p$. The Squire equation is obtained by taking the curl of the linearized NS equations. This yields an evolution model, in the form of two PDEs, for the wall-normal velocity and vorticity~\citep{kimmoimos87}, $\bpsi \DefinedAs (v,\eta)$. All other velocity and vorticity components can be expressed in terms of ($v,\eta$) \mbox{via kinematic relations~\citep{jovbamJFM05}.}

Standard stability analysis of a laminar Poiseuille flow predicts modal instability for $Re = 5772$. The discrepancy with experiments, in which transition occurs for $Re \approx 1000$, can be explained using nonmodal analysis~\citep{sch07} which reveals significant transient growth of fluctuations~\citep{gus91,butfar92} and strong amplification of disturbances~\citep{tretrereddri93,farioa93,bamdah01}. 

% Heading 3
	\vspace*{-2ex}
\subsubsection{Resolvent, transfer function, impulse and frequency response operators}

While the governing equations and geometry determine the dynamical generator $\Ak$, there is flexibility in selecting the operators $\Bk$ and $\Ck$ and different choices can reveal different aspects of flow physics~\citep{jovbamJFM05}. All of these operators play a role in the response of System~\ref{eq.lnse1} which arises from the initial condition $\bpsi_\bk (0)$ and the exogenous input~$\bd_\bk (t)$,
	\be
	\ba{rcccc}
	& \; & 
    	\mbox{\tc{RoyalBlue}{\bf natural response}}    
    	& & 
    	\mbox{\tc{dred}{\bf forced response}}     
        	\\
         \bxi_\bk (t)
%    	\;  = \;
%    	\Ck \bpsi_\bk (t)
	& \; = & 
	\tikz[baseline]{
            \node[fill=blue!20,anchor=base] (t1)
            {$\Ck
	\mre^{\Ak t} 
	\,
	\bpsi_\bk (0)$};
            }
    	& + & 
	\tikz[baseline]{
            \node[fill=red!20,anchor=base] (t1)
            {$
    	\ds{\int_{0}^{t}}
    	\Ck \, \mre^{\Ak ( t \, - \, \tau)}
    	\,
    	\Bk
    	\,
    	\bd_\bk (\tau)
    	\,
    	\mrd \tau,$};
         }
         \ea
	\label{eq.vc-t}
	\ee
where $\mre^{\Ak t}$ is the {\em state-transition operator\/} associated with $\Ak$. The Laplace transform can be utilized to rewrite Equation~\ref{eq.vc-t} as,
	\be
       	\hat{\bxi}_\bk (s)
	\; = \; 
	\tikz[baseline]{
            \node[fill=blue!20,anchor=base] (t1)
            {$\Ck
	\left(
    	s I \, - \, \Ak
    	\right)^{-1}
	\bpsi_\bk (0)$};
            }
    	\; + \;
	\tikz[baseline]{
            \node[fill=red!20,anchor=base] (t1)
            {$\Ck
    	\left(
    	s I \, - \, \Ak
    	\right)^{-1}   
	\! 	
	\Bk
    	\,
    	\hat{\bd}_\bk (s),$};
            }
	\label{eq.vc-s}
	\ee
where $s$ is the complex number, $I$ is the identity operator, $\hat{\bxi}_\bk (s)$ is the Laplace transform of $\bxi_\bk (t)$, and $(s I - \Ak)^{-1}$ is the {\em resolvent operator}. Equations~\ref{eq.vc-t} and~\ref{eq.vc-s} determine responses of System~\ref{eq.lnse1} and provide the basis for quantifying important dynamical features of the linearized flow equations. As the blue terms demonstrate, the natural (i.e., unforced) responses are characterized by the state-transition $\mre^{\Ak t} $ and resolvent 
	$
	( s I - \Ak )^{-1}  
	$
operators. On the other hand, the forced response is obtained by convolving an input $\bd_\bk (t)$ with the {\em impulse response operator\/} $\Tk (t)$; equivalently, the {\em transfer function\/} $\Tk (s)$ specifies an input-output mapping in the complex domain, i.e., $\hat{\bxi}_\bk (s) = \Tk (s) \, \hat{\bd}_\bk (s)$, where
	\be
	\ba{ccc}
	\mbox{\tc{dred}{\bf impulse response}}
	& ~~ &
	\mbox{\tc{dred}{\bf transfer function}}
	\\
	\tikz[baseline]{
            \node[fill=red!20,anchor=base] (t1)
            {$
            \Tk (t) \; \DefinedAs \; \Ck \, \mre^{\Ak t} \Bk$};
            }
	    &
	    \hspace*{-0.5cm}
	   \begin{tabular}{c}
            \\[-0.75cm]
    	   \scalebox{1}{%_______________________________________________________________________________
%
%   Block diagram of the periodic modification to the original dynamics
%   drawn from Right to Left
%
%   Mihailo Jovanovic, February 23, 2020
%_______________________________________________________________________________
%
% TikZ styles for drawing
%
%_______________________________________________________________
%
%     Tikz_common_styles
%
%     TikZ styles for drawing block diagrams.
%
%      Roy Smith,  15 July 2014
%
%_______________________________________________________________
\typeout{Reading Tikz styles:  Tikz_common_styles}
%
%   Block diagram styles
% added by MJ
\tikzstyle{input} = [coordinate]
\tikzstyle{output} = [coordinate]
\tikzstyle{block} = [draw,rectangle,thick,minimum height=2em,minimum width=1.0cm,
                                top color=blue!10, bottom color=blue!10]%
\tikzstyle{sum} = [draw,circle,inner sep=0mm,minimum size=2mm]%
\tikzstyle{connector} = [->,thick]%
\tikzstyle{line} = [thick]%
\tikzstyle{branch} = [circle,inner sep=0pt,minimum size=1mm,fill=black,draw=black]%
\tikzstyle{guide} = []%
%
%   colours for various blocks
\tikzstyle{deltablock} = [block, top color=red!10, bottom color=red!10]%
\tikzstyle{controlblock} = [block, top color=green!10, bottom color=green!10]%
\tikzstyle{weightblock} = [block, top color=orange!10, bottom color=orange!10]%
\tikzstyle{clpblock} = [block, top color=cyan!10, bottom color=cyan!10]%
%
%  styles for dimming block components.
\tikzstyle{block_dim} = [draw,rectangle,thick,minimum height=2em,minimum width=2em,
                                color=black!15]%
\tikzstyle{sum_dim} = [draw,circle,inner sep=0mm,minimum size=2mm,color=black!15]%
\tikzstyle{connector_dim} = [->,thick,color=black!15]%
%
%    Graph styles
\tikzstyle{smalllabel} = [font=\footnotesize]%
\tikzstyle{axiswidth}=[semithick]%
\tikzstyle{axiscolor}=[color=black!50]%
\tikzstyle{help lines} =[color=blue!40,very thin]%
\tikzstyle{axes} = [axiswidth,axiscolor,<->,smalllabel]%
\tikzstyle{axis} = [axiswidth,axiscolor,->,smalllabel]%
\tikzstyle{tickmark} = [thin,smalllabel]%
\tikzstyle{plain_axes} = [axiswidth,smalllabel]%
\tikzstyle{w_axes} = [axiswidth,->,smalllabel]%
\tikzstyle{m_axes} = [axiswidth,smalllabel]%
\tikzstyle{dataplot} = [thick]%
%
%_______________________________________________________________

%
%   set a filename for externalization
% \tikzsetnextfilename{clp_2dof_input_pert_config}
%
\noindent
\begin{tikzpicture}[scale=1, auto, >=stealth']

	% output node
	% starting point for uend
	% \node [input, name=uend] {};
	\node[] (ubegin) at (0,0) {};
	
	\node[] (uend) at ($(ubegin) + (3.35cm,0cm)$) {};
		
		% input dbegin to block plant
    	\draw [connector, ultra thick] (ubegin) -- node [midway, above] {$\ba{c} \mbox{\bf Laplace transform} \ea$} (uend);
	 	
\end{tikzpicture}
%_______________________________________________________________________________
}
    \end{tabular}
    	\hspace*{-0.25cm}
	   &
	   \tikz[baseline]{
            \node[fill=red!20,anchor=base] (t1)
            {$
            \Tk (s) \; \DefinedAs \; \Ck \left( s I \, - \, \Ak \right)^{-1} \! \Bk.$};
            }
	\ea
	\non
	\ee

For flows over perfectly smooth walls and in noise-free environments, study of natural responses aids in understanding the fundamental fluid mechanics. Specifically, the eigenvalue decomposition of $\Ak$ and the singular value decomposition of $\mre^{\Ak t}$, respectively, offer insights into modal and nonmodal aspects of the flow~\citep{sch07}. While such insights are valuable, engineering flows seldom exist in isolation and understanding the forced responses is equally important. In particular, input-output analysis examines forced responses with the objective of quantifying amplification of disturbances and impact of modeling imperfections on fluctuations' dynamics. In contrast to natural responses, study of forced responses requires specifying how disturbances enter into System~\ref{eq.lnse1}, through the operator $\Bk$.

In the special case when the input excites all degrees of freedom equally and the output is the entire state, $\Bk$ and $\Ck$ are the identity operators and the resolvent completely determines the transfer function. However, it is often of interest to confine the inputs to certain spatial regions and to examine outputs that are given by a linear combination of certain state variables. In such cases, the transfer function is determined by a ``compressed resolvent'' and its analysis can uncover important dynamical aspects that may be obscured by only paying attention to the ``standard resolvent''. This distinction played a key role in understanding how turbulent jets generate noise.~\citet{jeunicjovPOF16} utilized ``compressed resolvent'' analysis by restricting inputs to the vicinity of the jet turbulence and selecting far-field pressure as the output. In contrast to a standard resolvent analysis, which provides links to jet hydrodynamics but does not explain noise generation~\citep{garsanles13}, this approach identifies acoustic sources to be wavepackets that are in excellent agreement with experiments~\citep{jorcol13} and reveals \mbox{mechanisms for noise generation.} 
		
 	\vspace*{-3ex}
\subsubsection*{\tc{dred}{Singular Value Decomposition.}} In transient growth analysis, SVD identifies spatial structure of initial conditions that maximize energy at a given time. SVD also provides the tool for quantifying responses to unsteady deterministic as well as stochastic inputs $\bd_{\bk} (t)$ that neither grow nor decay in time (on average). This allows us to set $s = \mri \omega$, and the frequency response $\Tk (\mri \omega)$ is obtained by evaluating the transfer function $\Tk (s)$ along the imaginary axis; see the sidebar FREQUENCY RESPONSE OPERATOR. 

SVD of $\Tk (\mri \omega)$ identifies fundamental input-output features across ($\bk,\omega$),
	\begin{subequations}
	\label{eq.svd-adjoint}
	\be
	\hat{\bxi}_\bk (\mri \omega)
	\; = \;
	\Tk (\mri \omega) \, \hat{\bd}_\bk (\mri \omega)
	\; = \;
	\sum_{j \, = \, 1}^{\infty}
	\sigma_{\bk,j} (\omega)
	\hat{\bu}_{\bk,j} (\omega)
	\inner{\hat{\bv}_{\bk,j} (\omega)}{\hat{\bd}_\bk ( \mri \omega)}.
	\ee
The left and right singular functions, $\hat{\bv}_{\bk,j} (\omega)$ and $\hat{\bu}_{\bk,j} (\omega)$, provide orthonormal bases of the input and output spaces, the singular value $\sigma_{\bk,j} (\omega)$ determines the corresponding amplification, and $\inner{\cdot}{\cdot}$ is the inner product. SVD requires computation of the adjoint~$\Tk^{\dagger} (\mri \omega)$,
	\be
	\inner{\Tk^{\dagger} (\mri \omega) \hat{\bxi}_\bk (\mri \omega)}{\hat{\bd}_\bk (\mri \omega)}
	\; = \;
	\inner{\hat{\bxi}_\bk (\mri \omega)}{\Tk (\mri \omega) \hat{\bd}_\bk (\mri \omega)},
	\label{eq.adjoint}
	\ee
	\end{subequations}
and the eigenvalue decomposition of $T T^{\dagger}$ and $T^{\dagger} T$,
	$
	\Tk (\mri \omega) \Tk^{\dagger} (\mri \omega)
	\hat{\bu}_{\bk,j} (\omega)
	=
	\sigma_{\bk,j}^2 (\omega)
	\hat{\bu}_{\bk,j} (\omega),
	$
	$
	\Tk^{\dagger} (\mri \omega) \Tk (\mri \omega)
	\hat{\bv}_{\bk,j} (\omega)
	=  
	\sigma_{\bk,j}^2 (\omega)
	\hat{\bv}_{\bk,j} (\omega).
	$
Unless noted otherwise, the $L_2$ inner product $\inner{\cdot}{\cdot}$, which induces energy norm, is taken over inhomogeneous spatial directions in Equation~\ref{eq.svd-adjoint}. 

% Margin Note - basic definitions
\begin{marginnote}[]
\entry{Inner product}{$\inner{\, \cdot \,}{\, \cdot \,}$}
\entry{Norm}{$\| \, \cdot \, \|$}
\entry{Tensor product}{$[ \bu \otimes \bv ] \, \bw \DefinedAs \bu \inner{\bv}{\bw}$}
\entry{Square-integrable function space}{$L_2$}
\entry{Complex-conjugate transpose}{$(\cdot)^*$}
\entry{Adjoint}{$(\cdot)^\dagger$}
\entry{Expectation operator}{$\bE (\cdot)$}
\entry{Supremum}{$\sup$}
\entry{Set of integers}{$\bbZ$}
\end{marginnote}

	%=========
	% Figure 2  %
	%=========
	\begin{figure}[h!]
    \centering
     \vspace*{-0.25cm}
    {
    \begin{tabular}{ccc}
    \hspace*{-4.75cm}
    \begin{tabular}{c}
   \subfigure[input: $d_3 (x,y,z,t)$]
   {\includegraphics[width=0.325\textwidth]{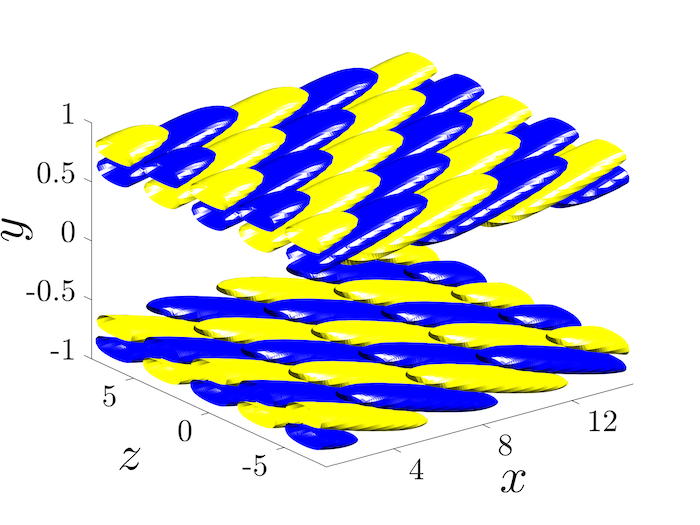}
           \label{fig.d3-dominant}}
           \end{tabular}   
           &    
            \hspace*{-6.5cm}
            \begin{tabular}{c}
            \\[-0.75cm]
    	   {\scalebox{1}{%_______________________________________________________________________________
%
%   Block diagram of the periodic modification to the original dynamics
%   drawn from Right to Left
%
%   Mihailo Jovanovic, February 23, 2020
%_______________________________________________________________________________
%
% TikZ styles for drawing
%
\input{figures/Tikz_common_styles}
%
%   set a filename for externalization
% \tikzsetnextfilename{clp_2dof_input_pert_config}
%
\noindent
\begin{tikzpicture}[scale=1, auto, >=stealth']

	% output node
	% starting point for uend
	% \node [input, name=uend] {};
	\node[] (ubegin) at (0,0) {};
	
	\node[] (uend) at ($(ubegin) + (3.cm,0cm)$) {};
		
		% input dbegin to block plant
    	\draw [connector, ultra thick] (ubegin) -- node [midway, above] {$\ba{c} \mbox{\bf linearized} \\ \mbox{\bf NS equations} \ea$} (uend);
	 	
\end{tikzpicture}
%_______________________________________________________________________________
}
           \label{fig.bd-arrow}}
    \end{tabular}
           &
   \hspace*{-6.25cm}
   \begin{tabular}{c}
    \subfigure[steady-state output: $u (x,y,z,t)$]
    {\includegraphics[width=0.325\textwidth]{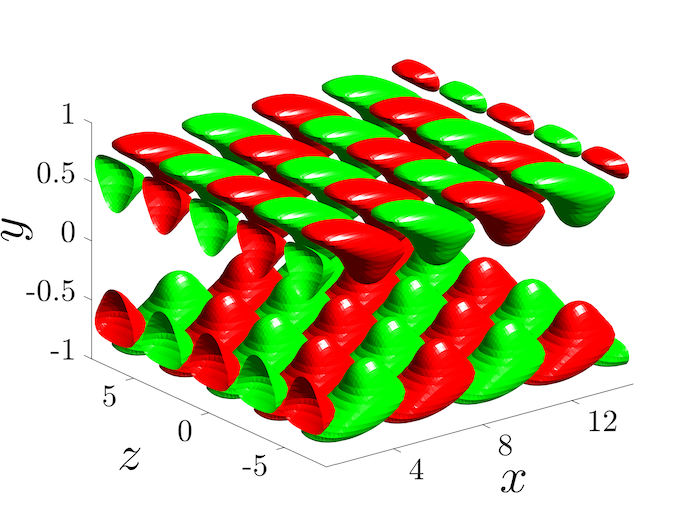}
           \label{fig.U-dominant}}
           \end{tabular}
	\end{tabular}
    }
    \caption{In linearly stable channel flows, the steady-state response of the linearized NS equations to a harmonic input in ($x,z,t$),
    	$
	\hat{\bd}_{\bk} (y,\mri \omega) \, \mre^{\mri (k_x x + k_z z + \omega t)},
	$
        is determined by 
        $
	\hat{\bxi}_{\bk} (y,\mri \omega) \, \mre^{\mri (k_x x + k_z z + \omega t)},
	$
	where 
	$\hat{\bxi}_{\bk} ( y ,\mri \omega) = [ \Tk (\mri \omega) \hat{\bd}_{\bk} ( \, \cdot \, ,\mri \omega) ] (y)$.
    Spatial structures of (a) spanwise forcing fluctuations; and (b) resulting streamwise velocity fluctuations at one time instant in Poiseuille flow with $Re = 2000$ for $4$ combinations of $(k_x,k_z,\omega) = (1,\pm 1, -0.385); (-1,\pm 1, 0.385)$.
    % ($k_x = 1,k_z = \pm 1, \omega = -0.385$), and~($k_x = -1,k_z = \pm 1, \omega = 0.385$).
    }
    \label{fig.dominant}
    \vspace*{-0.95cm}
    \end{figure}

	\vspace*{-4ex}
\subsection{Amplification of deterministic inputs}
	\label{sec.Hinf}

\begin{figure}[h!]
    \centering
    {
    \begin{tabular}{ccc}
     \hspace{-0.5cm}
	\begin{tabular}{c}
		\vspace{0.5cm}
		\normalsize{\rotatebox{90}{$\sigma_{\max}^2 (\Tk (\mri \omega))$}}
	\end{tabular}
	&
    \hspace*{-5.5cm}
    \begin{tabular}{c}
     \subfigure[]{\includegraphics[width=0.275\textwidth]{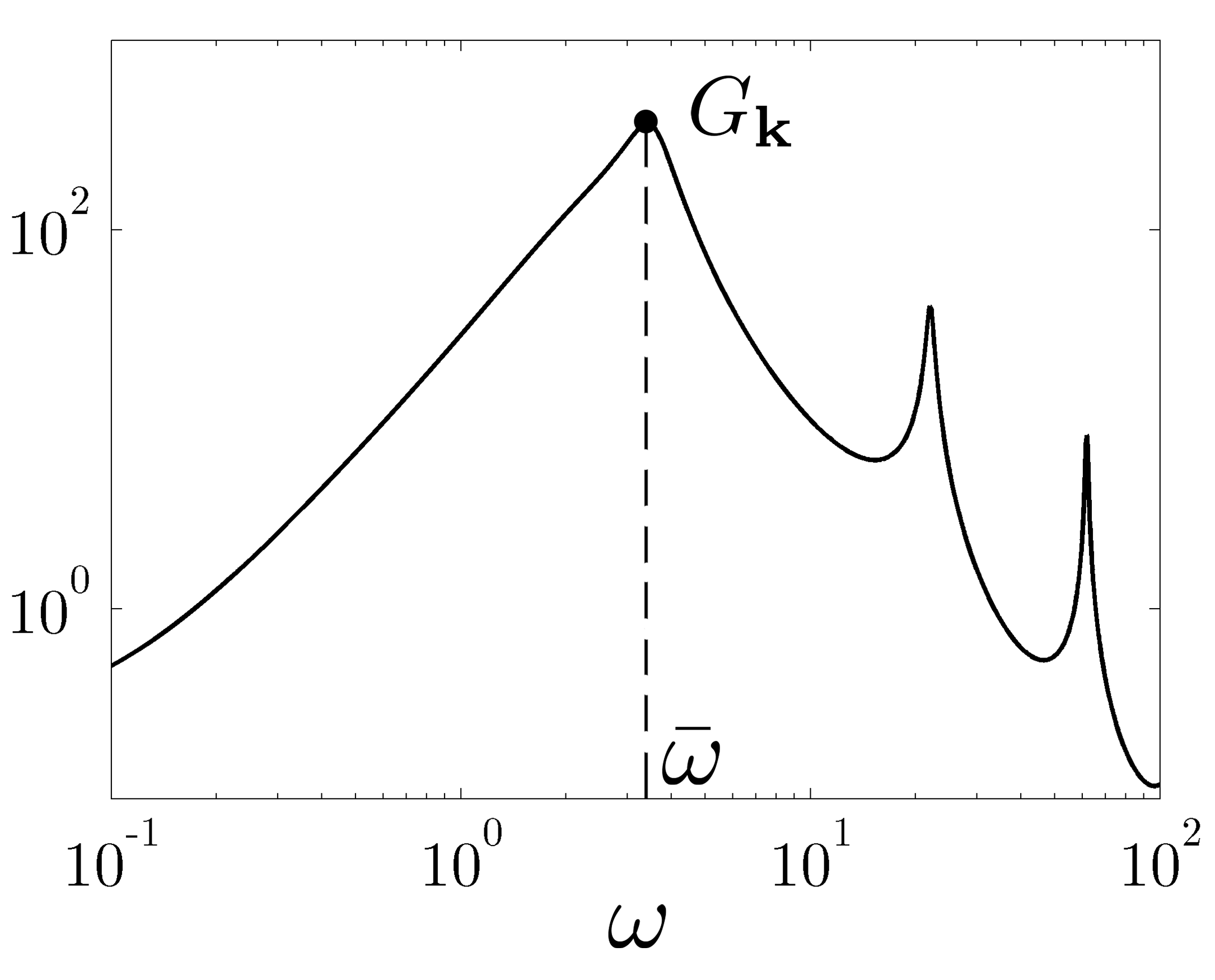}
     {\label{fig.Hinf-Gk}}}
            \end{tabular}   
           &    
   \hspace*{-6.5cm}
    \begin{tabular}{c}
    \subfigure[]{\scalebox{.65}{%_______________________________________________________________________________
%
%   Block diagram of the periodic modification to the original dynamics
%   drawn from Right to Left
%
%   Mihailo Jovanovic, February 17, 2020
%_______________________________________________________________________________
%
% TikZ styles for drawing
%
\input{figures/Tikz_common_styles}
%
%   set a filename for externalization
% \tikzsetnextfilename{clp_2dof_input_pert_config}
%
\noindent
\begin{tikzpicture}[scale=1, auto, >=stealth']

	% output node
	% starting point for uend
	% \node [input, name=uend] {};
	\node[] (end) at (0,0) {};
			
	% \node[] (midpoint1) at ($(uend) + (1.5cm,-1.75cm)$) {};
	 
   % original dynamics
    \node[block, minimum height = 1.5cm, top color=RoyalBlue!20, bottom color=RoyalBlue!20] (plant) at ($(end) + (6cm,0)$) {$\ba{rcl}\dot{\bpsi}_{\bk} (t) & = & A_{\bk} \, \bpsi_{\bk} (t) \, + \, B_{\bk} \, \bd_{\bk} (t) \\[0.075cm] \bxi_{\bk} (t) & = & C_{\bk} \, \bpsi_{\bk} (t) \ea$};
    
    \node[] (uend) at ($(plant.west) - (1.25cm,0cm)$) {};
    
    \node[] (uend1) at ($(plant.west) - (2.5cm,0cm)$) {};
    
    % periodic feedback
    \node[block, dashed, minimum height = 1cm, top color=red!20, bottom color=red!20] (Gamma) at ($(end) + (6cm,-1.5cm)$) {$\tc{red}{\Gamma_{\bk}}$};

	\node[] (ubegin) at ($(plant.east) + (1.25cm,0cm)$) {};
	
	\node[] (Gend) at ($(uend.center) - (0cm,1.5cm)$) {};

	\node[] (Gbegin) at ($(ubegin.center) - (0cm,1.5cm)$) {};

	% input nodes
	% inputs to plant
	% input dbegin to block plant
    	\draw [connector] (ubegin.center) -- node [midway, above] {$ \bd_\bk  (t)$} (plant.east);
		
	% connect plant with uenddown
         \draw [connector] (plant.west) -- node [midway, above] {$ \bxi_\bk  (t)$} (uend1.center);
         
         % connect plant with uenddown
         \draw [line, dashed] (uend.center) -- (Gend.center);

        \draw [connector, dashed] (Gend.center) -- (Gamma.west);
                 
        \draw [line, dashed] (Gamma.east) -- (Gbegin.center);
        
        \draw [line, dashed] (Gbegin.center) -- (ubegin.center);
        
        	 \node [below = -0.07cm of Gamma](extra){\tc{red}{$\ba{c} \mbox{\bf modeling uncertainty} \\
	 \mbox{\tc{black}{\bf (can be nonlinear or time-varying)}} \ea$}}; 
	 
	  \node [above = 0.cm of plant](extra){\tc{RoyalBlue}{$\ba{c} \mbox{\bf nominal linearized dynamics} \ea$}};  %

%         % connect plant with uenddown
%         \draw [line] (midpoint1.center) -- (midpoint2.center);
%
%	% connect midpoint1 with periodic block
%	 \draw [connector] (midpoint2.center) -- (periodic.west);
%	 
%	  \draw [line] (periodic.east) -- (midpoint3.center);
%
%         \draw [line] (midpoint3.center) -- (midpoint4.center);
%
%	 \draw [connector] (midpoint4.center) -- (plantdowneast.center);
%	 
%	 \node [below = -0.07cm of periodic](extra){\tc{red}{$\ba{c} \mbox{\bf sensor-less feedback} \ea$}};  %

\end{tikzpicture}
%_______________________________________________________________________________
} 
    {\label{fig.Hinf-bd}}}
    \end{tabular}
    \end{tabular}
     }
    \caption{(a) The $H_\infty$ norm is determined by the peak value of $\sigma_{\max} (\Tk (\mri \omega) )$ over $\omega$. 
    (b) A large $H_\infty$ norm of the linearized dynamics signals low robustness margins: modeling imperfections, captured by the operator $\Gamma_{\bk}$, with the $H_\infty$ norm $1/\sqrt{\Gk}$ can trigger instability of $\Ak + \Bk \Gamma_{\bk} \Ck$. This interpretation is related to the pseudo-spectra of linear operators~\citep{treemb05}.
}
    \label{fig.Hinf}
    \vspace*{-0.3cm}
    \end{figure}

	For a harmonic input $\bd_\bk (t) = \hat{\bd}_\bk (\mri \omega) \mathrm{e}^{\mri \omega t}$ with $\hat{\bd}_\bk (\mri \omega) = \hat{\bv}_{\bk,j} (\omega)$, where $\hat{\bv}_{\bk,j} (\omega)$ is the $j$th left singular function of $T_\bk (\mri \omega)$, the steady-state output $\bxi_\bk (t) = \hat{\bxi}_\bk (\mri \omega) \mathrm{e}^{\mri \omega t}$ of System~\ref{eq.lnse1} is in the direction of the $j$th right singular function, $\hat{\bxi}_\bk (\mri \omega) = \sigma_{\bk,j} (\omega) \hat{\bu}_{\bk,j} (\omega)$, and its energy is given by
	$
	\| \hat{\bxi}_\bk (\mri \omega) \|_2^2 
	\DefinedAs
	\langle {\hat{\bxi}_\bk (\mri \omega)},{\hat{\bxi}_\bk (\mri \omega)} \rangle = \sigma_{\bk,j}^2 (\omega). 
	$
The principal singular value, $\sigma_{\bk,1} (\omega) \DefinedAs \sigma_{\max} (\Tk (\mri \omega))$, determines the largest amplification at any ($\bk,\omega$) and the smallest upper bound over $\omega$ determines the $H_\infty$ norm of \mbox{System~\ref{eq.lnse1}~\citep{zhodoyglo96},} 
	$
		\Gk
		\DefinedAs  
		\sup_{\omega} \sigma_{\bk,1}^{2} ( \omega ).
	$
	This measure of input-output amplification has several appealing interpretations for any $\bk$.
	\vspace*{-4ex}
	\begin{itemize}
		\item[(a)] The $H_\infty$ norm represents the worst-case amplification of harmonic (in homogeneous directions and in time) deterministic (in inhomogeneous directions) inputs. This worst-case input-output gain is obtained by maximizing over spatial profiles (largest singular value of $\Tk$) and temporal frequency (supremum over $\omega$); see {\bf Figure~\ref{fig.Hinf-Gk}}.
		
		\item[(b)] The $H_\infty$ norm determines the induced gain from finite energy inputs to outputs,
		$
		 \Gk
	          = 
	    \sup_{\Ein \, \leq \, 1} 
	    ( \Eout / \Ein ),
		$
where $\Ein$ and $\Eout$ denote the $\bk$-parameterized energy of input and output, e.g.,
	    $
	    \Ein
	    \DefinedAs 
	    \int_{0}^{\infty} \| \bd_\bk (t) \|_2^2 \, \mrd t,
	    $
	    with 
	    $
	    \| \bd_\bk (t) \|_2^2
	    =
	    \inner{\bd_\bk (t)}{\bd_\bk (t)}.
	    $
For a unit-energy input $\bd_\bk (t)$ to stable System~\ref{eq.lnse1}, $\Gk$ quantifies the largest possible energy of the output $\bxi_\bk (t)$ across the \mbox{spatial wavenumber $\bk$.}

		\item[(c)] The $H_\infty$ norm quantifies robustness to modeling imperfections; see {\bf Figure~\ref{fig.Hinf-bd}}.
	\end{itemize}
	
	\vspace*{-2ex}
\begin{textbox}[h]
\section{FREQUENCY RESPONSE OPERATOR}
	\vspace*{-7ex}
	\subsubsection{Time-invariant systems} 
	The natural response of a stable Linear Time-Invariant (LTI) System~\ref{eq.lnse1} asymptotically decays to zero. 
	The frequency response operator determines the steady-state response to harmonic inputs with frequency $\omega$ and is obtained by evaluating the transfer function along the imaginary axis, 
	\be
	\Tk (\mri \omega)
	\; \DefinedAs \;
	\Tk (s) \Big|_{s \, = \, \mri \omega}
	\; = \;
	\Ck
    	\left(
    	\mri \omega I \, - \, \Ak
    	\right)^{-1}  
	\!   	
	\Bk.
	\label{eq.fr}
	\tag{FR}
	\ee
For $\bd_\bk (t) = \hat{\bd}_\bk (\mri \omega) \mathrm{e}^{\mri \omega t}$, the steady-state response of a stable System~\ref{eq.lnse1} is harmonic with the same frequency but with different amplitude and phase, i.e., $\bxi_\bk (t) = \hat{\bxi}_\bk (\mri \omega) \mathrm{e}^{\mri  \omega t}$. The frequency response $\Tk (\mri \omega)$ is an operator (in inhomogeneous spatial directions) that maps a spatial input profile $\hat{\bd}_\bk (\mri \omega)$ into the output~$\hat{\bxi}_\bk (\mri \omega)$, 
	$
	\hat{\bxi}_\bk (\mri \omega)
	= 
	T_\bk (\mri \omega) \hat{\bd}_\bk (\mri \omega),
	$		
thereby determining how amplitude and phase change across \mbox{$\bk$ and~$\omega$.} 

\vspace*{-2ex}
\subsubsection{Time-periodic systems} If the operator $\Ak$ in Equation~\ref{eq.lnse1} has time-periodic coefficients, i.e., $\Ak (t) = \Ak (t + 2\pi/\omega_t)$, the steady-state response to a harmonic input with frequency $\omega$ contains an infinite number of harmonics separated by integer multiplies of $\omega_t$, i.e., $\omega + n \omega_t$, $n \in \bbZ$.  The proper normal modes for frequency response analysis are no longer purely harmonic, $\mre^{\mri \omega t}$. Rather, they are determined by the {\em Bloch waves\/}~\citep{odekel64}, i.e., by a product of $\mre^{\mri \theta t}$ and the $2 \pi/\omega_t$ \mbox{periodic function in $t$,} 
	\be
    \bd_\bk (t)
    \; = \;
    \ds{\sum_{n \, = \, - \infty}^{\infty}}
    \hat{\bd}_{\bk,n} (\mri \theta)
    \,
    \mre^{\mri (\theta \, + \, n \omega_t) t},
    ~~
    \theta \in [0, \omega_t),
    \label{eq.BW}
    \tag{BW}
    \ee
where $\theta$ is the angular frequency and $\theta = 0$ and $\theta = \omega_t/2$ identify the fundamental and subharmonic modes, respectively. The steady-state response of a stable linear time-periodic system to a Bloch wave input~\ref{eq.BW} is also a Bloch wave, 
	$
    \bxi_\bk (t)
    = 
   \sum_{n}
    \hat{\bxi}_{\bk,n} (\mri \theta)
    \,
    \mre^{\mri (\theta + n \omega_t) t},
    $
and, for any ($\bk,\theta$), the frequency response operator $\Tk (\mri \theta)$ maps 
	$
	\hat{\bd}_\bk (\mri \theta) 
	 \DefinedAs
    	\col \, \{\hat{\bd}_{\bk,n} (\mri \theta)\}_{n \, \in \, \bbZ}
    	$
to 
	$
	\hat{\bxi}_\bk (\mri \theta) 
	 \DefinedAs
    	\col \, \{\hat{\bxi}_{\bk,n} (\mri \theta)\}_{n \, \in \, \bbZ},
    	$
	i.e., 
	$
	\hat{\bxi}_\bk (\mri \theta) 
	= 
	\Tk (\mri \theta)
	\hat{\bd}_\bk (\mri \theta).
	$
If the operators $\Bk$ and $\Ck$ in Equation~\ref{eq.lnse1} are time-invariant, for a system with $\Ak (t) = \sum_{m} A_{\bk,m} \mre^{\mri m \omega_t t}$ we have
	\beq
	\Tk (\mri \theta)
	\; = \;
	\cCk
	\left(
	{\cal E} (\mri \theta)	
	\, - \,
	\cAk
	\right)^{-1}
	\cBk,
	\label{eq.fr-ltp}
	\tag{FR1}
	\eeq
where ${\cal E} (\mri \theta) \DefinedAs \diag \{ \mri (\theta + n \omega_t) I\}_{n \, \in \, \bbZ}$, $\cBk$ and $\cCk$ are the block-diagonal operators with $\Bk$ and $\Ck$ on the main diagonal, and $\cAk \DefinedAs \toep \, \{ \ldots, A_{\bk,1}, A_{\bk,0}, A_{\bk,-1}, \ldots \}$ is the block-Toeplitz operator~\citep{jovPOF08}. 
	\vspace*{-0.35cm}
\end{textbox}

	%=========
	% Figure 3  %
	%=========

	\vspace*{-2ex}
\subsection{Amplification of stochastic inputs}
	\label{sec.H2}
	
A common criticism of transient growth and resolvent analyses is difficulty of implementing the worst-case initial conditions or inputs in the lab. An alternative approach introduces a random excitation to the NS equations that can account for background noise.   It identifies almost identical dominant flow structures and opens the door to turbulence modeling. 

\begin{figure}[h!]
    \centering
    \vspace*{-0.15cm}
    {
    \begin{tabular}{cccc}
     \hspace{-0.5cm}
	\begin{tabular}{c}
		\vspace{0.5cm}
		\normalsize{\rotatebox{90}{$(1/t) \int_{0}^{t} \| \bv_\bk (\tau) \|^2 \, \mrd \tau$}}
	\end{tabular}
	&
    \hspace*{-5.5cm}
    \begin{tabular}{c}
     \includegraphics[width=0.3\textwidth]{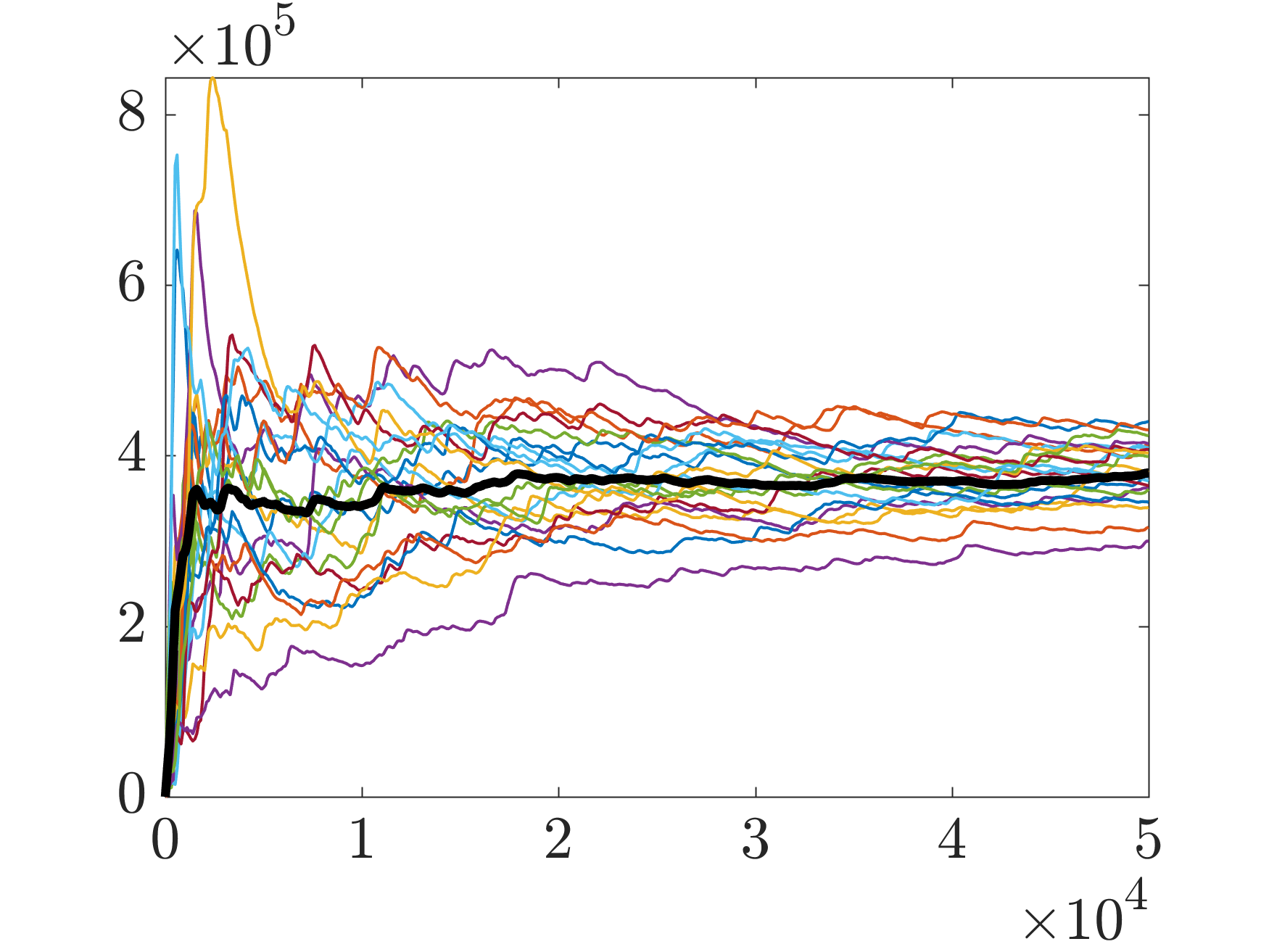}
           \\[-0.1cm]
           $t$
%           \\[-0.25cm]
%            \subfigure[streamwise]
%   {\label{fig.H2u}}
           \end{tabular}   
           &    
   \hspace*{-10.cm}
   \begin{tabular}{c}
   \includegraphics[width=0.3\textwidth]{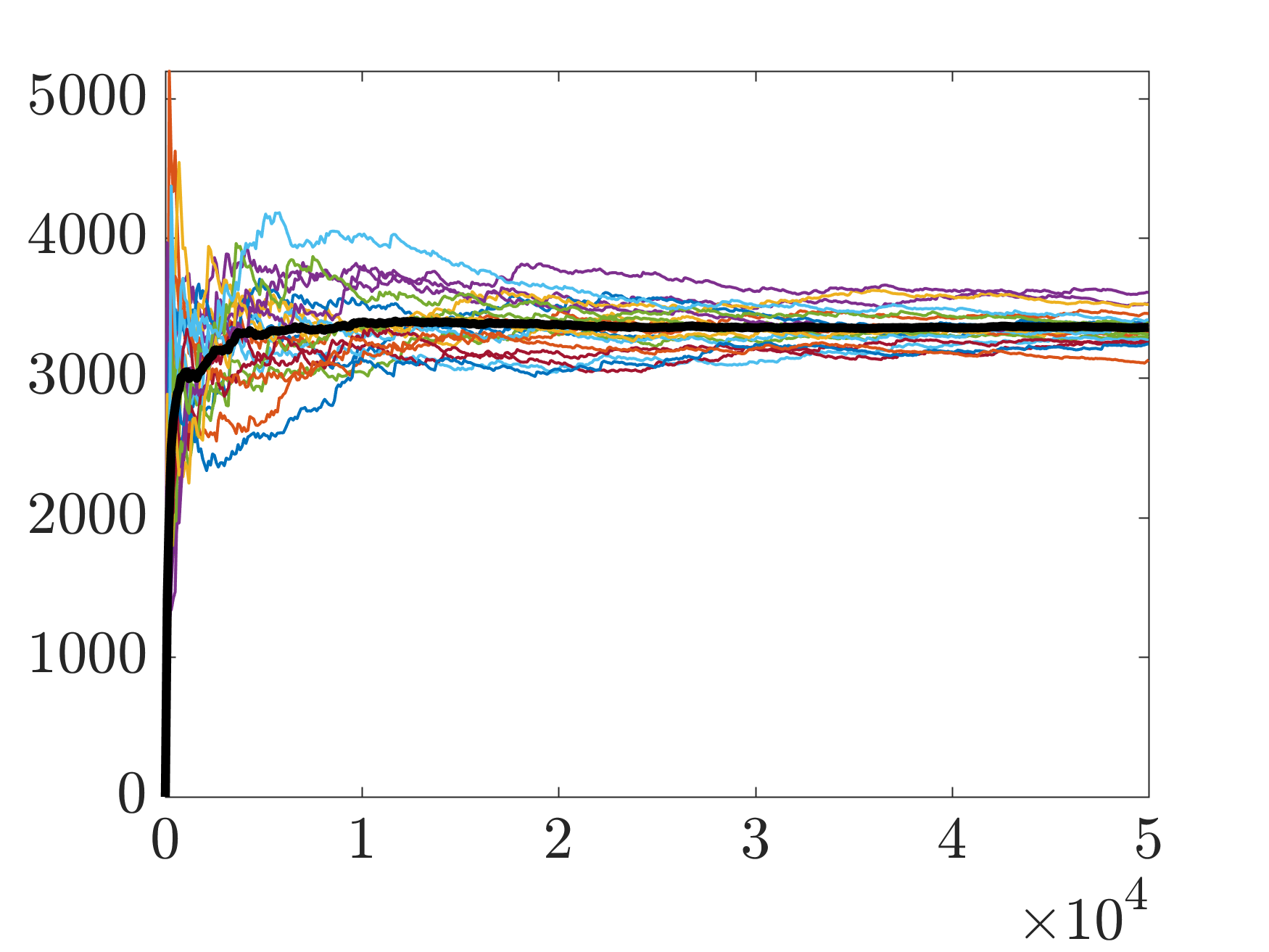}            
           \\[-0.1cm]
           $t$
%           \\[-0.25cm]
%            \subfigure[wall-normal]
%   {\label{fig.H2v}}
           \end{tabular}
           &    
   \hspace*{-10.cm}
   \begin{tabular}{c}
    \includegraphics[width=0.3\textwidth]{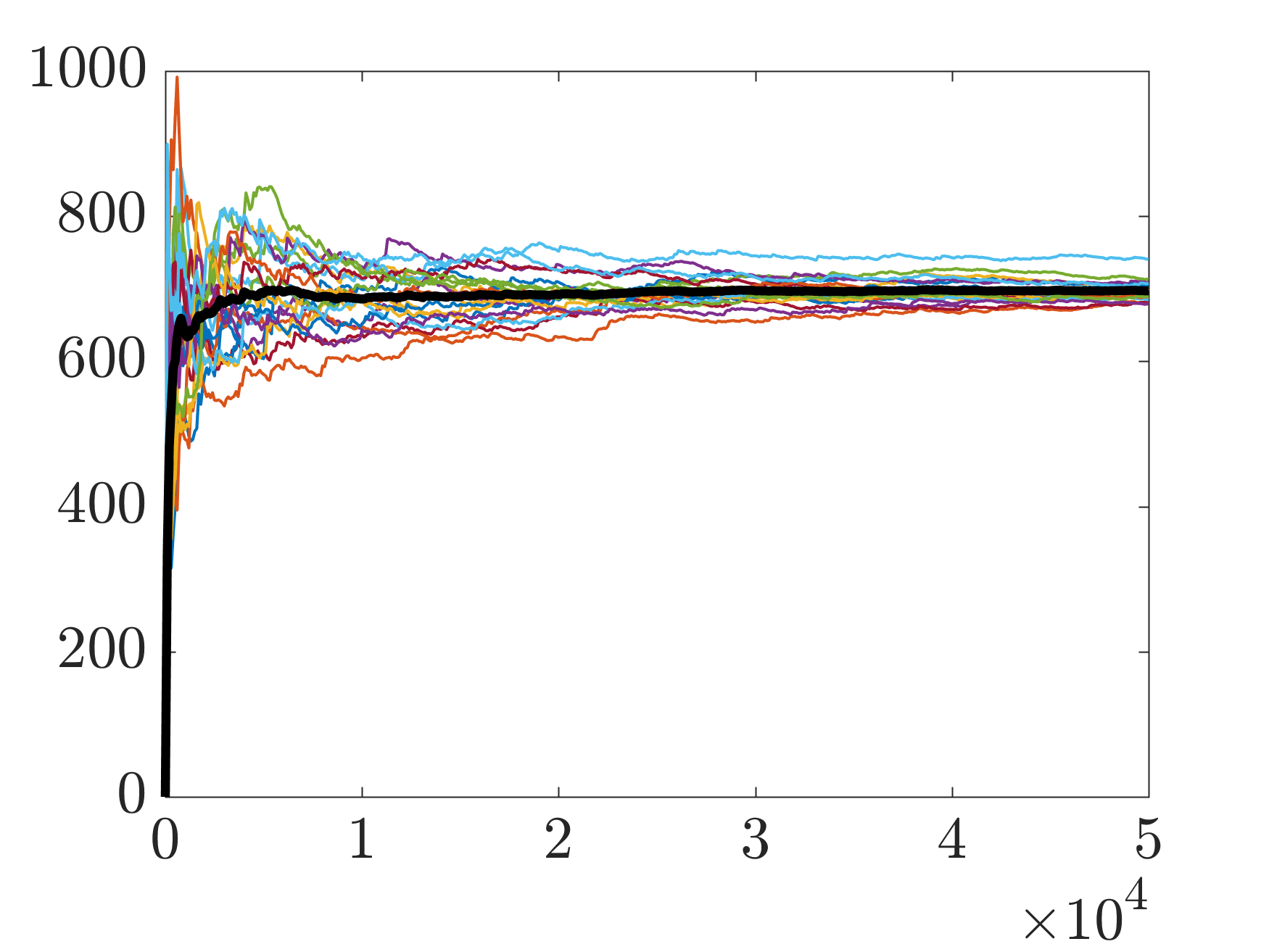}
           \\[-0.1cm]
           $t$
%           \\[-0.25cm]
%            \subfigure[spanwise]
%  {\label{fig.H2w}}
           \end{tabular}
	\end{tabular}
    }
   \caption{The variance of velocity fluctuations, $(1/t) \int_{0}^{t} \| \bv_\bk (\tau) \|_2^2 \, \mrd \tau$, for $20$ realizations of stochastic forcing to the linearized NS equations in Poiseuille flow with $Re = 2000$, $\bk = (0, 1.78)$; $(1, 1)$; and $(1, 0)$. The variance averaged over all simulations is shown by a thick black line.}
\label{fig.ssim}
	\vspace*{-0.5cm}
    \end{figure}

{\em Control-theoretic tools can be utilized to exploit the structure of the linearized Model~\ref{eq.lnse1}, avoid costly stochastic simulations, and offer insight into amplification mechanisms.} For $20$ realizations of persistent channel-wide temporally and spatially uncorrelated stochastic input $\bd_\bk (t)$ to Equation~\ref{eq.lnse1}, {\bf Figure~\ref{fig.ssim}} shows the variance of the velocity fluctuation vector $\bv_\bk (t) \DefinedAs ( u_{\bk} (t) ,v_{\bk} (t),w_{\bk} (t))$ in Poiseuille flow with $Re = 2000$. Although individual simulations display different responses, their average (marked by a thick black line) reaches the steady-state limit. In the absence of modal instability, viscosity asymptotically dissipates natural responses but a persistent excitation source maintains fluctuations for all times. 
	%=========
	% Figure 4  %
	%=========

\vspace*{-3ex}
	\subsubsection*{\tc{dred}{The Reynolds-Orr equation.}}
In channel flow with stochastic forcing, the kinetic energy $\Ek (t) \DefinedAs \mathbf{E} (\inner{\bv_\bk(t)}{\bv_\bk(t)})$ of fluctuations $\bv_\bk(t)$ around ($U(y),0,0$) obeys,
    \beq
    \dfrac{1}{2}
    \dfrac{\mrd {E}_\bk (t)}{\mrd t}
    \; = \;
    % 2
    \mathbf{E}
    \left(
    % (1/Re)
    \dfrac{1}{Re}
    \inprod{\bv_{\bk} (t)}{\Delta \bv_{\bk} (t)}
    \, - \,
    \inprod{u_{\bk} (t)}{U' v_{\bk} (t)}
    \, + \,
    \inprod{\bv_{\bk} (t)}{\bd_{\bk} (t)}
    \right).
    \label{eq.ro}
    \non
    \eeq
Here, $\bE$ is the expectation operator, $\inner{\cdot}{\cdot}$ is the $L_2 [-1,1]$ inner product, $U'(y) \DefinedAs \mrd U(y)/ \mrd y$, and the terms on the right-hand side denote the viscous energy dissipation, the energy exchange with the base shear, and the work done by the body forces, respectively. The nonlinear terms in the NS equations are conservative and the Reynolds-Orr equation takes the same form for nonlinear and linearized dynamics~\citep{schhen01}. Since it is driven by the terms that need to be determined by solving the equations for flow fluctuations, it is not in the form which allows for direct determination of its solution. For the linearized NS equations, both the kinetic energy and the terms on the right-hand side of the Reynolds-Orr equation can be computed using the solution to differential Lyapunov equation~\ref{eq.DL}, that we present below. This avoids the need for costly stochastic simulations and provides an alternative way for solving an important equation in fluid mechanics.

 \vspace*{-3ex}
	\subsubsection*{\tc{dred}{Time-invariant systems.}}
Let System~\ref{eq.lnse1}, with the output $\bxi_\bk (t) = \bv_\bk (t)$, be driven by a stochastic input $\bd_\bk (t)$ with the spectral density $\Sin$. The spectral density operator $\Sout = \Tk (\mri \omega) \Sin \Tk^\dagger (\mri \omega)$ quantifies the two-point correlations of $\bv_\bk (t)$ across the wavenumber $\bk$ and the frequency $\omega$, where $\Tk (\mri \omega)$ is the frequency response given in Equation~\ref{eq.fr}. The inverse Fourier transform of $\Sout$ yields the lagged \mbox{covariance operator,}
\beq
\label{eq.correlation-tensor}
	\Pk (\tau)
	\; \DefinedAs \;
	\lim_{t \, \to \, \infty} 
	\bE \left( \bv_\bk (t) \otimes \bv_\bk (t + \tau) \right)
	\; = \;
	\dfrac{1}{2 \pi}
	\,
	\ds{\int^{+\infty}_{-\infty}} \Sout \, \mre^{\mri \omega \tau} \, \mrd \omega,
	\non
\eeq
where $\otimes$ denotes the tensor product. Furthermore, the integration of $\Sout$ over $\omega$ yields the steady-state two-point correlation (i.e., covariance) operator $\Vk$ of the output $\bv_\bk (t)$,
\beq
\label{eq.relation-V-S}
	\Vk
	\; \DefinedAs \;
	\Pk (0)
	\; = \;
	\lim\limits_{t \, \to \, \infty}
	\Vk (t)
	\;=\;
	\dfrac{1}{2 \pi}
	\,
	\ds{\int^{+\infty}_{-\infty}} \Sout \, \mrd \omega,
	\non
\eeq
where $\Vk (t) \DefinedAs \bE \left( \bv_\bk (t) \otimes \bv_\bk (t) \right)$ is the time-dependent covariance operator of velocity fluctuations. For System~\ref{eq.lnse1}, $\Vk (t) = \Ck \Xk (t) \Ck^\dagger$, where $\Xk (t) \DefinedAs \bE \left( \bpsi_\bk (t) \otimes \bpsi_\bk (t) \right)$ is the covariance operator of the state $\bpsi_\bk (t)$ and $\Ck^\dagger$ is the adjoint of the operator~$\Ck$. In channel flow, for any $\bk$, $\Vk$ is an operator in the wall-normal direction, 
	$
	\mathbf{g}_\bk (y_1)
	\DefinedAs 
	\left[ \Vk \, \mathbf{f}_\bk (\cdot) \right] (y_1), 
	$
whose kernel representation determines all stationary two-point correlations of $\bv_\bk (t)$,
	\beq
	\mathbf{g}_\bk (y_1)
	\; = \;
	\int_{-1}^1
	\tikz[baseline]{
            \node[fill=RoyalBlue!20,anchor=base] (t1)
            {$
            \Vk^{\mathrm{ker}} (y_1,y_2)
	$};
            }
            \;
	\mathbf{f}_\bk (y_2) \, \mrd y_2
	\; = \;
	\int_{-1}^1
	\tikz[baseline]{
            \node[fill=RoyalBlue!20,anchor=base] (t1)
            {$
            \lim\limits_{t \, \to \, \infty} \bE \left( \bv_\bk (y_1,t) \bv_\bk^* (y_2,t) \right)
	    $};
            }
            \;
	\mathbf{f}_\bk (y_2) \, \mrd y_2.
	\non
	\eeq 
One- and two-point correlations in $y$ are obtained for $y_1 = y_2$ and $y_1 \neq y_2$, respectively; $\Vk^{\mathrm{ker}} (y_1,y_2)$ determines the two-point spectral density tensor and its inverse Fourier transform gives the two-point correlation tensor in $x$ and $z$~\citep{moimos89}.	

\begin{textbox}
	\vspace*{-1ex}
\section{LYAPUNOV EQUATION: TWO-POINT CORRELATIONS OF LINEAR SYSTEMS}
	 The Lyapunov equation can be used to propagate two-point correlations of the white stochastic input $\bd (t)$ into colored statistics of the state $\bpsi (t)$ of a linear systems~\citep[Appendix~A]{bamdah01}. Herein, we derive the Lyapunov equation for a finite-dimensional discrete-time LTI system,
	\beq
	\bpsi (t+1)
	\; = \;
	A \, \bpsi (t) 
	\; + \:
	B \, \bd (t),
	\label{eq.ss-dt}
	\tag{DT}
	\eeq
where time $t$ is a non-negative integer and $A$, $B$ are constant matrices. The derivation for continuous-time systems is standard~\citep[Chapter~1.11]{kwasiv72} but is more involved and less intuitive. Let 
	$
	X (t)
	 \DefinedAs 
	\bE 
	\left(  \bpsi (t) \bpsi^* (t) \right)
	$
be the covariance matrix of the state at time $t$, where $\bE$ is the expectation operator and $\bpsi^* (t)$ is the complex-conjugate transpose of the vector $\bpsi (t)$. Then, Equation~\ref{eq.ss-dt} can be used to write, 
	\beq
	\ba{rcl}
	\tcr{X (t + 1)}
	& \; = \;\, &
	\bE 
	\,
	\big(
	\left( 
	A \bpsi (t) 
	\, + \,
	B \bd (t) 
	\right)
	\left(
	\bpsi^* (t) A^* 
	\, + \,
	\bd^* (t) B^* 
	\right) 
	\big)
	\\[0.1cm]
	& \; = \;\, &
	A
	\;\! 
	\tcr{
	\bE \left( 
	\bpsi (t) 
	\bpsi^* (t)
	\right)}
	A^*
	\, + \,
	B
	\,
	\tcg{
	\bE \left( 
	\bd (t) 
	\bpsi^* (t)
	\right)}
	A^*
	\, + \,
	A
	\;\! 
	\tcg{
	\bE \left( 
	\bpsi (t) 
	\bd^* (t)
	\right)}
	B^*
	\, + \,
	B
	\,
	\tcb{
	\bE \left( 
	\bd (t) 
	\bd^* (t)
	\right)}
	B^*.
	\ea
	\label{eq.dlyap-derive}
	\eeq
If the stochastic input is white-in-time with the covariance matrix $W$, i.e., 
	$
	\bE \left( \bd (t) \bd^* (\tau) \right)
	=
	W 
	\delta (t - \tau),
	$
where $\delta$ is the Kronecker delta, the cross terms in Equation~\ref{eq.dlyap-derive} disappear and we obtain the Lyapunov~equation, 
	\beq
	\tcr{X (t + 1)}
	\; = \;
	A
	\,
	\tcr{X (t)}
	A^*
	\; + \;
	B
	\, 
	\tcb{W}
	B^*,
	~~
	\tcr{X (0)}
	\; = \;
	\tcr{X_0}.
	\label{eq.dlyap+}
	\eeq
If the matrices $A$ and $B$ in Equation~\ref{eq.ss-dt} as well as the matrices $W$ and $X_0$ are known, this deterministic equation can be propagated forward in time to obtain the covariance matrix $X (t)$. Even though the above derivation holds irrespective of stability properties of System~\ref{eq.ss-dt}, the steady-state limit, $X \DefinedAs \lim_{t \, \to \, \infty} X (t)$, only exists for stable systems. In this case, Equation~\ref{eq.dlyap+} converges to the \mbox{algebraic Lyapunov~equation,}
	\beq
	A
	\,
	\tcr{X}
	A^*
	\; - \;
	\tcr{X}
	\; = \;
	-
	B
	\, 
	\tcb{W}
	B^*,
	\label{eq.dlyap}
	\eeq
which is linear in $X$ and it is typically used to compute the stationary covariance matrix $X$ for given $A$, $B$, and $W$. For colored in-time stochastic inputs $\bd(t)$, the cross terms in Equation~\ref{eq.dlyap-derive} are non-zero and introducing the matrix 
	$
	H (t)
	\DefinedAs
	A 
	\,
	\bE \left( 
	\bpsi (t) 
	\bd^* (t)
	\right)
	+
	\tfrac{1}{2}
	B
	\,
	\bE \left( 
	\bd (t) 
	\bd^* (t)
	\right)
	$
 in Equation~\ref{eq.dlyap-derive} allows us to write it as,
	\beq
	\tcr{X (t + 1)}
	\; = \;
	A
	\;\! 
	\tcr{
	X (t)}
	A^*
	\, + \,
	B
	\tcb{H^* (t)}
	\, + \, 
	\tcb{H (t)} B^*,
	~~
	\tcr{X (0)}
	\; = \;
	\tcr{X_0}.
	\label{eq.dlyap+H}
	\eeq
For stable System~\ref{eq.ss-dt}, Equation~\ref{eq.dlyap+H} converges asymptotically to the algebraic Lyapunov-like equation,
	\beq
	A
	\,
	\tcr{X}
	A^*
	\; - \;
	\tcr{X}
	\; = \;
	-
	\left(
	B
	\tcb{H^*}
	\, + \, 
	\tcb{H} B^*
	\right),
	\label{eq.dlyapH}
	\eeq
where $H \DefinedAs \lim_{t \, \to \, \infty} H (t)$. For continuous-time systems, Equation~\ref{eq.dlyap+} takes the form of the differential Lyapunov equation~\ref{eq.DL} which, for stable systems, converges to the algebraic Lyapunov equation~\ref{eq.AL}. While Equations~\ref{eq.dlyap-derive},~\ref{eq.dlyap+}, and~\ref{eq.dlyap+H} also hold for systems in which the matrices $A (t)$ and $B (t)$ depend on time, their steady-state limits may not be~well-defined. Finally, for infinite-dimensional systems, the complex-conjugate transpose of a matrix, e.g.,~$A^*$, should be replaced with an adjoint of an operator, e.g.,~$A^\dagger$. 
	\vspace*{-0.35ex}
\end{textbox}

 \vspace*{-2ex}
	\subsubsection*{\tc{dred}{Lyapunov equation.}} For a zero-mean temporally uncorrelated input $\bd_\bk (t)$ with the covariance operator $\Wk$, i.e.,
	$
	\bE ( \bd_\bk (t) ) 
	= 
	0,
	$
	$
	\bE ( \bd_\bk (t) \otimes \bd_\bk (\tau) ) 
	= 
	\Wk \delta (t - \tau), 
	$
the input spectral density $\Sin$ is constant across $\omega$, i.e., $\Sin = \Wk$. In this case, as described in the sidebar LYAPUNOV EQUATION: TWO-POINT CORRELATIONS, the covariance operator $\Xk (t)$ of the state $\bpsi_\bk (t)$ in System~\ref{eq.lnse1} satisfies the differential Lyapunov~equation,
	\beq
	\tikz[baseline]{
            \node[fill=red!20,anchor=base] (t1)
            {$
            \dfrac{\mrd {\Xk (t)}}{\mrd t}
		\; = \;
		\Ak  \;\! {\Xk (t) } 
		\; + \; 
		{\Xk (t)} \;\! \Ak^\dagger 
    	 	\; + \;  
	 	\Bk \;\! {\Wk} \;\! \Bk^\dagger.
	$};
            }
%	\dfrac{\mrd {\Xk (t)}}{\mrd t}
%	\; = \;
%	\Ak  \;\! {\Xk (t) } 
%	\; + \; 
%	{\Xk (t)} \;\! \Ak^\dagger 
%    	 \; + \;  
%	 \Bk \;\! {\Wk} \;\! \Bk^\dagger.
	\label{eq.DL}
	\tag{DL}
	\eeq
For System~\ref{eq.lnse1} with the input covariance $\Wk$ and the initial condition $\Xk (0)$, this {\em operator-valued differential equation\/} can be used to compute $\Xk (t)$ and determine energy of fluctuations via 
	$\Ek (t) 
	= 
	\trace 
	\,
	(
	\Ck \Xk (t) \Ck^\dagger
	).
	$
For linearly unstable flows, the steady-state limit of $\Xk (t)$ is either unbounded or it does not exist. However, the solution of Equation~\ref{eq.DL} can still be computed, e.g., by forward marching in time or via \mbox{the following formula,}
	 \beq
    	\Xk (t)
    \; = \;
    \mre^{ \Ak t}
    \Xk (0)
    \mre^{ \Ak^\dagger t}
    \; + \;
    \obt{I}{~0}
    \exp
    \left(
    \left[
    \ba{cc}
    \Ak & ~\Bk \;\! {\Wk} \;\! \Bk^\dagger
    \\[0.05cm]
    0 & ~ -\Ak^\dagger 
    \ea
    \right]
    t
    \right)
    \left[
    \ba{c}
    0
    \\[0.05cm]
    I
    \ea
    \right]
    \mre^{ \Ak^\dagger t}.
    \non
    \eeq
In the absence of modal instability, $\Xk \DefinedAs \lim_{t \, \to \, \infty} \Xk (t)$ is well-defined and the steady-state limit of Equation~\ref{eq.DL} is given by,
	\beq
	\tikz[baseline]{
            \node[fill=red!20,anchor=base] (t1)
            {$
            \Ak
	\;\!
	{\Xk}
	\; + \; 
	{\Xk}
	\;\!
	\Ak^\dagger
	\; = \;
	-
	\Bk
	\;\!
	{\Wk}
	\;\!
	\Bk^\dagger.
	$};
            }
%	\Ak
%	\;\!
%	{\Xk}
%	\; + \; 
%	{\Xk}
%	\;\!
%	\Ak^\dagger
%	\; + \;
%	\Bk
%	\;\!
%	{\Wk}
%	\;\!
%	\Bk^\dagger
%	\; = \;
%	0.
	\label{eq.AL}
	\tag{AL}
	\eeq
In this case, $\Xk (t)$ can be computed from the solution $\Xk$ to the algebraic Lyapunov equation~\ref{eq.AL} and the initial condition $\Xk (0)$ via  
	$
    \Xk (t)
    = 
    \Xk 
    - 
    \mre^{ \Ak t}
    \!
    \left(
    \Xk 
    - 
    \Xk (0) 
    \right)
    \mre^{ \Ak^\dagger t},
   $	 
%	\beq
%    \Xk (t)
%    \; = \;
%    \Xk 
%    \; - \;
%    \mre^{ \Ak t}
%    \!
%    \left(
%    \Xk 
%    \, - \,
%    \Xk (0) 
%    \right)
%    \mre^{ \Ak^\dagger t},
%    \non
%    \eeq 		
and the steady-state limit of $\Ek (t)$ determines the energy amplification  
	$
	\Ek
	\DefinedAs
	\lim_{t \, \to \, \infty}
	\Ek (t) 
	= 
	\trace 
	\,
	(
	\Ck \Xk \Ck^\dagger
	)
	$; see the sidebar POWER SPECTRAL DENSITY AND ENERGY AMPLIFICATION. Finally, for colored-in-time input $\bd_\bk (t)$, $\Xk$ satisfies, 
	\beq
	\tikz[baseline]{
            \node[fill=red!20,anchor=base] (t1)
            {$
           \Ak
	\;\!
	{\Xk}
	\; + \; 
	{\Xk}
	\;\!
	\Ak^\dagger
	\; = \;
	-
	(
	\Bk
	\;\!
	\Hk^\dagger
	\; + \;
	\Hk 
	\;\!
	\Bk^\dagger
	),
	$};
            }
%	\Ak
%	\;\!
%	{\Xk}
%	\; + \; 
%	{\Xk}
%	\;\!
%	\Ak^\dagger
%	\; + \;
%	\Bk
%	\;\!
%	\Hk^\dagger
%	\; + \;
%	\Hk 
%	\;\!
%	\Bk^\dagger
%	\; = \;
%	0,
	\label{eq.ALc}
	\tag{ALc}
	\non
	\eeq
where the operator $\Hk$ determines the stationary cross-correlation between the input $\bd_\bk (t)$ and the state $\bpsi_\bk (t)$ in Equation~\ref{eq.lnse1}~\cite[Appendix B]{zarjovgeoJFM17}. 

Departure from the white-in-time restriction removes sign-definiteness requirement on the right-hand-side in Equation~\ref{eq.AL}: while the operator
	$
	\Bk
	\Wk
	\Bk^\dagger
	$
in Equation~\ref{eq.AL} has non-negative eigenvalues, 
	$
	\Bk
	\Hk^\dagger
	+
	\Hk 
	\Bk^\dagger
	$ 
in Equation~\ref{eq.ALc} is allowed to be sign-indefinite which provides additional flexibility. Furthermore, for a zero-mean white input $\bw_\bk (t)$ with the covariance operator $\Wk$, the stationary covariance operator of $\bpsi_\bk (t)$ in the system
	\begin{align}
 	\label{eq.cascade}
             	% \tbo{\dot{\bpsi}_\bk (t)}{\dot{\bphi}_\bk (t)}
	        \dfrac{\mrd}{\mrd t}  \tbo{\bpsi_\bk (t)}{\bphi_\bk (t)}
            	\; = \; 
            	\tbt{\Ak}{~-\Bk \Kk}{0}{~\Ak \, - \, \Bk \Kk}
            	\tbo{\bpsi_\bk (t)}{\bphi_\bk (t)}
            	\; + \;
            	\tbo{\Bk}{\Bk}
		\bw_\bk (t),
 \end{align}
with
	$
	\Kk
	\DefinedAs
	((1/2) \Wk \Bk^\dagger - \Hk^\dagger) \Xk^{-1},
	$ 
is given by $\Xk$~\citep{zarjovgeoJFM17}. The $\bphi_\bk$-subsystem in Equation~\ref{eq.cascade} maps the white input $\bw_\bk (t)$ to the colored input $\bd_\bk (t)$ in System~\ref{eq.lnse1} such that
	$
	\Xk 
	=
	\lim_{t \, \to \, \infty}
	\bE \left( \bpsi_\bk (t) \otimes \bpsi_\bk (t) \right).
	$
Equivalently, the mapping from $\bw_\bk (t)$ to $\bpsi_\bk (t)$ in Equation~\ref{eq.cascade} can be represented via,
	\begin{align}
	\tikz[baseline]{
            \node[fill=red!20,anchor=base] (t1)
            {$
            % \dot{\bpsi}_\bk (t)
            \dfrac{\mrd \bpsi_\bk (t)}{\mrd t}
        	    \; = \;
          ( \Ak \, - \, \tc{red}{\Bk \Kk} ) 
           \,
          \bpsi_\bk (t)
          \; + \; 
          \Bk 
          \bw_\bk (t),
	  $};
            }
%        \dot{\bpsi}_\bk (t)
%        \; = \;
%        ( \Ak \, - \, \tc{red}{\Bk \Kk} ) 
%        \,
%        \bpsi_\bk (t)
%        \; + \; 
%        \Bk 
%        \bw_\bk (t),
        	\label{eq.fbk-lnse1}
	\end{align}
and the algebraic Lyapunov Equation~\ref{eq.AL} can be used to verify that the stationary two-point correlation operator of $\bpsi_\bk (t)$ is indeed given by $\Xk$. Thus, the {\em impact of a colored-in-time input can be interpreted as a state-feedback modification\/} of the operator $\Ak$ in Equation~\ref{eq.lnse1}.

For a stable stochastically-forced System~\ref{eq.lnse1}, algebraic Relation~\ref{eq.ALc} identifies admissible steady-state covariance operators. This fundamental relation was recently utilized for low-complexity stochastic dynamical modeling of turbulent flows~\citep{zarjovgeoJFM17,zarchejovgeoTAC17,zargeojovARC20}.  

\begin{textbox}[h]
	\vspace*{-1ex}
\section{POWER SPECTRAL DENSITY AND ENERGY AMPLIFICATION}
	The power spectral density quantifies the energy of the output $\bxi_\bk (t)$ of stochastically-forced System~\ref{eq.lnse1} across the wavenumber $\bk$ and temporal frequency~$\omega$,
	\beq
	\Piout
	\; \DefinedAs \; 
	\trace
	\, 
	( \Sout )
	\; = \; 
	\trace 
	\left( 
	\Tk (\mri \omega) \Sin \Tk^\dagger (\mri \omega)
	\right),
%	\; = \;
%	\sum_{j \, = \, 1}^{\infty}
%	\lambda_j
%	\! 
%	\left( 
%	\Tk (\mri \omega) \Sin \Tk^\dagger (\mri \omega)
%	\right).
	\non
	\eeq
where $\Tk (\mri \omega)$ is the frequency response and $\Sin$ is the spectral density of $\bd_\bk (t)$. At any $\bk$, the temporal-average of $\Piout$ determines the energy (variance) amplification of harmonic (in homogeneous spatial directions) stochastic (in inhomogeneous directions and time) disturbances to the linearized NS equations, 	
    \beq
    \Ek
    \; \DefinedAs \;
    \dfrac{1}{2 \pi}
    \int_{-\infty}^{\infty}
    \Piout 
    \, \mrd \omega.
    \non
    \eeq
This quantity is also known as the {\em ensemble-average energy density\/} of the statistical steady-state, and it is hereafter referred to as the (steady-state) {\em energy amplification\/} (or energy density). For white-in-time inputs $\bd_\bk (t)$ with $\Sin = \Wk$, the solution to the algebraic Lyapunov equation~\ref{eq.AL} can be used to compute $\Ek$,
	\beq
	\Ek
	\; = \;
	\trace 
	\left(
	\Ck \Xk \Ck^\dagger
	\right),
	\label{eq.EA}
	\tag{EA}
	\eeq
thereby avoiding integration over $\omega$. When the input is uncorrelated in inhomogeneous spatial directions with $\Wk = I$, the sum of squares of the singular values of $\Tk (\mri \omega)$ gives the power spectral density, i.e., the Hilbert-Schmidt norm of $\Tk (\mri \omega)$. In this case, $\Ek$ determines the $H_2$ norm of \mbox{System~\ref{eq.lnse1} and Parseval's identity~yields,}
	\beq
	 \Ek
    	\; = \;
	\dfrac{1}{2 \pi}
    	\int_{-\infty}^{\infty}
	\sum_{j \, = \, 1}^{\infty}
	\sigma^2_{\bk, j} (\omega)
	\, 
	\mrd \omega
	\; = \;
    	\dfrac{1}{2 \pi}
    	\int_{-\infty}^{\infty}
    	\trace 
	\left( 
	\Tk (\mri \omega) \Tk^\dagger (\mri \omega)
	\right)
	\mrd \omega
	\; = \;
	\int_{-\infty}^{\infty}
    	\trace 
	\left( 
	\Tk (t) \Tk^\dagger (t)
	\right)
	\mrd t.
	\non
	\eeq
Thus, in addition to quantifying the steady-state variance of System~\ref{eq.lnse1} subject to spatially and temporally uncorrelated stochastic inputs, the $H_2$ norm also determines the $L_2$-norm of the impulse response and the same control-theoretic quantity enjoys both stochastic and deterministic interpretations. 

	\vspace*{-2ex}
	\subsubsection{Comparison of $H_2$ and $H_\infty$ norms}
For flows without homogeneous directions, these two quantities compress the dynamics into a single positive number; otherwise, they are parameterized by \mbox{the wavenumber~$\bk$.} Section~\ref{sec.Hinf} offers interpretations of the $H_\infty$ norm and this sidebar discusses the $H_2$ norm. Herein, we highlight how these measures of input-output amplification of System~\ref{eq.lnse1} compress information in inhomogeneous directions and in time; while the $H_\infty$ norm maximizes over both spatial profiles and frequency by computing the temporal supremum of $\sigma_{\max} ( \Tk (\mri \omega))$, the $H_2$ norm quantifies the aggregate effect of inputs by integrating the sum of squares of the singular values of $\Tk (\mri \omega)$ over~$\omega$. 		\vspace*{-0.35ex}
\end{textbox}

\vspace*{-2ex}
	\subsubsection*{\tc{dred}{Time-periodic systems.}}
The response of a linear time-periodic System~\ref{eq.lnse1} to a stationary stochastic input is a cyclo-stationary process~\citep{gar90}; the covariance operator of the state is $2 \pi/\omega_t$ periodic, i.e., 
	$
	\Xk (t) 
	\DefinedAs 
	\bE \left( \bpsi_\bk (t) \otimes \bpsi_\bk (t) \right)	
	=
	\sum_n 
	X_{\bk,n} \mre^{\mri n \omega_t t},
	$
with $X_{\bk,-n} = X_{\bk,n}^\dagger$, and the effect of the stationary input, over one period $T \DefinedAs 2 \pi/\omega_t$, is determined by 
	$
	(1/T) 
	\int_{0}^{T}
	\Xk (t) \, \mrd t
	= 
	X_{\bk,0}.
	$
If the stochastic input $\bd_{\bk} (t)$ is white-in-time with the spatial covariance $\Wk$, the harmonic Lyapunov equation, 
	\beq
	\tikz[baseline]{
            \node[fill=red!20,anchor=base] (t1)
            {$
	(\cAk \, - \, {\cal E} (\mri 0)) 
	\cXk 
	\; + \;
	\cXk 
	(\cAk \, - \, {\cal E} (\mri 0))^\dagger
	\; = \;
	-
	\cBk \;\! \cWk \;\! \cBk^\dagger,
	$};
            }
	\label{eq.HLE}
	\tag{HLE}
	\eeq 
can be used to compute the Fourier series coefficients $X_{\bk,n}$ of $\Xk (t)$. Here, $\cAk$, $\cBk$, and $\cal E$ are defined in the sidebar FREQUENCY RESPONSE OPERATOR, $\cWk$ is the block-diagonal operator with $\Wk$ on the main diagonal, and $\cXk$ is the self-adjoint block-Toeplitz operator whose elements are determined by $X_{\bk,n}$~\citep{jovPOF08,jovfarAUT08}. 
 
	\vspace*{-3ex}
\section{UNCOVERING MECHANISMS IN WALL-BOUNDED SHEAR FLOWS} 	
	\label{sec.mechanisms}

We next illustrate how the input-output approach provides insights into the physics of transitional and turbulent wall-bounded shear flows of Newtonian and viscoelastic fluids. In addition to offering a computational framework that quantifies impact of modeling imperfections on relevant flow quantities, a control-theoretic viewpoint also reveals influence of dimensionless groups on amplification of deterministic as well as stochastic disturbances and uncovers mechanisms that may initiate bypass transition. In Section~\ref{sec.lnse}, we  highlight how streamwise streaks, oblique waves, and Orr-Sommerfeld modes are identified as input-output resonances of the operator that maps forcing fluctuations to different velocity components in Newtonian fluids. In Section~\ref{sec.ob}, we demonstrate how a control-theoretic approach discovers a viscoelastic analogue of the familiar inertial lift-up mechanism, thereby identifying mechanisms that may trigger transition to elastic turbulence in rectilinear flows of viscoelastic fluids. Finally, in Section~\ref{sec.turbulent}, we offer a brief overview of the merits and the effectiveness of the input-output analysis in turbulent channel and pipe \mbox{flows of Newtonian fluids.} 

% Margin Note - non-dimensional NS equations
\begin{marginnote}[]
\entry{Dimensionless NS equations}{in incompressible Newtonian fluids with density $\rho$, we scale length with the channel half-height $h$, velocity with $\bar{u}$, time with the inertial time scale $h/\bar{u}$, pressure with $\rho \bar{u}^2$, and forcing per unit mass with $\bar{u}^2/h$. In analysis of transition, $\bar{u}$ is the largest velocity of the laminar base flow and, in analysis of turbulence, $\bar{u} $ is the friction velocity. 	
}
\end{marginnote}

	\vspace*{-3ex}
\subsection{Bypass transition in channel flows of Newtonian fluids}
	\label{sec.lnse}
 
\begin{figure}
    \centering
    {
    \begin{tabular}{cccc}
     \hspace{-0.5cm}
	\begin{tabular}{c}
		\vspace{0.5cm}
		\normalsize{\rotatebox{90}{$k_x$}}
	\end{tabular}
	&
    \hspace*{-5.5cm}
    \begin{tabular}{c}
     \includegraphics[width=0.3\textwidth]{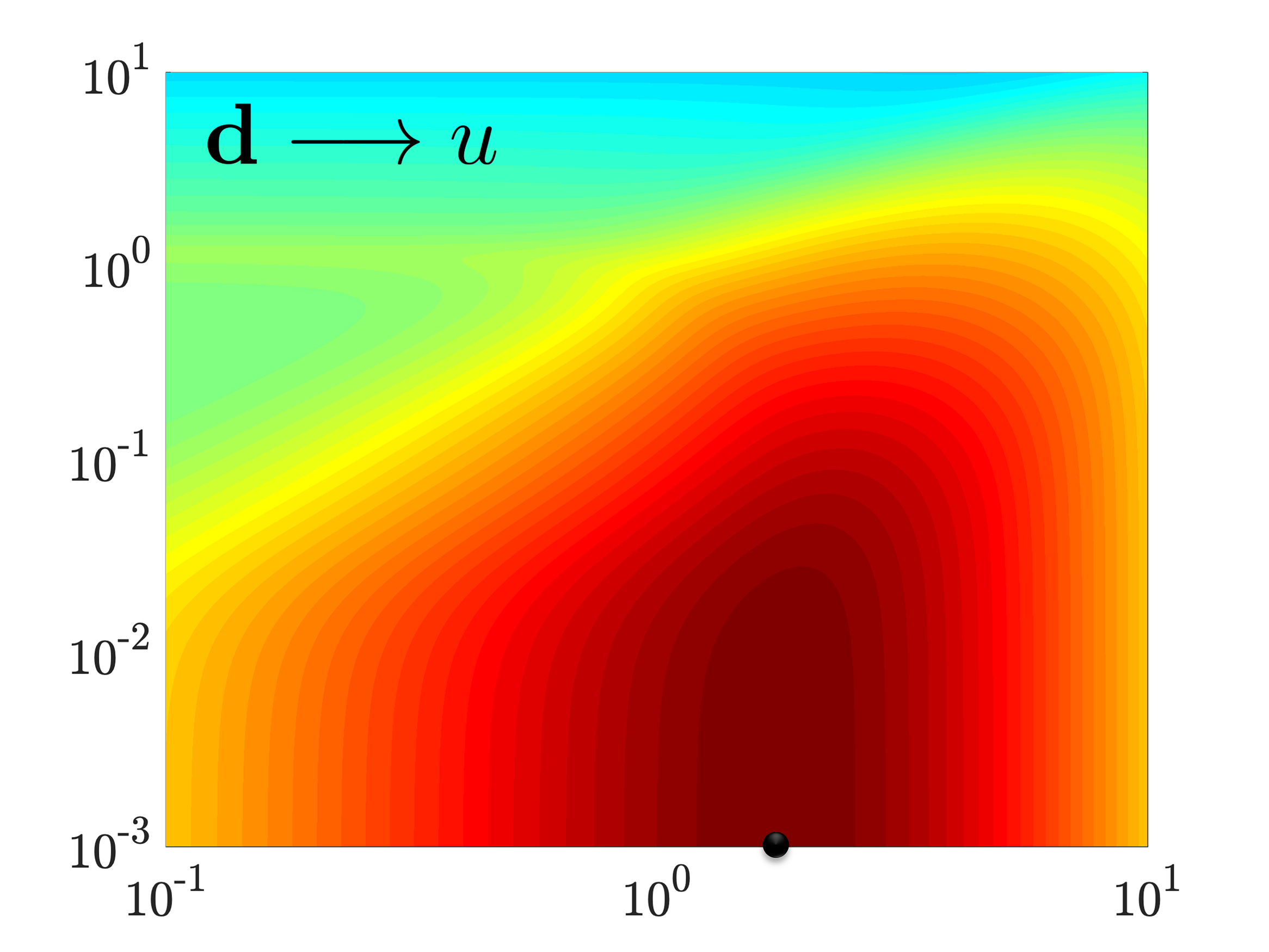}
           \\[-0.1cm]
           $k_z$
%           \\[-0.25cm]
%            \subfigure[streamwise]
%   {\label{fig.H2u}}
           \end{tabular}   
           &    
   \hspace*{-10.cm}
   \begin{tabular}{c}
   \includegraphics[width=0.3\textwidth]{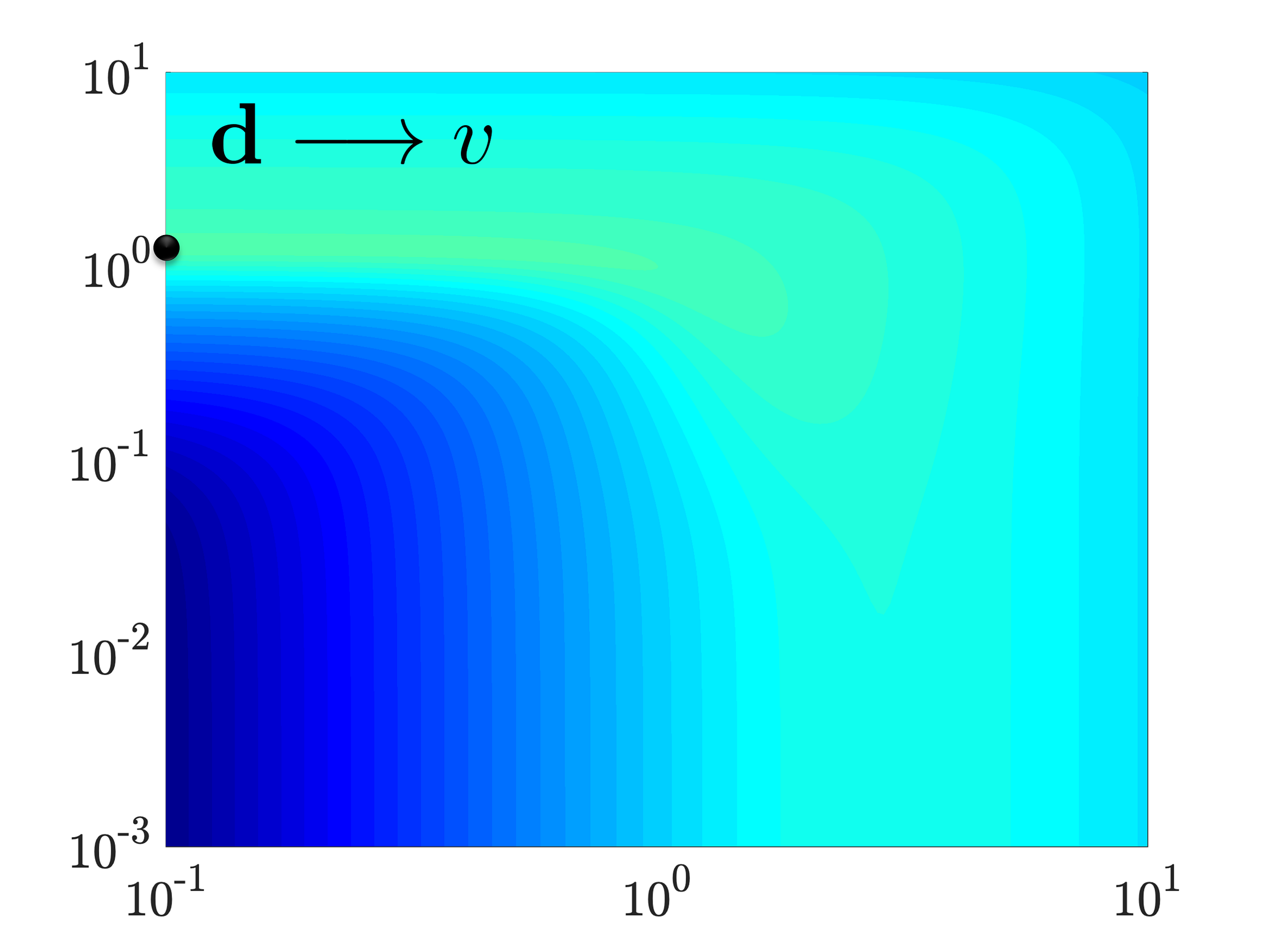}            
           \\[-0.1cm]
           $k_z$
%           \\[-0.25cm]
%            \subfigure[wall-normal]
%   {\label{fig.H2v}}
           \end{tabular}
           &    
   \hspace*{-10.cm}
   \begin{tabular}{c}
    \includegraphics[width=0.3\textwidth]{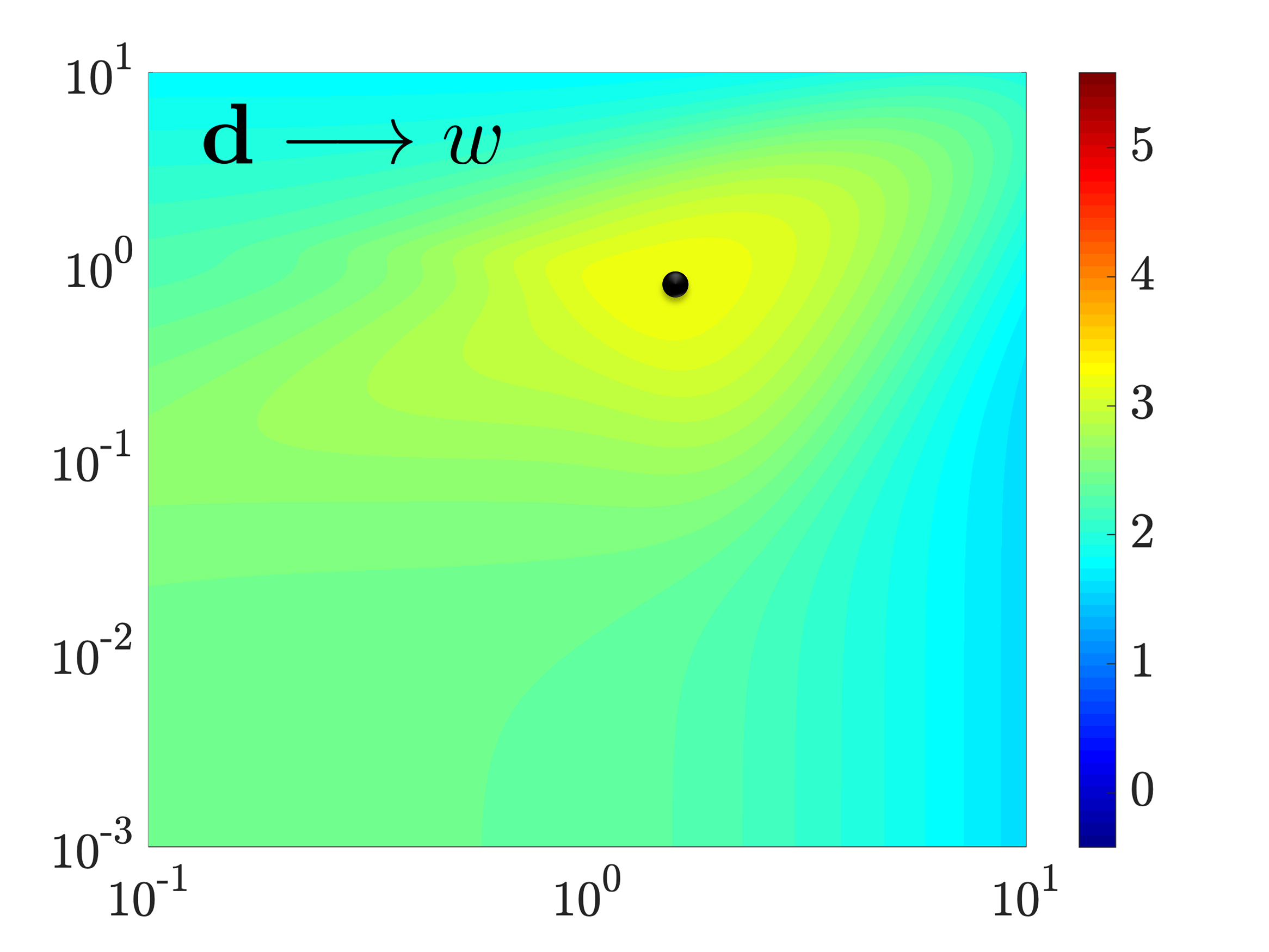}
           \\[-0.1cm]
           $k_z$
%           \\[-0.25cm]
%            \subfigure[spanwise]
%  {\label{fig.H2w}}
           \end{tabular}
	\end{tabular}
    }
    \caption{Energy amplification of streamwise (left), wall-normal (middle), and spanwise (right) velocity fluctuations for the linearized NS equations subject to channel-wide stochastic forcing in Poiseuille flow with $Re = 2000$. The largest value in each plot is marked by a black dot and a logarithmic scaling with the same color map is employed. The streamwise velocity contains most energy and the dominant flow structures are given by the streamwise elongated spanwise periodic streaks.}
    \label{fig.H2channel}
    \vspace*{-0.75cm}
    \end{figure}

For Poiseuille flow with $Re = 2000$, {\bf Figure~\ref{fig.ssim}} shows that the streamwise constant flow structures with $k_z = 1.78$ are much more energetic than the oblique waves ($k_x = k_z = 1$) and the Orr-Sommerfeld modes ($k_x = 1$, $k_z = 0$). We next illustrate how the tools of Section~\ref{sec.H2} offer insights into the physics of transitional flows while avoiding need for stochastic simulations. {\bf Figure~\ref{fig.H2channel}} displays the joint impact of forcing fluctuations in all three spatial directions on the individual velocity components. For a channel-wide forcing $\bd_\bk (t)$, we utilize Equation~\ref{eq.EA} to evaluate the impact of the wavenumbers $k_x$ and $k_z$ on the steady-state variance of $u$, $v$, and $w$. The streamwise velocity component $u$ contains most energy and the strongest amplification occurs in the dark red region that corresponds to small values of $k_x$ and $O(1)$ values of $k_z$. The oblique modes (i.e., the flow structures with $O(1)$ values of $k_x$ and $k_z$) emerge as input-output resonances in the response of the spanwise velocity $w$ and they are significantly less amplified than the streamwise elongated flow structures with $k_x \approx 0$. On the other hand, the least-stable Orr-Sommerfeld mode, which is the dominant source of amplification for the wall-normal velocity $v$, creates only a local peak around ($k_x \approx 1, k_z = 0$) in the response of $u$. Thus, the flow structures that are deemed important in classical hydrodynamic stability play a marginal role in amplification of stochastic disturbances. This identifies shortcomings of modal stability theory, highlights the utility of componentwise input-output analysis~\citep{jovbamJFM05}, and demonstrates that significant insight can be gained by examining linearized dynamics in the presence of modeling imperfections (in this case, additive stochastic disturbances). 

	%=========
	% Figure 5  %
	%=========
 	
	\vspace*{-3ex}
\subsubsection{Streamwise constant model:~lift-up mechanism}
	\label{sec.lnse-kx0}

In addition to computational advantages, a control-theoretic viewpoint also uncovers mechanisms for subcritical transition and quantifies impact of the Reynolds number on amplification of deterministic as well as stochastic disturbances~\citep{mj-phd04,jovbamJFM05}. By considering how the disturbances propagate through the linearized dynamics, important insight can be gained~{\em without any computations\/}. Since the streamwise constant fluctuations experience the largest amplification (see {\bf Figure~\ref{fig.H2channel}}), we examine System~\ref{eq.lnse1} for $\bk \DefinedAs (k_x,k_z) = (0,k_z)$, 
	\beq
    \ba{rcl}
    \dfrac{\mrd}{\mrd t}
    \tbo{v (t)}{\eta (t)}
    % \tbo{\dot{v} (t)}{\dot{\eta} (t)}
    & = &
    \overbrace{{\tbt{\tc{red}{\tfrac{1}{Re}} \, \Aos}{0}{\tikz[baseline]{
            \node[fill=red!20,anchor=base] (t1)
            {$\Acpn$};
            }}
            {\tc{red}{\tfrac{1}{Re}} \, \Asq}}}^{\mbox{\tc{dred}{\bf non-normal}}}
    \tbo{v (t)}{\eta (t)} 
    \; + \;
    \tbth{0}{B_2}{B_3}{B_1}{0}{0}
    \thbo{d_1 (t)}{d_2 (t)}{d_3 (t)},
    \\[-0.1cm]
    \thbo{u (t)}{v (t)}{w (t)}
    & = &
    \thbt{0}{C_{u}}{C_v}{0}{C_w}{0}
    \tbo{v (t)}{\eta (t)},
    \ea
    \label{eq.SC}
    \tag{2D3C}
    % \tag{SC}
    \eeq
where we suppress the dependence on the spanwise wavenumber $k_z$. Here, $v$ and $\eta$ denote the wall-normal velocity and vorticity fluctuations, whereas ($d_1,d_2,d_3$) and ($u,v,w$) are the forcing and velocity fluctuations in ($x,y,z$). The Orr-Sommerfeld, Squire, and Coupling operators are given by $\Aos \DefinedAs \Delta^{-1} \Delta^2$, $\Asq \DefinedAs \Delta$, and $\Acpn \DefinedAs - \mri k_z U'(y)$, where $\Delta = \partial_{yy} - k_z^2 I$ is a Laplacian with homogeneous Dirichlet boundary conditions, $\Delta^{-1}$ is the inverse of the Laplacian, $\Delta^2 = \partial_{yyyy} - 2 k_z^2 \partial_{yy} + k_z^4 I$ with homogeneous Dirichlet and Neumann boundary conditions, and $U'(y) = \mrd U(y)/ \mrd y$. We refer the reader to~\citet[Section~4]{jovbamJFM05} for a definition of the input and output operators $B$ and $C$.

% Margin Note - normal opearator
\begin{marginnote}[]
\entry{Normal operator}{an operator is normal if it commutes with its adjoint. A normal operator is unitarily diagonalizable (i.e., it has a complete set of orthogonal eigenfunctions).}
\end{marginnote}

As described in the sidebar BLOCK DIAGRAMS, this control-theoretic tool decomposes complex systems into essential pieces, abstract unnecessary details, and highlight the flow of information. A graphical representation of the frequency response operator in {\bf Figure~\ref{fig.bd-lnse-kx0}} illustrates that the wall-normal and spanwise forcing fluctuations ($d_2,d_3$) produce $O (Re)$ fluctuations in $v$ and $w$. Although these are dissipated by viscosity, the resulting spanwise variations in $v$, $\mri k_z v$, tilt the spanwise vorticity of the laminar base flow, $- U'(y)$, in the wall-normal direction $y$, thereby triggering $O (Re^2)$ fluctuations in $\eta$ and, consequently, in $u = \eta/(\mri k_z)$. This {\em lift-up mechanism\/}~\citep{lan75} is a dominant source of amplification in wall-bounded shear flows of Newtonian fluids. The operator $\Acpn$ acts as a {\em source in the wall-normal vorticity equation\/} and it accounts for {\em vortex tilting\/} which arises from {\em linearization of the convective terms\/} in the NS equations. Since $\Aos$ and $\Asq$ are self-adjoint, in the absence of vortex tilting the dynamics are characterized by viscous~dissipation. 

	\vspace*{-3ex}
\subsection{Early stages of transition to elastic turbulence:~viscoelastic lift-up}
	\label{sec.ob}

In complex fluids and complex flows, it is even more important to explicitly account for modeling imperfections by quantifying their influence on transient and asymptotic dynamics. Herein, we illustrate how input-output analysis discovers mechanisms that may initiate {\em bypass transition\/} in channel flows of viscoelastic fluids {\em in the absence of inertia\/}. Transition in fluids that contain polymer chains can impact polymer processing and enhance micro-fluidic mixing. In contrast to Newtonian fluids, viscoelastic liquids can deviate from laminar profiles even when inertia is negligible~\citep{groste00} and, in curvilinear flows, a purely elastic instability triggers transition~\citep{larshamul90}. In low inertial regimes, rectilinear flows are asymptotically stable but the dynamics associated with polymer stress fluctuations can still induce complex responses~\citep{qinsalhudarrPRL19,qinsalhudarrJFM19}. Since no single constitutive equation {\em fully describes the range of phenomena in viscoelastic fluids\/}, it is important to understand how modeling imperfections may adversely affect their dynamics.

Newtonian fluids are characterized by a static-in-time linear relation between stresses and velocity gradients. In viscoelastic fluids, constitutive equations determine the influence of velocity gradients on the dynamics of stress tensor. For dilute polymer solutions, polymer molecules are treated as springs that connect spherical beads~\citep{Bird1987}; the Oldroyd-B (infinitely extensible linear spring) and the FENE-type (finitely extensible nonlinear elastic) models are most commonly used. In the absence of inertia we can set $Re = 0$ and the Weissenberg number, $\We = \lambda \bar{u}/h$, and the viscosity ratio, $\beta = \mu_s/(\mu_s + \mu_p)$, characterize channel flows of Oldroyd-B fluids. The Weissenberg number quantifies the ratio between the elastic and viscous forces and it is given by the product of the polymer relaxation time $\lambda$ and the velocity gradient $\bar{u}/h$. The steady solution determines the laminar base flow ($\bar{\bu},\bar{\btau}$), where $\bar{\bu} = (U(y),0,0)$, 
	$
    U(y) = 1 - y^2
    $
in pressure-driven Poiseuille flow, 
    $
    U(y) = y
    $
in shear-driven Couette flow,~and the non-zero components of the base polymer stress tensor $\bar{\btau}$ are $\bar{\tau}_{11} = 2 \We ( U'(y) )^2$ and $\bar{\tau}_{12} = \bar{\tau}_{21} = U'(y)$. Equations for infinitesimal velocity, pressure, and stress fluctuations are obtained by linearization around ($\bar{\bu},\bar{\btau}$).

% Margin Note - Oldroyd-B model
\begin{marginnote}[]
\entry{Dimensionless Oldroyd-B model}{in channel flows of Oldroyd-B fluids (with density $\rho$, solvent and polymer viscosities $\mu_s$ and $\mu_p$), equations can be brought to a non-dimensional form by scaling length with the channel half-height $h$, velocity with the largest velocity of the base flow $\bar{u}$, time with the polymer relaxation time $\lambda$, polymer stresses with $\mu_p \bar{u} /h$, pressure with $(\mu_s + \mu_p) \bar{u}/h$, and forcing per unit mass with $(\mu_s + \mu_p) \bar{u}/\rho h^2$.
	}
\end{marginnote}

\cite{hodjovkumJFM08,hodjovkumJFM09} were the first to investigate nonmodal amplification of disturbances in channel flows of viscoelastic fluids and demonstrate high sensitivity of the laminar flow in both inertia- and elasticity-dominated regimes. \citet{jovkumPOF10} showed that velocity and stress fluctuations experience significant transient growth even in the absence of inertia.~\cite{jovkumJNNFM11} identified a new slow-fast decomposition of the governing equations and used singular-perturbation techniques to {\em analytically\/} establish unfavorable scaling of the energy amplification with the Weissenberg number in weakly-inertial flows.~\citet{liejovkumJFM13} quantified the role of finite extensibility of polymers on the worst-case amplification of disturbances in FENE-type models and~\cite{harjovkumJNNFM18} studied amplification of localized body forces. The combined effects of inertia and elasticity on streak evolution was examined in~\citet{pagzakJFM14,agabrazakJFM14}. 

For streamwise-constant flows of Oldroyd-B fluids the block diagram in {\bf Figure~\ref{fig.bd-OB-kx0}} reflects the structure of the frequency response operator that maps disturbances to the momentum equation (inputs) to the velocity fluctuations (outputs) and eliminates all unnecessary variables. Apart from the operator 
	$
	\Acpv
	\DefinedAs 
	 \mri k_z
	 (
        	 U'(y) \Delta + 2 U''(y) \py
	 ),
	 $
which accounts for {\em stretching of polymer stress fluctuations by a base shear\/}, all other operators are the same as in Newtonian fluids; see Section~\ref{sec.lnse-kx0}. The block diagrams reveal striking structural similarity between streamwise-constant inertial flows of Newtonian fluids and inertialess flows of viscoelastic fluids. In the absence of base shear $U'(y)$ and spanwise variations in fluctuations, the responses of viscoelastic fluids are governed by viscous dissipation and all velocity components are $\We$-independent. However, in contrast to Newtonian fluids, spanwise variations in fluctuations and their interactions with $U'(y)$ provide a source in the vorticity equation even in the absence of inertia. In particular, the influence of $d_2$ and $d_3$ on $u$ can be understood by analyzing the wall-normal \mbox{vorticity equation~\citep{jovkumJNNFM11},}
	\beq
	\ba{rcc}
	\beta \Delta \;\! \dot{\eta} (t)
	\; = \;
	-
	\Delta \;\! \eta (t)
	& - \; &
	\tikz[baseline]{
            \node[fill=red!20,anchor=base] (t1)
            {$(1 - \beta) \, \tc{red}{\We} \, ( U'(y) \Delta \, + \, 2 U''(y) \py ) \, \mri k_z \vartheta (t)$};
            },
	\ea
	\non
	\eeq
 where $\vartheta$ denotes a low-pass version of the wall-normal velocity $v$, 
 	$
    \hat{\vartheta} \DefinedAs \hat{v} / (\mri \omega + 1).
    $
The source term arises from stretching of polymer stress fluctuations by a base shear and it introduces a lift-up of fluctuations in a similar way as vortex tilting in inertia-dominated flows of Newtonian fluids. Thus, the wall-normal and spanwise inputs give rise to an energy transfer from the base flow to fluctuations and generate streamwise velocity fluctuations that are proportional to the Weissenberg number. Responses from all other inputs to all other velocities are $\We$-independent and they are governed by viscous~dissipation.~\citet{jovkumJNNFM11} also demonstrated that $d_2$ and $d_3$ induce a quadratic scaling with the Weissenberg number of the streamwise component of the polymer stress tensor, $\tau_{11}$.

\vspace*{-3ex}
	\subsubsection*{\tc{dred}{Summary.}}
Elementary control-theoretic analysis identifies key physical mechanisms and demonstrates that the wall-normal and spanwise body forces have the largest impact on the streamwise velocity fluctuation in inertia-dominated channel flows of Newtonian fluids and elasticity-dominated flows of viscoelastic fluids. These conclusions are derived {\em without any computations\/} by examining the frequency responses of streamwise constant fluctuations and showing that $d_2$ and $d_3$ induce a quadratic scaling of $u$ with the Reynolds number (in Newtonian fluids) and a linear scaling of $u$ with the Weissenberg number (in inertialess Oldroyd-B fluids). At $k_x = 0$, $d_1$ does not influence $v$ and $w$ and the mappings from all other forcing to all other velocity components are proportional to the Reynolds number (in Newtonian fluids) and are $\We$-independent (in viscoelastic fluids); see~\citet[Section~4]{jovbamJFM05} and~\citet{jovkumJNNFM11} for additional~details. In spite of these structural similarities, amplification in Newtonian and viscoelastic fluids originates from different physical mechanisms; vortex tilting and polymer stretching, respectively.

 \vspace*{-2ex}
	\begin{textbox}[h]
	\vspace*{-1ex}
\section{BLOCK DIAGRAMS: A TOOL FOR REVEALING STRUCTURE W/O COMPUTATIONS}
Block diagrams decompose complex systems into essential pieces, abstract unnecessary details, and highlight the flow of information. This control-theoretic tool reveals structure without any computations and allows to make useful analogies. The circles denote summation of input signals and the boxes represent different parts of the system. Inputs into each box/circle are represented by lines with arrows directed toward the box/circle, and outputs are represented by lines with arrows leading away from the box/circle. The inputs specify the signals affecting subsystems, and the outputs specify the signals of interest or signals affecting other parts of the system. In streamwise-constant channel flows of Newtonian and inertialess viscoelastic fluids,  the block diagrams in {\bf Figure~\ref{fig.comparison}} illustrate influence of disturbances ($d_1,d_2,d_3$) to the momentum equation on the velocity fluctuations $(u,v,w)$. The blue boxes represent resolvent operators associated with $\Aos$ and $\Asq$, and the red boxes represents the coupling operators $\Acpn \DefinedAs - \mri k_z U'(y)$ and $\Acpv \DefinedAs \mri k_z ( U'(y) \Delta + 2 U''(y) \py).$ In Newtonian fluids, $\Omega \DefinedAs \omega Re$ is the frequency scaled with the diffusive time scale, $h^2/\nu$, and in viscoelastic fluids, $\omega$ is the frequency scaled with the polymer relaxation time, $\lambda$.
	\vspace*{-0.3ex}
		\end{textbox}

	%=========
	% Figure 6  %
	%=========
	\begin{figure}
	 \vspace*{-0.75cm}
    \centering
    {
    \begin{tabular}{c}
    \begin{tabular}{c}
    {\hspace*{-1.cm} \subfigure[]{\scalebox{.77}{%_______________________________________________________________________________
%
%   Block diagram of the streamwise constant fluctuations in channel flow of Newtonian fluids
%   drawn from Right to Left
%
%   Mihailo Jovanovic, January 12, 2020
%_______________________________________________________________________________
%
% TikZ styles for drawing
%
\input{figures/Tikz_common_styles}
%
%   set a filename for externalization
% \tikzsetnextfilename{clp_2dof_input_pert_config}
%
\noindent
\begin{tikzpicture}[scale=1, auto, >=stealth']

	% output node
	% starting point for uend
	% \node [input, name=uend] {};
	\node[] (uend) at (0,0) {};
	
	%	 % end point for v
	 \node[] (vend) at ($(uend) - (0.cm,1.5cm)$) {};
	 
	 % end point for w
	 \node[] (wend) at ($(vend) - (0,1.5cm)$) {};
	 
   % operator $C_u$
    \node[block, minimum height = 1.cm, top color=white!20, bottom color=white!20] (Cu) at ($(uend) + (1.5cm,0)$) {$C_u$};
    
    % operator $B_1$
    \node[block, minimum height = 1.cm, top color=white!20, bottom color=white!20] (Cv) at ($(Cu) - (0,1.5cm)$) {$C_v$};

     % operator $B_1$
    \node[block, minimum height = 1.cm, top color=white!20, bottom color=white!20] (Cw) at ($(Cv) - (0,1.5cm)$) {$C_w$};
    
     % Squire operator 
     \node[block, minimum height = 1.cm, top color=RoyalBlue!20, bottom color=RoyalBlue!20] (sq) at ($(Cu.east) + (2.25cm,0cm)$) {$\tc{red}{Re} \left( \mri \Omega I \, - \, \Asq \right)^{-1}$};
     
      % second summation
    \node[sum] (sum2) at ($(sq.east) + (1.cm,0)$) {$+$};
    
    % coupling operator
  \node[block, minimum height = 1.cm, top color=red!20, bottom color=red!20] (cp) at ($(sum2.east) + (1.25cm,0cm)$) {$\Acpn$};

   % OS operator 
     \node[block, minimum height = 1.cm, top color=RoyalBlue!20, bottom color=RoyalBlue!20] (os) at ($(cp.east) + (2.5cm,0cm)$) {$\tc{red}{Re} \left( \mri \Omega I \, - \, \Aos \right)^{-1}$}; 
  
  % first summation
    \node[sum] (sum1) at ($(os.east) + (1.cm,0)$) {$+$};
    
    % operator B2
  	\node[block, minimum height = 1.cm, top color=white!20, bottom color=white!20] (B2) at ($(sum1.east) + (1.5cm,0cm)$) {$B_2$};
     
	% operator B1
         \node[block, minimum height = 1.cm, top color=white!20, bottom color=white!20] (B1) at ($(B2) + (0,1.5cm)$) {$B_1$};
         
         % operator B3
         \node[block, minimum height = 1.cm, top color=white!20, bottom color=white!20] (B3) at ($(B2) - (0,1.5cm)$) {$B_3$};
  
	% input nodes
	% starting point for d2
     	\node[] (d2begin) at ($(B2.east) + (1.cm,0)$) {};

	% starting point for d1
     	\node[] (d1begin) at ($(B1.east) + (1.cm,0)$) {};
	
	% starting point for d3
	\node[] (d3begin) at ($(B3.east) + (1.cm,0)$) {};
	
	 % mid nodes
	 % midsum1 - 1.5cm below center of sum1 node
	  \node[] (midsum1) at ($(os.east) + (1.cm,-1.5cm)$) {};
	  
	   % midsum2
 	   \node[] (midsum2) at ($(sq.east) + (1.cm,1.5cm)$) {};

%%	  \node[] (midsum1a) at ($(B3.east) + (1.12cm,0)$) {};
%%	  \node[] (midsum1b) at ($(B3.east) + (1.0cm,-0.13cm)$) {};
%%	  \node[] (midsum1a) at ($(d2begin) + (4.12cm,-1.5cm)$) {};
%%	  \node[] (midsum1b) at ($(B3.east) + (4.12cm,-1.5cm)$) {};
%	  
%	  % \node[]  (midsum2b) at ($(sum2.center) + (0,1.63cm)$) {};
	  
	  % midpoint1
 	   \node[] (midpoint1) at ($(cp.east) + (0.6cm,0)$) {};
	   \node[] (midpoint2) at ($(cp.east) + (0.6cm,-1.5cm)$) {};
	   \node[] (midpoint3) at ($(cp.east) + (0.6cm,-3cm)$) {};

%%     \node[] (output-node2) at ($(sys1.east) + (1.5cm,0)$) {};
%%
%%     \node[] (u0) at ($(output-node1.north) + (0cm,1.0cm)$) {};
%%
%%     \node[] (u0w) at ($(u0.west) + (-0.02cm,0)$) {};
%%
%%     \node[] (midpt1) at ($(u0.west) + (-1.82cm,0)$) {};
%%     
%%     \node[] (midpt11) at ($(os.north) + (.9cm,0)$) {};
%%
%%     \node[] (midpt2) at ($(output-node1.east) + (3.1cm,0)$) {};
%%     
%%     \node[] (midpt21) at ($(sys1.north) + (0.9cm,0)$) {};
%%
%%
%%     \node[] (u0end2) at ($(sys1.south) + (-0.5cm,-0.8cm)$) {};

        % now link the nodes
 	% input d1 to block B1
    	\draw [connector] (d1begin.west) -- node [midway, above] {$\hat{d}_1$} (B1.east);
    
	% input d2 to block B2
    	\draw [connector,ultra thick] (d2begin.west) -- node [midway, above] {$\hat{d}_2$} (B2.east);
	
	% input d3 to block B3
    	\draw [connector,ultra thick] (d3begin.west) -- node [midway, above] {$\hat{d}_3$} (B3.east);

	% connect block B2 with sum1
         \draw [connector,ultra thick] (B2.west) -- node [midway, above] {$$} (sum1.east);

	% connect sum1 with block OS
         \draw [connector,ultra thick] (sum1.west) -- node [midway, above] {$$} (os.east);

	% connect block OS with block CP
         \draw [connector,ultra thick] (os.west) -- node [midway, above] {$\hat{v}$} (cp.east);
          
          % connect block cp with sum2
          \draw [connector,ultra thick] (cp.west) -- node [midway, above] {$$} (sum2.east);
    
    	% connect sum2 with block SQ
        \draw [connector,ultra thick] (sum2.west) -- node [midway, above] {$$} (sq.east);

	% connect block SQ with block Cu
         \draw [connector,ultra thick] (sq.west) -- node [midway, above] {$\hat{\eta}$} (Cu.east);
    
    	% connect block Cu with uend
         \draw [connector,ultra thick] (Cu.west) -- node [midway, above] {$\hat{u}$} (uend);
         
         % connect block Cv with vend
         \draw [connector] (Cv.west) -- node [midway, above] {$\hat{v}$} (vend);
         
         % connect block Cu with uend
         \draw [connector] (Cw.west) -- node [midway, above] {$\hat{w}$} (wend);

     % connect block B3 with sum1
%     \draw [line] (B3.east) -- (midsum1a);
%     \draw [connector] (midsum1b) -- (sum1.south);
	\draw [line,ultra thick] (B3.west) -- (midsum1.center);
     \draw [connector,ultra thick] (midsum1.center) -- (sum1.south);
     
     % connect block B1 with sum2
     \draw [line] (B1.west) -- (midsum2.center);
     \draw [connector] (midsum2.center) -- (sum2.north);
     
     % connect midpoint1 with midpoint2
     \draw [line] (midpoint1.center) -- (midpoint2.center);
     
      % connect midpoint2 with midpoint3
     \draw [line] (midpoint2.center) -- (midpoint3.center);
    
    % connect midpoint2 with Cv
     \draw [connector] (midpoint2.center) -- (Cv.east);

     % connect midpoint3 with Cw
     \draw [connector] (midpoint3.center) -- (Cw.east);
     
     	% label OS, CP, and SQ blocks - physical interpretation
	\node [above = -0.07cm of os](extra){\tc{RoyalBlue}{$\ba{c} \mbox{\bf generalized} \\[-0.05cm] \mbox{\bf diffusion} \ea$}};  %
        \node [above = -0.07cm of cp](extra){\tc{red}{$\ba{c} \mbox{\bf vortex} \\[-0.05cm] \mbox{\bf tilting} \ea$}};  %
        \node [below = -0.07cm of cp](extra){\tc{dred}{$\ba{c} \mbox{\bf source of} \\[-0.05cm] \mbox{\bf amplification} \ea$}};  %
	\node [above = -0.07cm of sq](extra){\tc{RoyalBlue}{$\ba{c} \mbox{\bf viscous} \\[-0.05cm] \mbox{\bf dissipation} \ea$}};  %
	
\end{tikzpicture}
%_______________________________________________________________________________
}
           \label{fig.bd-lnse-kx0}}}
    \end{tabular}
    \\
    \begin{tabular}{c}
    {\hspace*{-1.cm} \subfigure[]{\scalebox{.77}{%_______________________________________________________________________________
%
%   Block diagram of the streamwise constant fluctuations in channel flow of Oldroyd-B fluids
%   drawn from Right to Left
%
%   Mihailo Jovanovic, January 12, 2020
%_______________________________________________________________________________
%
% TikZ styles for drawing
%
\input{figures/Tikz_common_styles}
%
%   set a filename for externalization
% \tikzsetnextfilename{clp_2dof_input_pert_config}
%
\noindent
\begin{tikzpicture}[scale=1, auto, >=stealth']

	% output node
	% starting point for uend
	% \node [input, name=uend] {};
         \node[] (uend) at (0,0) {};
         	
	%	 % end point for v
	 \node[] (vend) at ($(uend) - (0.cm,1.5cm)$) {};
	 
	 % end point for w
	 \node[] (wend) at ($(vend) - (0,1.5cm)$) {};
	 
   % operator Cu
    \node[block, minimum height = 1.cm, top color=white!20, bottom color=white!20] (Cu) at ($(uend) + (1.5cm,0)$) {$C_u$};
    % (1 + \mri \omega) - filter for Cv
     
     % (1 + \mri \omega) - filter for Cu
     \node[block, minimum height = 1.cm, top color=white!20, bottom color=white!20] (filterCu) at ($(cp.east) + (0.7cm,1.5cm)$) 
     {$\mri \omega + 1$};

    % operator Cv
    \node[block, minimum height = 1.cm, top color=white!20, bottom color=white!20] (Cv) at ($(Cu) - (0,1.5cm)$) 
    {$C_v$};
    % {$(1 + \mri \omega) \, C_v$};
    
     % (1 + \mri \omega) - filter for Cv
     \node[block, minimum height = 1.cm, top color=white!20, bottom color=white!20] (filterCv) at ($(Cv.east) + (2.25cm,0cm)$) 
     {$\mri \omega + 1$};

     % operator Cw
    \node[block, minimum height = 1.cm, top color=white!20, bottom color=white!20] (Cw) at ($(Cv) - (0,1.5cm)$) 
        {$C_w$};
    % {$(1 + \mri \omega) \, C_w$};
    
    % (1 + \mri \omega) - filter for Cw
     \node[block, minimum height = 1.cm, top color=white!20, bottom color=white!20] (filterCw) at ($(Cw.east) + (2.25cm,0cm)$) 
     {$\mri \omega + 1$};

     % Squire operator 
     \node[block, minimum height = 1.cm, top color=RoyalBlue!20, bottom color=RoyalBlue!20] (sq) at ($(Cu.east) + (2.25cm,0cm)$) {$\dfrac{-(1 - \beta)}{\beta ( \mri \omega ) + 1} \,  \Asq^{-1}$};
     
      % second summation
    \node[sum] (sum2) at ($(sq.east) + (1.cm,0)$) {$+$};
    
    % coupling operator
  \node[block, minimum height = 1.cm, top color=red!20, bottom color=red!20] (cp) at ($(sum2.east) + (1.25cm,0cm)$) 
  {$\tc{red}{\We} \, \Acpv$};

   % OS operator 
     \node[block, minimum height = 1.cm, top color=RoyalBlue!20, bottom color=RoyalBlue!20] (os) at ($(cp.east) + (2.5cm,0cm)$) 
     {$\dfrac{-1}{\beta ( \mri \omega ) + 1} \, \Aos^{-1}$};  
  % first summation
    \node[sum] (sum1) at ($(os.east) + (1.cm,0)$) {$+$};
    
    % operator B2
  	\node[block, minimum height = 1.cm, top color=white!20, bottom color=white!20] (B2) at ($(sum1.east) + (1.5cm,0cm)$) 
	{$B_2$};
     
	% operator B1
         \node[block, minimum height = 1.cm, top color=white!20, bottom color=white!20] (B1) at ($(B2) + (0,1.5cm)$) {$B_1$};
         
         % operator B3
         \node[block, minimum height = 1.cm, top color=white!20, bottom color=white!20] (B3) at ($(B2) - (0,1.5cm)$) {$B_3$};
  
	% input nodes
	% starting point for d2
     	\node[] (d2begin) at ($(B2.east) + (1.cm,0)$) {};

	% starting point for d1
     	\node[] (d1begin) at ($(B1.east) + (1.cm,0)$) {};
	
	% starting point for d3
	\node[] (d3begin) at ($(B3.east) + (1.cm,0)$) {};
	
	 % mid nodes
	 % midsum1 - 1.5cm below center of sum1 node
	  \node[] (midsum1) at ($(os.east) + (1.cm,-1.5cm)$) {};
	  
	   % midsum2
 	   \node[] (midsum2) at ($(sq.east) + (1.cm,1.5cm)$) {};

%%	  \node[] (midsum1a) at ($(B3.east) + (1.12cm,0)$) {};
%%	  \node[] (midsum1b) at ($(B3.east) + (1.0cm,-0.13cm)$) {};
%%	  \node[] (midsum1a) at ($(d2begin) + (4.12cm,-1.5cm)$) {};
%%	  \node[] (midsum1b) at ($(B3.east) + (4.12cm,-1.5cm)$) {};
%	  
%	  % \node[]  (midsum2b) at ($(sum2.center) + (0,1.63cm)$) {};
	  
	  % midpoints
 	   \node[] (midpoint1) at ($(cp.east) + (0.7cm,0)$) {};
	   \node[] (midpoint2) at ($(cp.east) + (0.7cm,-1.5cm)$) {};
	   \node[] (midpoint3) at ($(cp.east) + (0.7cm,-3cm)$) {};
	   % \node[] (midpoint4) at ($(cp.east) + (0.7cm,1.5cm)$) {};

        % now link the nodes
 	% input d1 to block B1
    	\draw [connector] (d1begin.west) -- node [midway, above] {$\hat{d}_1$} (B1.east);
    
	% input d2 to block B2
    	\draw [connector,ultra thick] (d2begin.west) -- node [midway, above] {$\hat{d}_2$} (B2.east);
	
	% input d3 to block B3
    	\draw [connector,ultra thick] (d3begin.west) -- node [midway, above] {$\hat{d}_3$} (B3.east);

	% connect block B2 with sum1
         \draw [connector,ultra thick] (B2.west) -- node [midway, above] {$$} (sum1.east);

	% connect sum1 with block OS
         \draw [connector,ultra thick] (sum1.west) -- node [midway, above] {$$} (os.east);

	% connect block OS with block CP
         \draw [connector,ultra thick] (os.west) -- node [midway, above] {$\hat{\vartheta}$} (cp.east);
          
          % connect block cp with sum2
          \draw [connector,ultra thick] (cp.west) -- node [midway, above] {$$} (sum2.east);
    
    	% connect sum2 with block SQ
        \draw [connector,ultra thick] (sum2.west) -- node [midway, above] {$$} (sq.east);

	% connect block SQ with block Cu
         \draw [connector,ultra thick] (sq.west) -- node [midway, above] {$\hat{\eta}$} (Cu.east);
    
    	% connect block Cu with uend
         \draw [connector,ultra thick] (Cu.west) -- node [midway, above] {$\hat{u}$} (uend);
         
         % connect block Cv with vend
         \draw [connector] (Cv.west) -- node [midway, above] {$\hat{v}$} (vend);
         
         % connect block Cu with uend
         \draw [connector] (Cw.west) -- node [midway, above] {$\hat{w}$} (wend);

     % connect block B3 with sum1
%     \draw [line] (B3.east) -- (midsum1a);
%     \draw [connector] (midsum1b) -- (sum1.south);
	\draw [line,ultra thick] (B3.west) -- (midsum1.center);
     \draw [connector,ultra thick] (midsum1.center) -- (sum1.south);
     
     % connect block B1 with sum2
     \draw [connector] (B1.west) -- (filterCu.east);
     \draw [line] (filterCu.west) -- (midsum2.center);
     \draw [connector] (midsum2.center) -- (sum2.north);
     
     % connect midpoint1 with midpoint2
     \draw [line] (midpoint1.center) -- (midpoint2.center);
     
     % connect midpoint2 with midpoint3
     \draw [line] (midpoint2.center) -- (midpoint3.center);
    
    % connect midpoint2 with Cv
     \draw [connector] (midpoint2.center) -- (filterCv.east);
     \draw [connector] (filterCv.west) -- (Cv.east);

     % connect midpoint3 with Cw
     \draw [connector] (midpoint3.center) -- (filterCw.east);
     \draw [connector] (filterCw.west) -- (Cw.east);

     	% label OS, CP, and SQ blocks - physical interpretation
	\node [above = -0.07cm of os](extra){\tc{RoyalBlue}{$\ba{c} \mbox{\bf generalized} \\[-0.05cm] \mbox{\bf diffusion} \ea$}};  %
        \node [above = -0.07cm of cp](extra){\tc{red}{$\ba{c} \mbox{\bf polymer} \\[-0.05cm] \mbox{\bf stretching} \ea$}};  %
        \node [below = -0.07cm of cp](extra){\tc{dred}{$\ba{c} \mbox{\bf source of} \\[-0.05cm] \mbox{\bf amplification} \ea$}};  %
	\node [above = -0.07cm of sq](extra){\tc{RoyalBlue}{$\ba{c} \mbox{\bf viscous} \\[-0.05cm] \mbox{\bf dissipation} \ea$}};  %

\end{tikzpicture}
%_______________________________________________________________________________
}
        \label{fig.bd-OB-kx0}}}
    \end{tabular}
    \end{tabular}
    }
    \caption{Block diagrams of the frequency response operators that map the forcing to the velocity fluctuations in streamwise-constant channel flows of
    (a) Newtonian fluids; and 
    (b) inertialess Oldroyd-B fluids.
    The thick black lines indicate the part of the system responsible for large amplification. 
    In Newtonian fluids amplification originates from vortex tilting, i.e., the operator
    $
    \Acpn
    $
    in Equation~\ref{eq.SC} and in viscoelastic fluids it originates from polymer stretching, i.e., the operator
    $
    \Acpv.
    $
    In Newtonian fluids, singular values of the frequency responses from $d_l$ to $r$ are proportional to $Re^2$, for $r = u$ and $l = \{2, 3\}$; to $Re$, for $\{r = u$, $l = 1$; $r = \{ v, w \}$, $l = \{ 2, 3 \} \}$; and are equal to zero for $\{ r = \{ v, w \}$, $l = 1\}$; in inertialess flows of viscoelastic fluids, they are proportional to $\We$, for $\{ r = u$, $l = \{2, 3\} \}$; all other singular values are $\We$-independent. 
    }
    \label{fig.comparison}
    \vspace*{-0.35cm}
    \end{figure}

	\vspace*{-1.5ex}
\subsection{Turbulent channel and pipe flows of Newtonian fluids}
	\label{sec.turbulent}

% As demonstrated in Section~\ref{sec.lnse}, input-output analysis of the linearized NS equations captures the early stages of transition in channel flows and identifies mechanisms for subcritical transition.
\citet{leekimmoi90} used DNS of homogeneous turbulence to demonstrate that the linear amplification of eddies that interact with large mean shear induces streamwise streaks even in the absence of a solid boundary. This study also employed linear rapid distortion theory~\citep{pop00} to predict the lack of isotropy and the structure of turbulence at high shear rate. Furthermore,~\citet{kimlim00} used DNS of a turbulent channel flow to show decay of near-wall turbulence in the absence of the linear vortex tilting term.

In contrast to the laminar base flow, the time-averaged turbulent mean velocity is not a solution of the NS equations and even the question of what to linearize around can be contentious~\citep{bensiparndanles16}. Since the linearized NS equations around the turbulent mean flow are stable~\citep{mal56,reytie67}, they are well-suited for input-output analysis.~\citet{butfar93} utilized transient growth analysis over a horizon determined by the eddy turnover time to show that the streak spacing of approximately $100$ wall units represents the optimal response of the NS equations linearized around the turbulent mean flow.~\cite{mcksha10} employed a gain-based decomposition of fluctuations around mean velocity in turbulent pipe flow to characterize energetic structures in terms of their convection speeds and wavelengths. This study highlighted the role of critical layers in wall-normal localization of experimentally identified energetic modes and related the wave speed, $c \DefinedAs \omega/k_x$, to the wall-normal localization of the dominant flow structures.~\cite{moashatromck13} leveraged the role of wave speed to formally determine three different scalings for the most amplified modes; showed that these scales are consistent with inner, logarithmic, and outer layers in the turbulent mean velocity; and established dependence of the dominant resolvent \mbox{modes on the spatial coordinates.}

Other classes of linearized models have also been utilized to identify the spatio-temporal structure of the most energetic fluctuations in turbulent flows. In particular, the turbulent mean flow can be obtained as the steady-state solution of the NS equations in which molecular viscosity is augmented with turbulent eddy-viscosity~\citep{reytie67,reyhus72-3}. \cite{deljim06,cospujdep09,pajgarcosdep09,hwacosJFM10a,hwacosJFM10b} demonstrated that transient growth and input-output analyses of the resulting linearization qualitatively capture features of turbulent flows.

\begin{figure}[b!]
    \centering
    \vspace*{-0.15cm}
    {
    \begin{tabular}{cccccc}
    \hspace*{-0.5cm}
    \begin{tabular}{c}
		\vspace{0.5cm}
		\normalsize{\rotatebox{90}{$k_z E_{k_z}$}}
	\end{tabular}
	&
    \hspace*{-5.65cm}	
    \begin{tabular}{c}
     \includegraphics[width=0.28\textwidth]{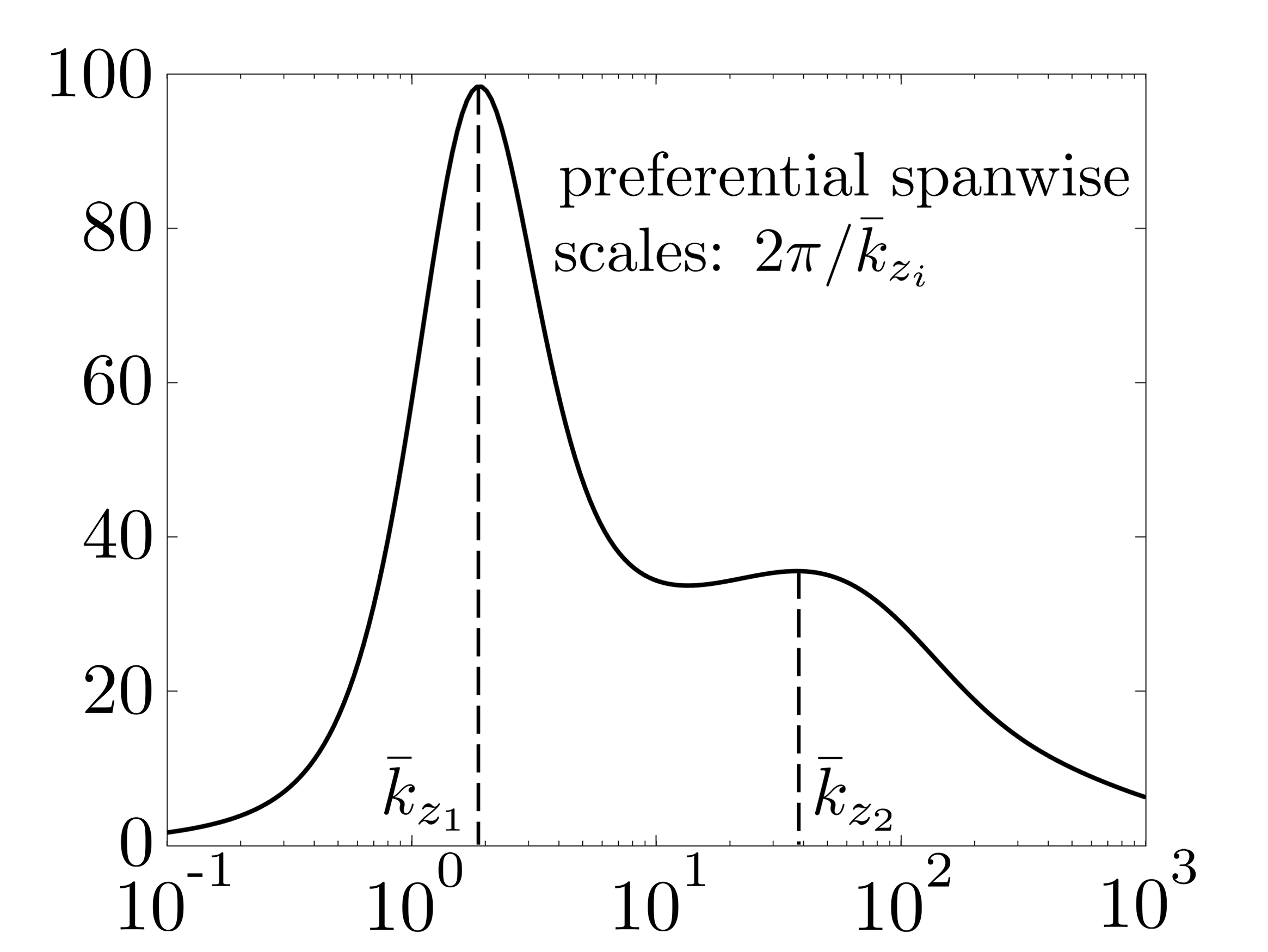}
           \\[-0.1cm]
           $k_z$
%           \\[-0.25cm]
%            \subfigure[streamwise]
%   {\label{fig.H2u}}
           \end{tabular}   
           &    
           \hspace*{-9.85cm}
           \begin{tabular}{c}
		\vspace{0.5cm}
		\normalsize{\rotatebox{90}{$y$}}
	\end{tabular}
	&
   \hspace*{-9.85cm}
   \begin{tabular}{c}
   \includegraphics[width=0.28\textwidth]{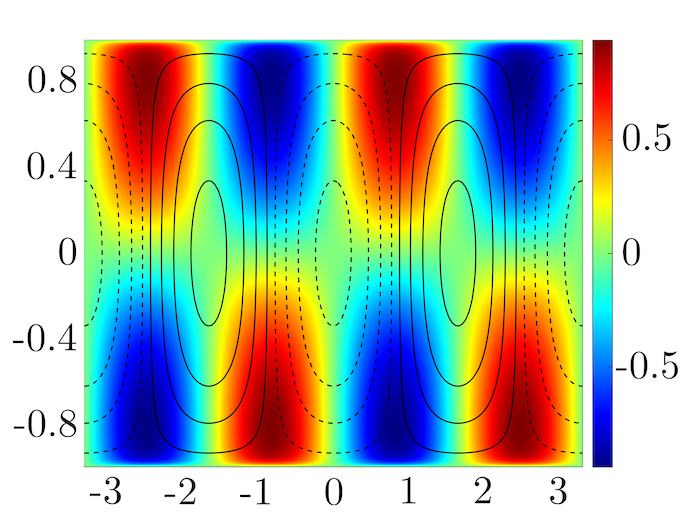}            
           \\[-0.1cm]
           $z$
%           \\[-0.25cm]
%            \subfigure[wall-normal]
%   {\label{fig.H2v}}
           \end{tabular}
           &    
           \hspace*{-9.85cm}
           \begin{tabular}{c}
		\vspace{0.65cm}
		\normalsize{\rotatebox{90}{$y^+$}}
	\end{tabular}
	& 
   \hspace*{-9.75cm}
   \begin{tabular}{c}
    \includegraphics[width=0.28\textwidth]{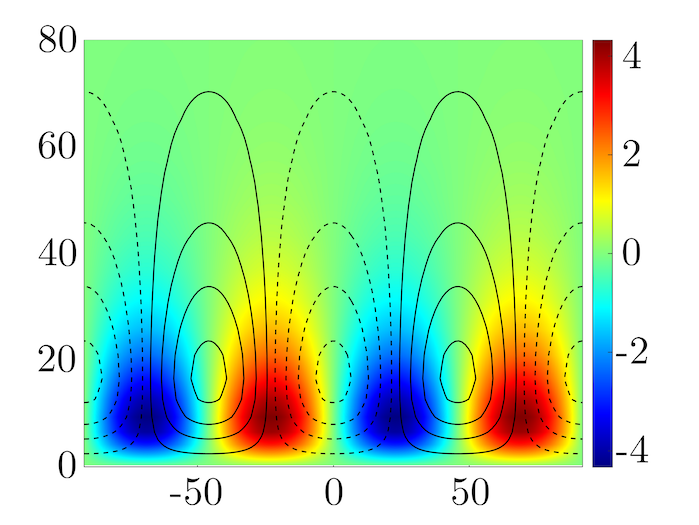}
           \\[-0.1cm]
           $z^{+}$
           \end{tabular}
           \\[-0.25cm]
            \hspace*{-0.5cm}
    &
     \hspace*{-5.65cm}	
     \subfigure[]
   {\label{fig.a}}
	&  
           \hspace*{-9.85cm}
           &
            \hspace*{-9.85cm}
    \subfigure[]
   {\label{fig.b}}
   &
    \hspace*{-9.85cm}
    &
    \hspace*{-9.75cm}
   \subfigure[]
   {\label{fig.c}}
	\end{tabular}
    }
    \caption{(a) Premultiplied energy spectrum, $k_z E_{k_z}$; and (b,c) dominant flow structures resulting from stochastically-forced eddy-viscosity enhanced linearization around the turbulent mean flow with $Re = 547$ (based on friction velocity) and $k_x = 0$. Color plots display most energetic streamwise velocity fluctuations $u (z,y)$ and contour lines show streamfunction fluctuations with the spanwise wavelength determined by (b) $2 \pi / \bar{k}_{z1}$ (channel-wide streaks); and (c) $2 \pi / \bar{k}_{z2}$ (near-wall streaks).}
    \label{fig.nuT-R547}
    \end{figure}

For a turbulent channel flow with $Re = 547$ and $k_x = 0$, {\bf Figure~\ref{fig.nuT-R547}} demonstrates the emergence of channel-wide and near-wall streaks in a stochastically-forced eddy-viscosity-enhanced linearized model. The values of $k_z$ where the two peaks in the premultiplied energy spectrum $k_z E_{k_z}$ emerge determine the spanwise length scales of the most energetic response of velocity fluctuations to stochastic forcing (left plot). Streamwise velocity fluctuations that contain the most variance are harmonic in $z$ and their wall-normal shapes are determined by the principal eigenfunctions of the stationary covariance operator $\Vk$ (middle and right plots). Pairs of counter-rotating streamwise vortices (contour lines) distribute momentum in the ($y,z$)-plane and promote amplification of high (hot colors) and low (cold colors) speed streamwise streaks. The most energetic flow structures occupy the entire channel width and the second set of strongly amplified fluctuations is determined by near-wall streaks. 

	%=========
	% Figure 7 %
	%=========

	%=========
	% Figure 8  %
	%=========
	\begin{figure}[b!]
    \centering
    \vspace*{-0.35cm}
    {
    \begin{tabular}{cc}
    \begin{tabular}{c}
    {\hspace*{-1.cm} \subfigure[]{\scalebox{.7}{%_______________________________________________________________________________
%
%   Block diagram of the periodic modification to the original dynamics
%   drawn from Right to Left
%
%   Mihailo Jovanovic, February 17, 2020
%_______________________________________________________________________________
%
% TikZ styles for drawing
%
\input{figures/Tikz_common_styles}
%
%   set a filename for externalization
% \tikzsetnextfilename{clp_2dof_input_pert_config}
%
\noindent
\begin{tikzpicture}[scale=1, auto, >=stealth']

	% output node
	% starting point for uend
	% \node [input, name=uend] {};
	\node[] (uend) at (0,0) {};
	
	\node[] (uendup) at ($(uend) + (0.cm,0.25cm)$) {};
		
	% \node[] (midpoint1) at ($(uend) + (1.5cm,-1.75cm)$) {};
	 
   % original dynamics
    \node[block, minimum height = 1.5cm, top color=RoyalBlue!20, bottom color=RoyalBlue!20] (plant) at ($(uend) + (4cm,0)$) {$\ba{c} \mbox{\bf original} \\ \mbox{\bf dynamics} \ea$};
    
    % periodic feedback
    \node[block, minimum height = 1cm, top color=red!20, bottom color=red!20] (periodic) at ($(uend) + (4cm,-1.75cm)$) {$\ba{c} \mbox{\bf periodic} \\ \mbox{\bf modification} \ea$};

    	% outputs to plant
	\node[] (plantupwest) at ($(plant.west) + (0.cm,0.25cm)$) {};
	\node[] (plantdownwest) at ($(plant.west) - (0.cm,0.25cm)$) {};
	
	\node[] (midpoint1) at ($(plantdownwest) - (1.5cm,0cm)$) {};
	\node[] (midpoint2) at ($(plantdownwest) - (1.5cm,1.5cm)$) {};

	\node[] (midpoint3) at ($(periodic.east) + (1.25cm,0cm)$) {};
	\node[] (midpoint4) at ($(midpoint3) + (0cm,1.5cm)$) {};

	% input nodes
	% inputs to plant
	\node[] (plantupeast) at ($(plant.east) + (0.cm,0.25cm)$) {};
	\node[] (plantdowneast) at ($(plant.east) - (0.cm,0.25cm)$) {};
	
	% starting point for flow disturbances	
     	\node[] (dbegin) at ($(plantupeast) + (3cm,0.cm)$) {};
	
	% input dbegin to block plant
    	\draw [connector] (dbegin.west) -- node [midway, above] {$\ba{c} \mbox{\bf flow} \\ \mbox{\bf disturbances} \ea$} (plantupeast.center);
	
	% input dbegin to block plant
    	\draw [connector] (plantupwest.center) -- node [midway, above] {$\left[ \ba{c} \mbox{\bf kinetic energy} \\ \mbox{\bf skin-friction} \ea \right]$} (uendup.center);
	
	% connect plant with uenddown
         \draw [line] (plantdownwest.center) -- (midpoint1.center);
         
         % connect plant with uenddown
         \draw [line] (midpoint1.center) -- (midpoint2.center);

	% connect midpoint1 with periodic block
	 \draw [connector] (midpoint2.center) -- (periodic.west);
	 
	  \draw [line] (periodic.east) -- (midpoint3.center);

         \draw [line] (midpoint3.center) -- (midpoint4.center);

	 \draw [connector] (midpoint4.center) -- (plantdowneast.center);
	 
	 \node [below = -0.07cm of periodic](extra){\tc{red}{$\ba{c} \mbox{\bf sensor-less feedback} \ea$}};  %

\end{tikzpicture}
%_______________________________________________________________________________
}
           \label{fig.bd-periodic}}}
    \end{tabular}
    &    
    \begin{tabular}{c}
    {\hspace*{-1.35cm} \subfigure[]{\scalebox{.7}{%_______________________________________________________________________________
%
%   Block diagram of the nuT modification to the original dynamics
%   drawn from Right to Left
%
%   Mihailo Jovanovic, February 17, 2020
%_______________________________________________________________________________
%
% TikZ styles for drawing
%
\input{figures/Tikz_common_styles}
%
%   set a filename for externalization
% \tikzsetnextfilename{clp_2dof_input_pert_config}
%
\noindent
\begin{tikzpicture}[scale=1, auto, >=stealth']

	% output node
	% starting point for uend
	% \node [input, name=uend] {};
	\node[] (uend) at (0,0) {};
	
	\node[] (uendup) at ($(uend) + (0.cm,0.25cm)$) {};
		
	% \node[] (midpoint1) at ($(uend) + (1.5cm,-1.75cm)$) {};
	
   % linear dynamics
    \node[block, minimum height = 1.5cm, top color=RoyalBlue!20, bottom color=RoyalBlue!20] (plant) at ($(uend) + (4cm,0)$) {$\ba{c} \mbox{\bf linearized} \\ \mbox{\bf dynamics} \ea$};

    % nuT
    \node[block, minimum height = 1.cm, top color=red!20, bottom color=red!20] (nuT) at ($(uend) + (1.85cm,-1.9cm)$) {$\nu_T =  cRe^2 (k^2/\epsilon)$};

    % mean flow
    \node[block, minimum height = 1.5cm, top color=red!20, bottom color=red!20] (meanflow) at ($(uend) + (6.5cm,-1.9cm)$) {\begin{tabular}{c} {\bf mean-flow} \\ {\bf equations} \end{tabular}};

    	% outputs to plant
	%\node[] (plantupwest) at ($(plant.west) + (0.cm,0.25cm)$) {};
	\node[] (plantdownwest) at ($(plant.west) - (0.cm,0.0cm)$) {};
	
	\node[] (midpoint1) at ($(plantdownwest) - (3cm,0cm)$) {};
	\node[] (midpoint2) at ($(plantdownwest) - (3cm,1.9cm)$) {};
	
	% input nodes
	% inputs to plant
	\node[] (plantupeast) at ($(plant.east) + (0.cm,0.25cm)$) {};
	\node[] (plantdowneast) at ($(plant.east) - (0.cm,0.25cm)$) {};
	
	% output of mean
	\node[] (meanupeast) at ($(meanflow.east) + (0.cm,0.25cm)$) {};
	\node[] (meandowneast) at ($(meanflow.east) - (0.cm,0.25cm)$) {};
	
	\node[] (midpoint4) at ($(meanflow.east) + (2.5cm,-0.25cm)$) {};
	
	% starting point for flow disturbances	
     	\node[] (dbegin) at ($(plantupeast) + (3.3cm,0.cm)$) {};
	
	\node[] (midpoint3) at ($(plantdowneast) + (3.2cm,0cm)$) {};

	% input dbegin to block plant
    	\draw [connector] (dbegin.west) -- node [midway, above] {$\ba{c} \mbox{\bf stochastic} \\ \mbox{\bf forcing} \ea$} (plantupeast.center);
	
	% connect plant with uenddown
         \draw [line] (plantdownwest.center) --  node [pos=0.5, above] {\begin{tabular}{l}{\bf second-order} \\ {\bf statistics:}~($k,\epsilon$) \end{tabular}} (midpoint1.center);

         % connect plant with uenddown
         \draw [line] (midpoint1.center) -- (midpoint2.center);

	% connect midpoint1 with nuT block
	 \draw [connector] (midpoint2.center) -- (nuT.west);
    
    % connect nuT block with meanflow block
     \draw [connector,double] (nuT.east) -- node [pos=0.5, above] {$\ba{c} \mbox{\bf turbulent} \\ \mbox{\bf viscosity} \ea$} (meanflow.west);
	
    % connect meanflow block with midpoint3
	 \draw [line,double] (meanupeast.center) -| node [pos=0.75, right] {\begin{tabular}{l}{\bf turbulent} \\ {\bf mean} \\ {\bf velocity} \end{tabular}} (midpoint3.center);
    
     % connect midpoint3 with plant block
	 \draw [connector,double] (midpoint3.center) -- (plantdowneast.center);
	 
	 % connect midpoint3 with plant block
	 \draw [connector] (meandowneast.center) -- node [midway, below] {\bf skin-friction} (midpoint4.center);
	
	% \node [below = -0.07cm of nuT](extra){\tc{red}{$\ba{c} \mbox{\bf sensor-less feedback} \ea$}};  %

\end{tikzpicture}
%_______________________________________________________________________________
}
        \label{fig.bd-nuT}}}
    \end{tabular}
    %\subfigure[]{ \setlength{\unitlength}{1.cm} \input{figures/fig6a}
    \end{tabular}
    }
    \caption{Block diagrams of (a) a modification to the dynamics introduced by spatio-temporal oscillations which introduce a {\em sensor-less feedback\/} by changing a base flow $U_0(y)$ to a periodic profile; and (b) a simulation-free approach for determining the influence of control on skin-friction drag in turbulent flows. The hollow arrows represent coefficients into the mean-flow and linearized equations. In~\citet{moajovJFM12}, the turbulent mean velocity is updated once.
}
    \label{fig.periodic}
    \end{figure}

	\vspace*{-5ex}
\section{CONTROL OF TRANSITIONAL AND TURBULENT FLOWS}

Flow control by sensor-less means is often inspired by the desire to bring the efficiency of birds and fish to engineering systems. Control of conductive fluids using the Lorentz force, periodic blowing and suction, wall oscillations, and geometry modifications (e.g., riblets, superhydrophobic surfaces, and jet-engine chevrons) are characterized by the absence of sensing capabilities and implementation of control without measurement of the relevant flow quantities or disturbances. Rather, as illustrated in {\bf Figure~\ref{fig.bd-periodic}}, the dynamics are impacted by spatio-temporal oscillations through geometry or base velocity modifications.

\cite{minsunspekim06} used DNS to show that a blowing and suction in the form of an upstream traveling wave can provide a {\em sustained sub-laminar drag\/} in a fully-developed turbulent channel flow. This paper inspired other researchers to examine fundamental limitations of streamwise traveling waves for control of transitional and turbulent flows~\citep{marjosmah07,hoefuk09,bew09,fuksugkas09,moajovJFM10,liemoajovJFM10}. Furthermore, simulations and experiments showed that spanwise wall oscillations can reduce skin-friction drag by as much as $45\%$~\citep{junmanakh92,laaskamor94,chodebcla98,cho02,ric04}. While these and other studies (e.g.,~\citet{fratalbracos06}) demonstrate the potential of sensor-less periodic strategies, until recently a model-based design for transitional and turbulent flows \mbox{remained elusive.} 

In Section~\ref{sec.stw}, we highlight the utility of the input-output framework in the design of traveling waves for controlling the onset of turbulence while achieving positive net efficiency~\citep{moajovJFM10}; and, in Section~\ref{sec.swo}, we describe how stochastic analysis in conjunction with control-oriented turbulence modeling quantifies the effect of control on turbulent flow dynamics and identifies the optimal period of oscillations for drag reduction~\citep{moajovJFM12}. Apart from demonstrating the merits of the input-output approach in the design of periodic strategies for controlling laminar and turbulent flows, we also illustrate how to overcome challenges that arise in ``secondary receptivity analysis'', i.e., in nonmodal analysis of the dynamics associated with spatially- or \mbox{time-periodic base flows.}

	\vspace*{-3ex}
\subsection{Controlling the onset of turbulence by streamwise traveling waves} 
	\label{sec.stw}

Let channel flow be subject to a uniform pressure gradient and a zero-net-mass-flux blowing and suction along the walls,
	$
        V(y = \pm 1)
        =
        \mp 2 \alpha
        \cos \, (\omega_x (\bar{x} - c t));
	$
see {\bf Figure~\ref{fig.channel-stw}}. Here, $\alpha$, $\omega_x$, and $c$ denote amplitude, frequency, and speed of the wave that travels in the streamwise direction $\bar{x}$. Positive/negative values of $c$ identify downstream/upstream waves, and $c = 0$ gives a standing wave. The Galilean transformation,
    $
    x \DefinedAs \bar{x} - c t,
    $
eliminates the time dependence in $V(\pm 1)$ and the steady-state solution of the NS equations, 
	$
    \bar{\bf u}
    =
    (U(x,y), V(x,y), 0),
    $ 
does not depend on $t$ in the frame of reference that travels with the wave. The new laminar base flow $\bar{\bf u}$ is no longer a parabola:~it is periodic in $x$, with frequency $\omega_x$, and it contains both streamwise and wall-normal~components, $U(x,y)$ and $V(x,y).$ 

	\vspace*{-3ex}
\subsubsection{Net efficiency of modified base flow}
	\label{sec.stw-net}
Blowing and suction induces a bulk flux in the direction opposite to the direction in which the wave travels~\citep{hoefuk09}. This {\em pumping mechanism\/} occurs even in the absence of the pressure gradient and a weakly-nonlinear analysis explains it. For small amplitude $\alpha$, $U (x,y)$ is given by 
	\[
    \ba{cccccclcc}
    & \; &
    \mbox{\tc{RoyalBlue}{\bf parabola}}    
    & \; &
    \mbox{\tc{dred}{\bf mean drift}}     
    & \; &
    \mbox{\tc{RoyalBlue}{\bf ~~~~~~\;oscillatory components}}
    & \; &
    \\
    U(x,y)
    & = \; &
    \tikz[baseline]{
            \node[fill=blue!20,anchor=base] (t1)
            {$U_{0}(y)$};
            } 
    & + \; &
    \tikz[baseline]{
            \node[fill=red!20,anchor=base] (t1)
            {$\tc{red}{\alpha^2} \, U_{20} (y)$};
            }
     & + \; &
      \tikz[baseline]{
            \node[fill=blue!20,anchor=base] (t1)
            {$ \tc{red}{\alpha}
   	 \left(
    	U_{1p}(y) \, \mre^{\mri \omega_x x }
    	\, + \,
	U_{1m}(y) \, \mre^{-\mri \omega_x x }
	\right)
	\phantom{~~\;\,}
            $};
            }
    & + \; &
    % O (\tc{red}{\alpha^3})
    \\
    & \; &    
    & \; &
    & \; &
      \tikz[baseline]{
            \node[fill=blue!20,anchor=base] (t1)
            {$    	 
            \tc{red}{\alpha^2}
    	\left(
    	U_{2p}(y) \, \mre^{\mri 2 \omega_x x }
	\, + \,
	U_{2m}(y) \, \mre^{-\mri 2 \omega_x x }
	\right)
	$};
            }
    & + \; &
    O (\tc{red}{\alpha^3}).
    \ea
    \]
In addition to an oscillatory $O(\alpha)$ correction to $U_0 (y)$ with frequency $\omega_x$, both the second harmonic $2 \omega_x$ and the mean flow correction $U_{20}(y)$ are induced by the quadratic nonlinearity in the NS equations at the level of $\alpha^2$. For the fixed pressure gradient, the skin-friction drag coefficient of the base flow is inversely proportional to the square of the bulk flux. Since the integral of $U_{20}(y)$ is positive for the upstream and negative for the downstream waves~\citep{hoefuk09,moajovJFM10}, upstream/downstream waves reduce/increase skin-friction drag coefficient relative to the laminar uncontrolled~flow. 

The net efficiency of wall-actuation is given by the difference of the produced and required powers~\citep{quaric04}. These two quantities, respectively, determine increase/decrease in bulk flux relative to the flow with no control and the control effort exerted at the walls. Compared to laminar uncontrolled flow, any strategy based on blowing and suction reduces net efficiency~\citep{bew09,fuksugkas09}. However, if uncontrolled flow becomes turbulent, both upstream and downstream waves of small enough amplitudes can improve net efficiency~\citep[Section 2.4]{moajovJFM10}.~\cite{moajovJFM10} also demonstrated that, apart from the net efficiency, {\em the dynamics of fluctuations around the modified base flow have to be evaluated when designing the traveling waves.}

	\vspace*{-3ex}
\subsubsection{Dynamics of velocity fluctuations}

The laminar base flow induced by the traveling waves is periodic in $x$, with frequency $\omega_x$, and the resulting linearization is not translationally-invariant in the streamwise direction. The normal modes in $z$ are still harmonic, $\mre^{\mri k_z z}$, but in $x$ they are given by the Bloch waves, which are determined by a product of $\mre^{\mri \theta x}$ and the $2 \pi/\omega_x$ periodic function in $x$, 
	$
	\tilde{\bd}_{\bk} (x,y,t)
         = 
    	\tilde{\bd}_{\bk} (x + 2 \pi/\omega_x,y,t),
	$	
	\be
    \bd(x,y,z,t)
    \; = \;
    \tilde{\bd}_{\bk} (x,y,t)
    \,
    \mre^{\mri (\theta x \, + \, k_z z)}
    \; = \;
    \ds{\sum_{n \, = \, - \infty}^{\infty}}
    \tilde{\bd}_{\bk,n} (y, t)
    \,
    \mre^{\mri ((\theta \, + \, n \omega_x) x \, + \, k_z z)},
    ~~
    \theta \in [0, \, \omega_x),
    \non
    \ee
where $\bk \DefinedAs (\theta,k_z)$ and $\tilde{\bd}_{\bk,n} (y,t)$ are the coefficients in the Fourier series expansions of $\tilde{\bd}_{\bk} (x,y,t)$. In this case, signals in Equation~\ref{eq.lnse1} are the $\bk$-parameterized bi-infinite column vectors whose components are determined by the corresponding Fourier series coefficients, e.g., 
    $
    \bd_{\bk} (t)
    \DefinedAs
    \col \, \{\tilde{\bd}_{\bk,n} (y,t)\}_{n \, \in \, \bbZ},
    $
and similarly for $\bpsi_{\bk} (t)$ and $\bxi_{\bk} (t)$. Thus, for each $\bk$, $\Ak$, $\Bk$, and $\Ck$ in Equation~\ref{eq.lnse1} are bi-infinite matrices whose entries are operators in the wall-normal direction $y$~\citep{moajovJFM10}, and the frequency response operator $\Tk (\mri \omega)$ in Equation~\ref{eq.fr} maps 
	$
    \hat{\bd}_{\bk} (\mri \omega)
    \DefinedAs
    \col \, \{\hat{\tilde{\bd}}_{\bk,n} (y,\mri \omega)\}_{n \, \in \, \bbZ}
    $
    to 
    $
    \hat{\bxi}_{\bk} (\mri \omega)
    \DefinedAs
    \col \, \{\hat{\tilde{\bxi}}_{\bk,n} (y,\mri \omega)\}_{n \, \in \, \bbZ}.
    $

\begin{figure}[hb!]
    \centering
    \vspace*{-0.25cm}
    {
    \begin{tabular}{cc}
    \hspace*{-4cm}
    \begin{tabular}{c}
   \subfigure[downstream waves: ($\omega_x = 2, c = 5$)]
   {\includegraphics[width=0.35\textwidth]{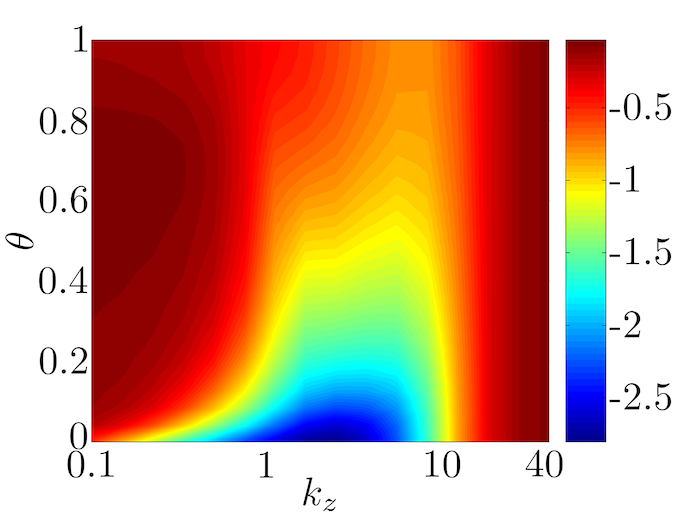}
           \label{fig.g-dtw}}
           \end{tabular}   
           &    
   \hspace*{-8.5cm}
   \begin{tabular}{c}
    \subfigure[upstream waves: ($\omega_x = 0.5, c = -2$)]
    {\includegraphics[width=0.35\textwidth]{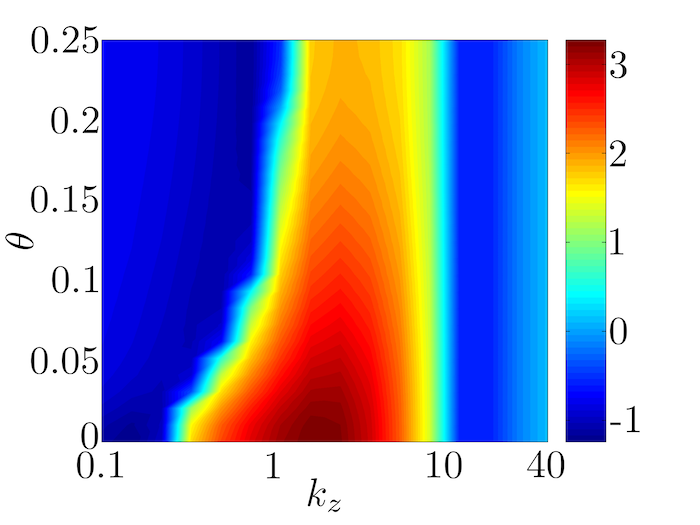}
           \label{fig.g-utw}}
           \end{tabular}
	\end{tabular}
    }
    \caption{The second order correction to the energy amplification in Equation~\ref{eq.g} visualized using a sign-preserving logarithmic scale, 
    $
    {\rm sign}(g_\bk)
    \log_{10} (1 + |g_\bk|),
    $
    in channel flow with $Re = 2000$ (based on the centerline velocity of the parabolic laminar profile and the channel half-height). While the downstream waves with selected parameters reduce amplification for all values of $\theta$ and $k_z$, the upstream waves promote amplification for a broad range of $\theta$ and $k_z$.}
    \label{fig.g-stw}
    \vspace*{-0.5cm}
    \end{figure}

Since modal stability does not capture the {\em early stages of transition\/},~\citet{moajovJFM10} utilized input-output analysis of a linearization around ($U(x,y), V(x,y), 0$) to quantify the effect of control on amplification of stochastic disturbances and identify waves that reduce receptivity relative to the flow without control. A discretization in $y$ and truncation of bi-infinite matrices in Equation~\ref{eq.lnse1} yield a large-scale Lyapunov Equation~\ref{eq.AL}; computing its solution to assess impact of control parameters ($\alpha,\omega_x,c$), wavenumbers ($\theta,k_z$), and the Reynolds number $Re$ on the energy amplification is demanding. Motivated by the observation that large values of $\alpha$ introduce high cost of control,~\citet{moajovJFM10} employed a perturbation analysis to efficiently compute the solution to Equation~\ref{eq.AL}. This approach offers significant advantages relative to the approach based on truncation: {\em impact of small amplitude waves on energy amplification can be assessed via computations that are of the same complexity as computations in the uncontrolled flow.} In particular, for small amplitude waves, the following {\em explicit formula\/}, 
	\beq
	\tikz[baseline]{
            \node[fill=blue!20,anchor=base] (t1)
            {$ \dfrac{\mbox{\tc{black}{energy amplification with control}}}
    	   {\mbox{\tc{black}{energy amplification without control}}}$};
            } 
%	 \dfrac{\mbox{\tc{black}{energy amplification with control}}}
%    	 {\mbox{\tc{black}{energy amplification without control}}}
    	\; = \;
    	1
    	\; + \;
	\tikz[baseline]{
            \node[fill=red!20,anchor=base] (t1)
            {$\tc{red}{\alpha^{2}} \, g_{\bk} (\omega_x,c,Re)$};
            } 
         \; + \;
         O(\alpha^4),
	\label{eq.g}
	\eeq
offers insights into the impact of control on energy amplification. For $\alpha \ll 1$, the analysis amounts to examining the dependence of the function $g_{\bk}$ in Equation~\ref{eq.g} on $\bk = (\theta,k_z)$, the frequency/speed ($\omega_x,c$) of the wave, and the Reynolds number $Re$. Positive (negative) values of $g_\bk$ identify parameters that increase (decrease) energy amplification. For channel flow with $Re = 2000$ and fixed values of $\omega_x$ and $c$, we use a sign-preserving logarithmic scale to visualize the $\bk$-dependence of the function $g_\bk$ in {\bf Figure~\ref{fig.g-stw}}. While the downstream waves with ($\omega_x = 2,c = 5$) reduce amplification for all values of $\theta$ and $k_z$, the upstream waves with ($\omega_x = 0.5,c = -2$) promote amplification for a broad range of $\theta$ and $k_z$. Thus, in addition to guaranteeing positive net efficiency relative to the uncontrolled flow that becomes turbulent (see Section~\ref{sec.stw-net}), the downstream waves also suppress energy of 3D fluctuations. On the other hand, the upstream waves with the parameters that provide favorable skin-friction coefficient of the modified laminar flow~\citep{minsunspekim06} increase amplification of the most energetic modes of the uncontrolled flow. In fact, since, at best, they exhibit receptivity similar to that of the uncontrolled flow~\citep{moajovJFM10} and can even induce modal instability of the modified laminar flow~\citep{leeminkim08}, they are not suitable for controlling the onset of turbulence. In contrast, {\em properly designed\/} downstream waves can {\em substantially reduce production of fluctuations' kinetic energy\/}~\citep{moajovJFM10} and they are an excellent candidate for preventing transition to turbulence. 

	\vspace*{-3ex}
	\subsubsection*{\tc{dred}{DNS~verification.}}
All theoretical predictions resulting from a simulation-free approach of~\citet{moajovJFM10} were verified by~\citet{liemoajovJFM10}. Their DNS confirmed that the downstream waves indeed provide an effective means for controlling the onset of turbulence and that the upstream waves promote transition even when the uncontrolled flow stays laminar. This demonstrates considerable predictive power of the input-output framework and suggests that reducing receptivity is a viable \mbox{approach to controlling transition.} 

	%=========
	% Figure 9  %
	%=========
   
	% \vspace*{-3ex}
\subsection{Turbulent drag reduction by spanwise wall oscillations} 
	\label{sec.swo}
	
Spanwise wall oscillations can reduce turbulent drag by as much as $45\%$. This observation was made in both simulations and experiments and theoretical studies~\citep{dhasi99,ban06,ricqua08} focused on explaining the effectiveness of this sensor-less strategy. Herein, we describe how input-output analysis in conjunction with a control-oriented turbulence modeling identifies the optimal period of oscillations for turbulence suppression in channel flow; see~\citet{moajovJFM12} for details.  

\vspace*{-3ex}
\subsubsection*{\tc{dred}{Modified mean flow.}}
In pressure-driven channel flow, the steady-state solution of the NS equations in which the molecular viscosity is augmented with the turbulent viscosity $\nu_{T 0} (y)$ is determined by the Reynolds-Tiederman profile, $U_0 (y)$. If the flow is also subject to     
	$
    W (y = \pm 1, t)
	=
	2 \alpha \sin \left( \omega_t t \right),
    $
the steady-state solution is given by 
	$
	(U_0 (y), 0, W_0 (y,t) 
	=
	 \alpha  
	(
	W_{p} (y) \mre^{\mri \omega_t t}
	+ 
	W_{p}^* (y) \mre^{-\mri \omega_t t}
	)).
	$  
Here, $U_0 (y)$ approximates the mean streamwise velocity of the uncontrolled turbulent flow and the {\em wall oscillations induce the time-periodic spanwise component\/} $W_0 (y,t)$ under the assumption that the turbulent viscosity is not modified by control. If this were the case, the oscillations would have no impact on $U_0$, which is at odds with simulations/experiments. In contrast, the estimates of the required power exerted by wall oscillations resulting from the use of $W_0$ closely match the DNS results of~\cite{quaric04} over a broad range of \mbox{oscillation periods~\citep[Section~2.2]{moajovJFM12}.}

\vspace*{-3ex}
\subsubsection*{\tc{dred}{Turbulence modeling.}}
The inability of the above approach to predict drag reduction arises from the fact that the wall oscillations change the turbulent viscosity of the flow with no control.~\citet{moajovJFM12} pioneered a method based on the stochastically-forced eddy-viscosity-enhanced NS equations linearized around ($U_0 (y), 0, W_0 (y,t)$) to capture the influence of control on turbulent viscosity. The approach utilizes the Boussinesq hypothesis but, in contrast to standard practice, the turbulent kinetic energy $k$ and its rate of dissipation $\epsilon$ are computed using the second-order statistics of velocity fluctuations in the linearized model. Using analogy with homogenous isotropic turbulence,~\citet[Section~3.1]{moajovJFM12} selected spatial correlations of white-in-time forcing to provide equivalence between the 2D energy spectra of the uncontrolled turbulent flow and the flow governed by the stochastically-forced linearization around ($U_0 (y),0,0$). This approach was the first to utilize available DNS data~\citep{deljim03,deljimzanmos04} of the uncontrolled turbulent flow to guide control-oriented modeling for design purposes; it takes advantage of the turbulent viscosity and the energy spectrum of the uncontrolled flow and determines the effect of control on the turbulent flow using a model-based approach. 

\vspace*{-3ex}
\subsubsection*{\tc{dred}{Dynamics of velocity fluctuations.}}
Linearization around ($U_0 (y), 0, W_0 (y,t)$) yields a time-periodic model with 
	$
	\Ak (t)
	= 
	A_{\bk,0} 
	+
	\alpha
	\,
    	(
    	A_{\bk,-1}  \mre^{-\mri \omega_t t}
	+ 
	A_{\bk,1} \mre^{\mri \omega_t t}
    	),
	$
and the solution to Equation~\ref{eq.HLE} provides two-point correlations. For small amplitude oscillations,~\cite{moajovJFM12} utilized perturbation analysis to efficiently solve this equation and identify the oscillation periods that yield the largest drag reduction and net efficiency. This approach quantifies the influence of velocity fluctuations on the turbulent viscosity in the flow with control,
	$
	\nu_T (y)
	= 
	\nu_{T0} (y)
	+
	\alpha^2 \;\! \nu_{T2} (y)
	+
	O (\alpha^4),
	$
where $\nu_{T0} (y)$ is the turbulent viscosity of the uncontrolled flow and $\nu_{T2} (y)$ is determined by the second-order corrections (in $\alpha$) to the kinetic energy $k_2 (y)$ and its rate of dissipation $\epsilon_2 (y)$. These quantities are obtained by averaging the second-order statistics resulting from a stochastically-forced linearization around ($U_0 (y), 0, W_0 (y,t)$) over the wall-parallel directions and one period of oscillations. 

\begin{figure}[t!]
    \centering
    {
    \begin{tabular}{cccc}
    \hspace*{-5cm}
    \begin{tabular}{c}
     \includegraphics[width=0.3\textwidth]{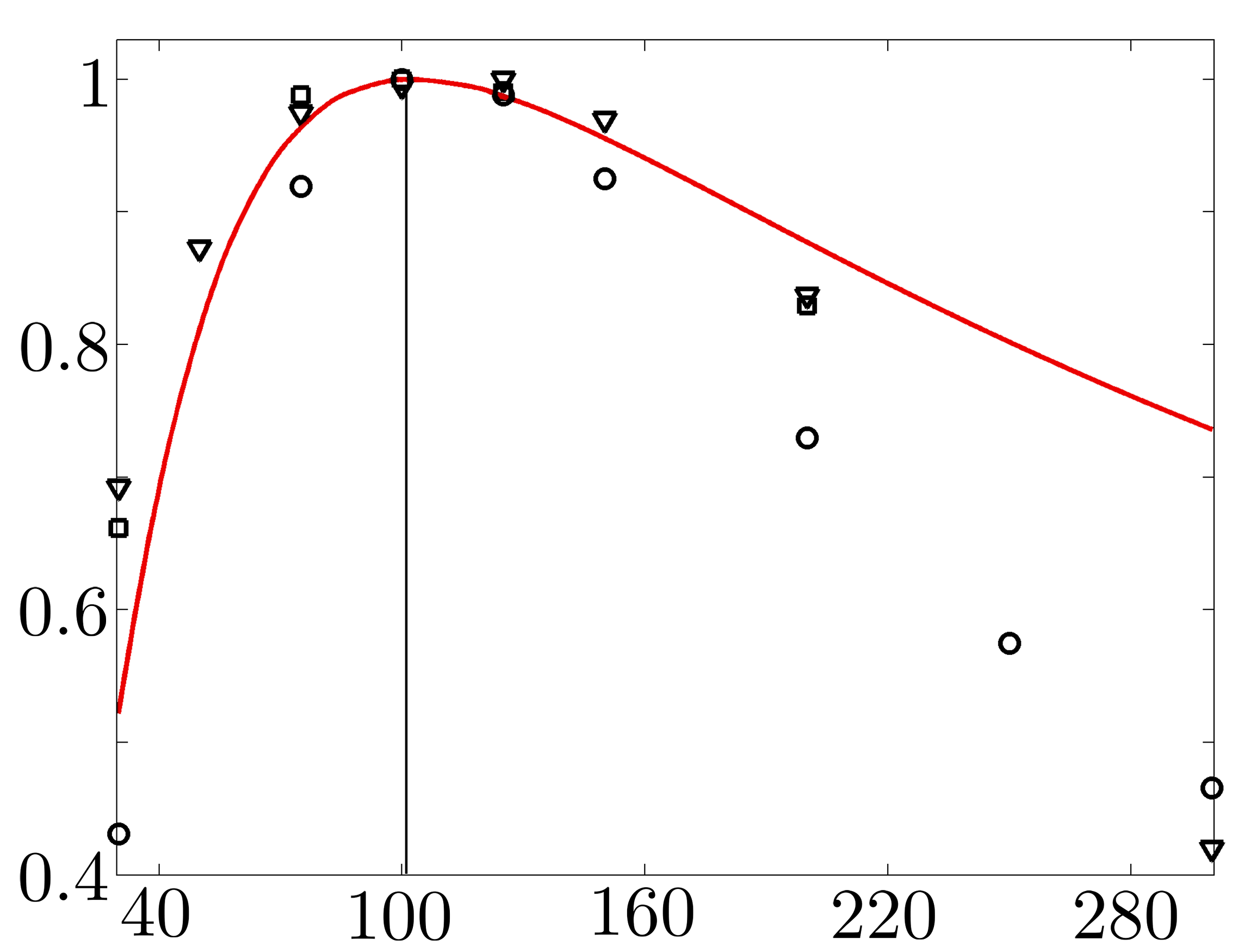}
           \\[-0.1cm]
           $T^+$
%           \\[-0.25cm]
%            \subfigure[streamwise]
%   {\label{fig.H2u}}
           \end{tabular}   
           &    
           \hspace*{-9.75cm}
           \begin{tabular}{c}
		\vspace{0.5cm}
		\normalsize{\rotatebox{90}{$k_x$}}
	\end{tabular}
	&
   \hspace*{-9.65cm}
   \begin{tabular}{c}
   \includegraphics[width=0.3\textwidth]{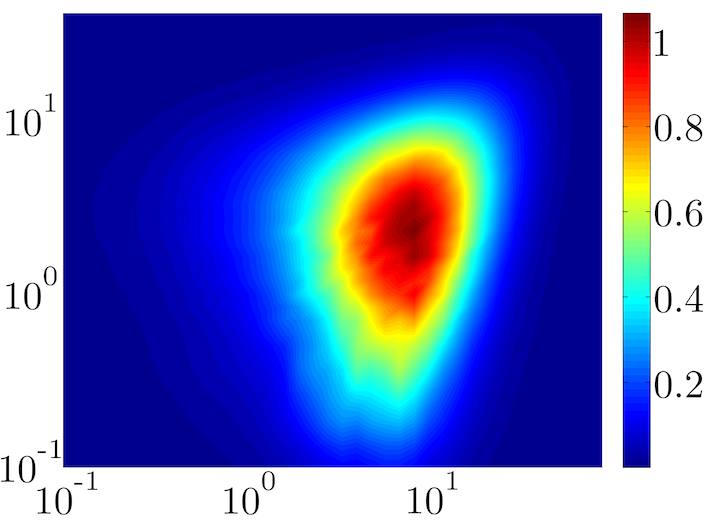}            
           \\[-0.1cm]
           $k_z$
%           \\[-0.25cm]
%            \subfigure[wall-normal]
%   {\label{fig.H2v}}
           \end{tabular}
           &    
   \hspace*{-10.cm}
   \begin{tabular}{c}
    \includegraphics[width=0.3\textwidth]{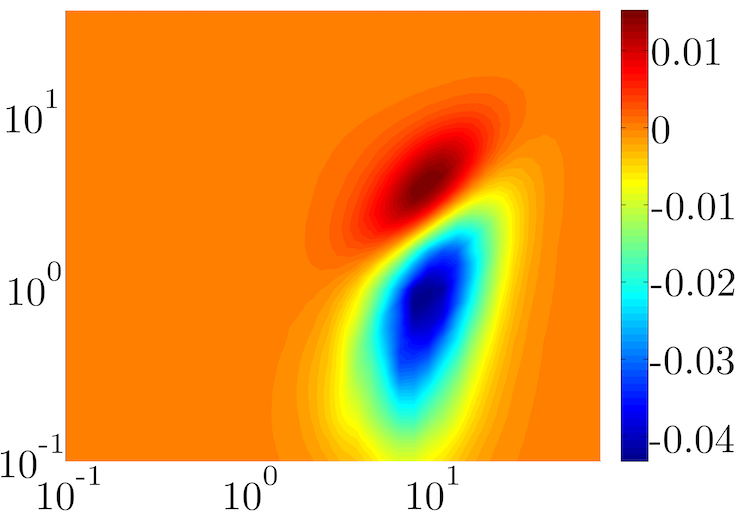}
           \\[-0.1cm]
           $k_z$
           \end{tabular}
 \\[-0.35cm]
          \hspace*{-5.cm}	
     \subfigure[]
   {\label{fig.swo-a}}
	&  
           \hspace*{-9.75cm}
           &
            \hspace*{-9.65cm}
    \subfigure[]
   {\label{fig.swo-b}}
   &
    \hspace*{-10cm}
   \subfigure[]
   {\label{fig.swo-c}}	
   \end{tabular}
    }
    \caption{(a) Second-order correction (in $\alpha$) to the skin-friction drag coefficient relative to the uncontrolled flow, $\% C_{f2} (T^+)$, normalized by $\max_{T^+} \% C_{f2} (T^+)$, as a function of the period of oscillations $T^+$ for the flow with $Re = 186$ (solid curve); DNS-based $\% C_{f} (T^+)$ normalized by the corresponding largest values at $Re = 200$~\citep{quaric04} for control amplitudes $\alpha = 2.25,\circ$; $\alpha = 6,\square$; and $\alpha = 9,\triangledown$. (b) Premultiplied DNS-based energy spectrum, $k_x k_z E_{\bk,0}$, of the uncontrolled turbulent flow with $Re = 186$; and (c) second-order correction to the energy spectrum, $k_x k_z E_{\bk,2}$, for the flow subject to wall oscillations with optimal drag-reducing period $T^+ = 102.5$.}
    \label{fig.swo}
    \vspace*{-0.9cm}
    \end{figure}

The solution to Equation~\ref{eq.HLE} and the above expression for $\nu_T$ are used to assess the influence of small amplitude oscillations on the dynamics of velocity fluctuations and to identify the optimal period of oscillations for drag reduction. For the controlled flow with constant bulk flux and the friction Reynolds number $Re = 186$, solid curve in {\bf Figure~\ref{fig.swo}} shows the second-order correction to the skin-friction coefficient $\%C_{f2} (T^+)$, normalized by its largest value, and symbols display normalized DNS results at $Re = 200$~\citep{quaric04}. A close agreement between a theoretical prediction for the optimal period resulting from input-output analysis ($T^+ = 102.5$) and DNS results ($T^+ \approx 100$) is observed. Middle and right plots in {\bf Figure~\ref{fig.swo}} show the premultiplied 2D energy spectrum of the uncontrolled flow, $k_x k_z E_{\bk,0}$, and the second-order correction, $k_x k_z E_{\bk,2}$, triggered by small amplitude oscillations with the optimal period $T^+ = 102.5$. The most energetic modes of the uncontrolled flow occur at ($k_x \approx 2.5,k_z \approx 6.5$). The red region in the right plot shows that the wall oscillations increase amplification of the modes with small streamwise wavelengths, and the blue region indicates suppression of energy of large streamwise wavelengths. This observation agrees with the study of the impact of wall oscillations on free-stream vortices in a pre-transitional boundary layer~\citep{ric11}.~\citet{moajovJFM12} also showed that the optimal wall-oscillations minimize the turbulent viscosity near the interface of the buffer and log-law layers and that oscillations are less effective at higher Reynolds numbers.

	%==========
	% Figure 10  %
	%==========
    
   	%==========
	% Figure 11  %
	%==========
    \begin{figure}[t]
	\begin{center}
	\begin{tabular}{cc}
    \begin{tabular}{c}
    {\hspace*{-1.cm} \subfigure[]{\scalebox{.5}{%_______________________________________________________________________________
%
%   tikz figure for block-diagrams in presentations and papers:
%
%   Mihailo Jovanovic, 02/20/2020
%  
%_______________________________________________________________________________
%
% TikZ styles for drawing
%
\input{figures/Tikz_common_styles}
\noindent
\begin{tikzpicture}[scale=1, auto, >=stealth']
  
    % \small
    
     \node[block, minimum height = 1cm, top color=RoyalBlue!20, bottom color=RoyalBlue!20] (sys1) {$\ba{c} \mbox{\bf linearized} \\ \mbox{\bf dynamics}\ea$};
     
     \node[block, minimum height = 1cm, top color=red!20, bottom color=red!20] (sys2) at ($(sys1.south) - (0cm,1cm)$) {$\ba{c} \mbox{\tc{red}{\bf nonlinear terms}} \\ -(\bv \cdot \nabla)\,\bv \ea$};
     
     \node[] (input-node) at ($(sys1.east) + (1cm,0)$) {}; 
     
     \node[] (output-node) at ($(sys1.west) - (2.3cm,0)$) {};
     
     \node[] (mid-node1) at ($(sys1.west) + (-1cm,1cm)$) {}; 
                     
%     \node[branch] (R1) at ($(sys1.east) + (1cm,0.0cm)$){};
     
      \node[] (R2) at ($(sys1.east) - (-2.5cm,0.7cm)$){};
      
      \node[] (R3) at ($(sys1.west) + (-2.5cm,-0.7cm)$){};
     
    % now link the nodes
                    	
    \draw [line] (sys1.west) -- ($(sys1.west) + (-1cm,0cm)$);
    
    \draw [line] (sys2.east) -|  (input-node.west);
    
    \draw [connector] (input-node.west) --  (sys1.east);
    
    \draw [connector] ($(sys1.west) + (-1cm,0cm)$) -- node [midway, above] {$\bv$} (output-node);
    
    \draw [connector] ($(sys1.west) + (-1cm,0cm)$) |- (sys2.west);
    
    \draw [dashed] (R2) -- (R3);
                                       
\end{tikzpicture}
%_______________________________________________________________________________
}
           \label{fig.ns}}}
    \end{tabular}
    &    
    \begin{tabular}{c}
    {\hspace*{-1.35cm} \subfigure[]{\scalebox{.5}{%_______________________________________________________________________________
%
%   tikz figure for block-diagrams in presentations and papers:
%
%   Mihailo Jovanovic, 02/20/20
%
%_______________________________________________________________________________
%
% TikZ styles for drawing
%
\input{figures/Tikz_common_styles}
\noindent
\begin{tikzpicture}[scale=1, auto, >=stealth']

    % \small

     \node[block, minimum height = 1.2cm, top color=RoyalBlue!20, bottom color=RoyalBlue!20] (sys1) {$\ba{rcl}\dot{\bpsi}_{\bk} & = & A_{\bk} \,\bpsi_{\bk} \, + \, B_{\bk} \, \bw_{\bk} \, + \, \tc{red}{B_{\bk} \, \bd_{\bk}} \\[0.075cm] \bv_{\bk} & = & C_{\bk} \, \bpsi_{\bk} \ea$};

     \node[block, minimum height = 1cm, top color=red!20, bottom color=red!20] (sys2) at ($(sys1.south) - (0cm,1cm)$) {$\tc{red}{-K_{\bk}}$};

     \node[] (R1) at ($(sys1.west) + (-1.5cm,-0.3cm)$){};

      \node[] (R2) at ($(sys1.east) - (-2.5cm,0.7cm)$){};

      \node[] (R3) at ($(sys1.west) + (-2.5cm,-0.7cm)$){};

    % now link the nodes

    \draw [connector] ($(sys1.east) + (2cm,.3cm)$) -- node [midway, above] {$\ba{l} \mbox{\bf white} \\ \mbox{\bf noise}~\bw_{\bk} \ea$} 	% node [midway, below] {$\bw_{\bk}$} 
    ($(sys1.east) + (0cm,.3cm)$);
          	
%    \draw [line] (sys1.east) -- (R1);

    \draw [line] (sys2.east) -|  ($(sys1.east) + (1.5cm,-.3cm)$);

    \draw [connector] ($(sys1.east) + (1.5cm,-.3cm)$) -- node [midway, above] {$\bd_{\bk}$} ($(sys1.east) + (0cm,-.3cm)$);

    \draw [connector] ($(sys1.west) + (0cm,.3cm)$) -- node [midway, above] {$\bv_{\bk}$} ($(sys1.west) + (-2cm,.3cm)$);

    \draw [line] ($(sys1.west) + (0cm,-.3cm)$) -- node[midway, above] {$\bpsi_{\bk}$} (R1.east);

    \draw [connector] (R1.east) |- (sys2.west);

%    \draw [dashed] (R2) -- (R3);

\end{tikzpicture}
%_______________________________________________________________________________
}
        \label{fig.lnse-modified}}}
    \end{tabular}
    \end{tabular}
	\end{center}
	\caption{{
(a) The NS equations can be viewed as a feedback interconnection of the linearized dynamics with the nonlinear term; (b) Stochastically-driven linearized NS equations with low-rank state-feedback modification. At the level of second-order statistics, the two representations can be made equivalent by proper selection of $\Bk$ and $\Kk$; cf.\ Equation~\ref{eq.fbk-lnse1}.}}
	\label{fig.nonlinear-vs-sf}
	\vspace*{-0.5cm}
\end{figure}

	%\vspace*{-3ex}
	\newpage
	\subsubsection*{\tc{dred}{Summary.}}
Traveling waves and wall oscillations introduce a {\em sensor-less feedback\/} via periodic modifications to the dynamics (see {\bf Figure~\ref{fig.bd-periodic}}) by changing a base flow $U_0(y)$ to a spatially- or time-periodic profile. Depending on the actuation waveform and the parameters, the properties can be improved or made worse relative to the flow without control. In contrast to a standard approach, which employs DNS and experiments to assess sensor-less periodic strategies,~\citet{moajovJFM10,moajovJFM12} developed a model-based framework for determining the influence of control on transitional and turbulent flows. These references demonstrate the critical importance of the dynamics  associated with the modified base flows for the design of effective strategies for controlling the onset of turbulence and drag reduction. The developed simulation-free method enables computationally-efficient design by merging receptivity analysis and control-oriented turbulence modeling with techniques from control theory and its utility goes beyond the case studies presented here. Recently, input-output approach was used to quantify the effect of riblets on kinetic energy and turbulent drag in channel flow~\citep{chaluh20,ranzarjovJFM20} and it is expected to enable optimal design of periodic strategies for control of transitional and turbulent flows.
	Input-output framework is also at the heart of the optimal and robust $H_2$ and $H_\infty$ feedback control strategies~\citep{zhodoyglo96} and it has recently found use in the model-based design of opposition control~\citep{luhshamck14,toeluhmck19}.

 \vspace*{-4ex}
\section{PHYSICS-AWARE DATA-DRIVEN TURBULENCE MODELING}  
	\label{sec.color}
	
	The advances in high-performance computing and measurement techniques provide abundance of data for a broad range of flows. Thus, turbulence modeling can be formulated as an inverse problem where the objective is to identify a parsimonious model that explains available and generalizes to unavailable data. Techniques from machine learning and statistical inference were recently employed to reduce uncertainty and improve predictive power of models based on the Reynolds-Averaged NS (RANS) equations~\citep{duriacxia19}. Large data sets can also be exploited to develop reduced dynamical representations~\citep{rowdaw17} but an exclusive reliance on data makes such models agnostic to physical constraints and can yield subpar performance in regimes that are not contained in the training data set. Moreover, sensing and actuation can significantly change the identified model, thereby making its use for flow control challenging~\citep{noamortad11,tadnoa11}. A compelling alternative for model-based optimization and control is to leverage data in conjunction with a prior model that arises from first principles. 

	As demonstrated in Section~\ref{sec.mechanisms}, the linearized NS equations in the presence of stochastic excitation can be used to {\em qualitatively predict structural features\/} of {\em transitional and turbulent shear flows\/}. In most prior studies excitations were restricted to white-in-time stochastic processes but {\em this assumption is often too narrow to fully capture observed statistics of turbulent flows\/}~\citep{jovgeoAPS10}. To overcome these limitations,~\citet{zarjovgeoJFM17,zarchejovgeoTAC17} developed a framework to allow for colored-in-time \mbox{inputs to the linearized NS equations.}
	
	We next briefly summarize how strategic use of data enhances predictive power of the linearized NS equations in order to capture second-order statistics of turbulent flows~\citep{zarjovgeoJFM17,zarchejovgeoTAC17,zargeojovARC20}. Since machine learning tools are physics-agnostic, the power spectrum of stochastic forcing is identified by merging tools from control theory and convex optimization. The resulting stochastic model, given by Equation~\ref{eq.fbk-lnse1}, accounts for neglected nonlinear interactions via a low-rank perturbation to the original \mbox{dynamics; see {\bf Figure~\ref{fig.nonlinear-vs-sf}}.}

	 \vspace*{-3ex}
\subsection{Completion of partially available channel flow statistics}
\label{sec.CCP}
	
Herein, we examine linearization around mean velocity in turbulent channel flow and highlight the utility of the framework developed in~\citet{zarjovgeoJFM17,zarchejovgeoTAC17}. A pseudo-spectral method~\citep{weired00} yields a finite-dimensional approximation of the operators in $y$ and a change of variables~\citep[Appendix A]{zarjovgeoJFM17} leads to an evolution model in which the kinetic energy at any $\bk$ is determined by the Euclidean norm of the state vector $\bpsi_\bk$. For given ($\Ak,\Bk$) and input statistics ($\Wk$ or $\Hk$), algebraic Relations~\ref{eq.AL} and~\ref{eq.ALc} can be used to compute the stationary covariance matrix $\Xk$ of the state $\bpsi_\bk$ in System~\ref{eq.lnse1}. However, in turbulence modeling, the converse question arises: starting from Model~\ref{eq.lnse1} and the covariance matrix $\Xk$ (resulting from DNS or experiments), can we identify the power spectrum of the stochastic input $\bd_{\bk} (t)$ that generates such statistics? For the NS equations linearized around turbulent mean velocity with white-in-time stochastic forcing, the answer to this question is negative~\citep[Figure~6]{zarjovgeoJFM17}. This limitation can be overcome by allowing for colored-in-time stochastic inputs to the linearized Model~\ref{eq.lnse1}.

The positive-definite matrix $\Xk$ is the stationary covariance of the state $\bpsi_\bk (t)$ of linear time-invariant System~\ref{eq.lnse1} with controllable pair $(\Ak,\Bk)$ if and only if~\citep{geo02b,geo02a}
\begin{align}
	\label{eq.RankConstraint-Sigma}
	\rank
	\left[
	\ba{cc}
	\Ak \Xk \,+\, \Xk \Ak^*  & ~\Bk
	\\
	\Bk^* & ~0
	\ea
	\right]
	\, = \;
	\rank 
	\left[
	\ba{cc}
	0 & ~\Bk
	\\
	\Bk^* & ~0
	\ea
	\right].
\end{align}
This fundamental condition guarantees that, for given $\Ak$, $\Bk$, and $\Xk$,  Equation~\ref{eq.ALc} can be solved for $\Hk$. It also implies that {\em any\/} $\Xk = \Xk^* \succ 0$ is admissible as a stationary covariance of $\bpsi_\bk (t)$ in Equation~\ref{eq.lnse1} if the input matrix $\Bk$ is full rank. In particular, for $\Bk = I$, Equation~\ref{eq.ALc} is satisfied with $\Hk^* = -\Ak \Xk$ and stochastically-forced System~\ref{eq.fbk-lnse1}, which for this choice of $\Hk$ and $\Wk = I$ simplifies to 
	$
	\dot{\bpsi}_\bk (t)
        = 
        - (1/2)
        \Xk^{-1} 
        \bpsi_\bk (t)
        + 
        \bw_\bk (t),
        $
can be used to generate $\Xk$. This implies that a colored-in-time input which excites all degrees of freedom in Equation~\ref{eq.lnse1} can completely {\em cancel relevant physics\/} contained in $\Ak$. Thus, in data-driven turbulence modeling, it is critically important to restrict the rank of the matrix $\Bk$ which specifies the number of inputs to the linearized NS equations.

To address this challenge,~\citet{zarjovgeoJFM17,zarchejovgeoTAC17} formulated and solved the problem of completing a subset of entries $V_{\bk,ij}^{\mathrm{dns}}$ of the stationary covariance matrix $\Vk^{\mathrm{dns}}$ of velocity fluctuations using stochastically-forced linearization around the turbulent mean velocity. The approach utilizes algebraic Relation~\ref{eq.ALc} with 
	$\Zk 
	\DefinedAs 
	\Bk
	\;\!
	\Hk^*
	+ 
	\Hk 
	\;\!
	\Bk^*
	$ 
and a maximum entropy formalism along with a convex surrogate for rank minimization to limit the number of inputs to the linearized model and identify spectral \mbox{content of the colored-in-time forcing,}	
	\begin{align} 
	\ba{clrr}
	\minimize\limits_{\tc{red}{\Xk}, \, \tc{red}{\Zk}}
	& 
	-\logdet\left( \tc{red}{\Xk} \right) 
	\; + \; 
	\gamma \sum_i \sigma_i ( \tc{red}{\Zk} )
	% \norm{ \tc{red}{\Zk} }_\star
	&
	~~~
	&
	\mbox{\tc{RoyalBlue}{\bf objective function}}
	\\[.15cm]
	\subject 
	&
	~\Ak \tc{red}{\Xk} \; + \; \tc{red}{\Xk} \Ak^* \; + \; \tc{red}{\Zk}  \;=\; 0
	&
	~~~
	&
	\mbox{\tc{dred}{\bf physics}}
	\\[0.05cm]
	&
	\, (\Ck \tc{red}{\Xk} \Ck^*)_{ij} 
	\;=\;
	V_{\bk,ij}^{\mathrm{dns}}, 
	~ (i,j) \, \in \; \mathcal{I}.
	&
	~~~
	&
	\mbox{\tc{RoyalBlue}{\bf available data}}
	 \ea
	 \tag{CC}
	\label{eq.CC}
\end{align} 
The Hermitian matrices $\Xk \succ 0$ and $\Zk$ are the optimization variables, whereas the matrices ($\Ak,\Ck$), the available entries $V_{\bk,ij}^{\mathrm{dns}}$ of $\Vk^{\mathrm{dns}}$ for a selection of indices $(i,j) \in \mathcal{I}$, and $\gamma > 0$ are known problem parameters. The first constraint in~\ref{eq.CC} comes from physics and it imposes the requirement that second-order statistics are consistent with linearization around turbulent mean velocity; and the second constraint requires that the available elements of the matrix $\Vk^{\mathrm{dns}}$ are exactly reproduced by the linearized model. The logarithmic barrier function is introduced to ensure positive-definiteness of $\Xk$~\citep{boyvan04} and the sum of singular values of the matrix $\Zk$, which reflects contribution of the stochastic input, is used as a convex proxy to restrict the rank of $\Zk$~\citep{faz02,recfazpar10}. 

The convexity of the objective function and the linearity of the constraint set in~\ref{eq.CC} imply the existence of a unique globally optimal solution ($\Xk^\star, \Zk^\star$). This solution reproduces all available entries of the stationary covariance matrix $\Vk^{\mathrm{dns}}$ resulting from DNS (or experiments) and completes unavailable second-order statistics via low-complexity stochastic dynamical model given by Equation~\ref{eq.fbk-lnse1}. In particular, the factorization of $\Zk^\star$ can be used to determine $\Bk^\star$ and $\Hk^\star$, which along with $\Xk^\star$ yield a low-rank modification $\Bk^\star \Kk^\star$ to $\Ak$ in Equation~\ref{eq.fbk-lnse1}. This approach provides a {\em model which refines predictive power of the linearized NS equations by employing data while preserving relevant \mbox{physics of turbulent flows.}} 

\begin{figure}
	\begin{tabular}{crccrc}
	\subfigure[]{\label{fig.Ruu_DNS}}
	&&&
	\subfigure[]{\label{fig.Ruu_NM}}
	&&
	\\[-.5cm]
	&
	\hspace{-2cm}
	\begin{tabular}{c}
		\vspace{.2cm}
		\normalsize{\rotatebox{90}{$y$}}
	\end{tabular}
	&
	\hspace{-4.5cm}
	\begin{tabular}{c}
		\normalsize{$V_{\bk,uu}^{\mathrm{dns}}$} 
		\\
		\includegraphics[width=5.5cm]{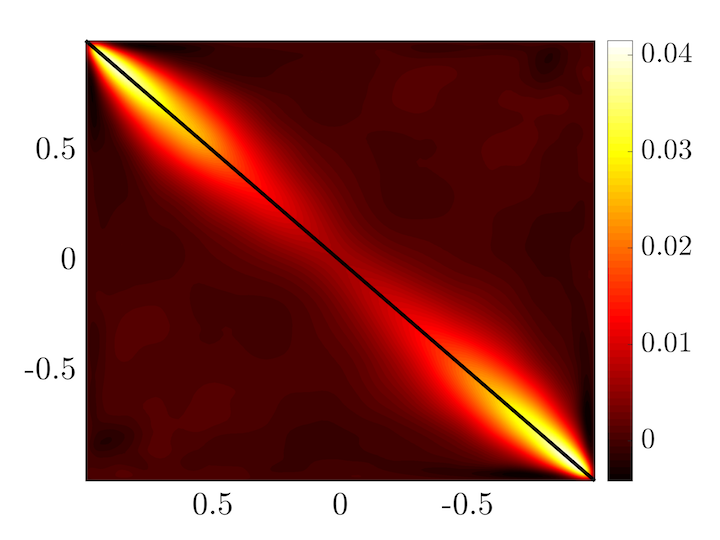}
	\end{tabular}
	\hspace{-4.15cm}
	&&&
	\hspace{-4.6cm}
	\begin{tabular}{c}
		\normalsize{$V_{\bk,uu}^\star$}
		\\
		\includegraphics[width=5.5cm]{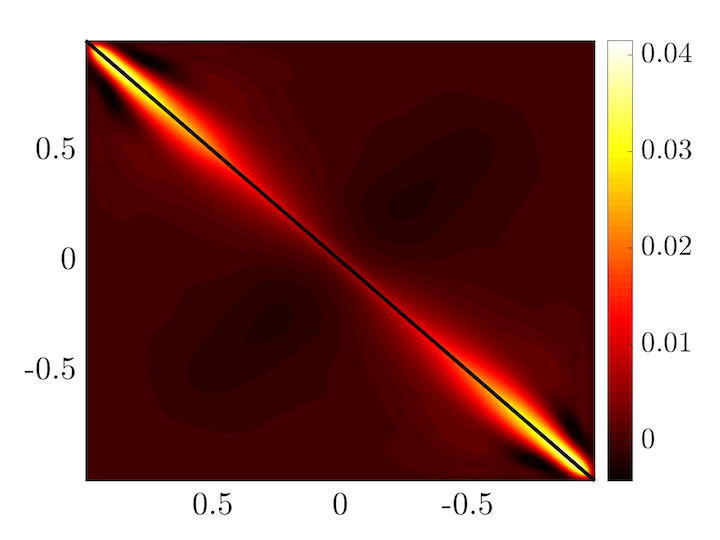}
	\end{tabular}
	\\[-.2cm]
	\subfigure[]{\label{fig.Ruv_DNS}}
	&&&
	\subfigure[]{\label{fig.Ruv_NM}}
	&&
	\\[-.5cm]
	&
	\hspace{-2cm}
	\begin{tabular}{c}
		\vspace{.4cm}
		\normalsize{\rotatebox{90}{$y$}}
	\end{tabular}
	&
	\hspace{-4.5cm}
	\begin{tabular}{c}
		\normalsize{$V_{\bk,uv}^{\mathrm{dns}}$} 
		\\
		\includegraphics[width=5.5cm]{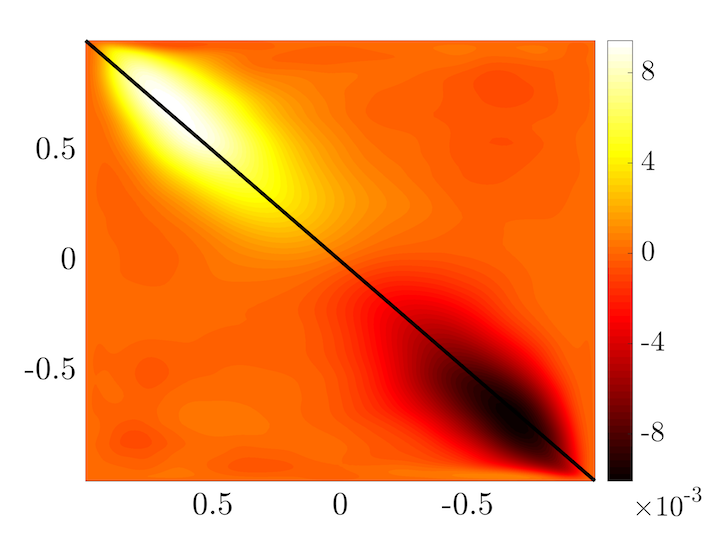}
		\\[-.1cm]
		\hspace{-.5cm}
		\normalsize{$y$}
	\end{tabular}
	\hspace{-4.15cm}
	&&&
	\hspace{-4.6cm}
	\begin{tabular}{c}
		\normalsize{$V_{\bk,uv}^\star$}
		\\
		\includegraphics[width=5.5cm]{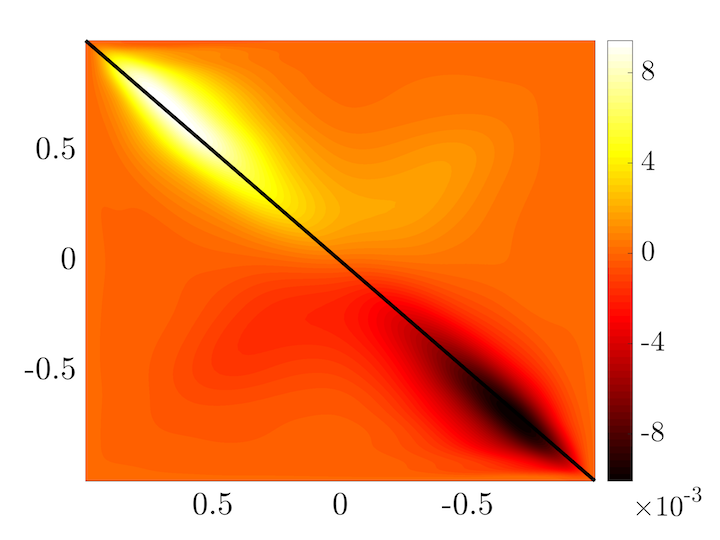}
		\\[-.1cm]
		\hspace{-.5cm}
		\normalsize{$y$}
	\end{tabular}
	\end{tabular}
		\caption{(a, b) Streamwise $V_{\bk,uu}$ and (c,d) streamwise/wall-normal $V_{\bk,uv}$ covariance matrices resulting from DNS of turbulent channel flow with $Re=186$ (left plots); and the solution to optimization problem~\ref{eq.CC} with $\gamma = 300$ (right plots) for $\bk = (2.5, 7)$. Black lines along the main diagonals mark the one-point correlations that are used as available data in~\ref{eq.CC} and are perfectly matched.}
	\label{fig.covariance_DNS_NM}
	\vspace*{-0.65cm}
\end{figure}

{\bf Figure~\ref{fig.covariance_DNS_NM}} shows covariance matrices $V_{\bk,uu}$ and $V_{\bk,uv}$ resulting from DNS of turbulent channel flow with $Re=186$ (left plots) and the solution to optimization problem~\ref{eq.CC} with $\gamma = 300$ (right plots) for $\bk = (2.5, 7)$. Black lines along the main diagonals mark the one-point correlations (in $y$) that are used as {\em available data\/} in~\ref{eq.CC} and are {\em perfectly matched}. Using a Frobenius norm measure, $\norm{\Vk^\star - \Vk^{\mathrm{dns}}}_F / \norm{\Vk^{\mathrm{dns}}}_F$, approximately $60\%$ of $\Vk^{\mathrm{dns}}$ can be recovered by the stationary covariance matrix $\Vk^\star = \Ck \Xk^\star \Ck^*$ of velocity fluctuations resulting from the solution of problem~\ref{eq.CC}~\citep{zarjovgeoJFM17}. The high-quality recovery of two-point correlations is attributed to the Lyapunov-like structural constraint in~\ref{eq.CC}, which {\em keeps physics in the mix and enforces consistency between data and \mbox{the linearized dynamics.}}

\vspace*{-4ex}
	\subsubsection*{\tc{dred}{Alternative formulations.}}

Covariance completion problem~\ref{eq.CC} can be cast as an optimal control problem aimed at establishing a trade-off between control energy and the number of feedback couplings that are required to modify $\Ak$ in System~\ref{eq.fbk-lnse1} and achieve consistency with available data~\citep{zarmohdhigeojovTAC19,zargeojovARC20}. Depending on modeling purpose and available data, many alternative turbulence modeling formulations are possible.~\cite{jovbamCDC01} showed that portions of one-point correlations in ($x,y,z$), resulting from the integration of DNS-based $\Vk^{\mathrm{ker}} (y,y)$ over $\bk$, can be approximated by the appropriate choice of covariance of white-in-time forcing to the NS equations linearized around turbulent mean velocity. This early success inspired the development of optimization algorithms for approximation of full covariance matrices using stochastic dynamical models~\citep{hoe05,linjovTAC09}. For the eddy-viscosity enhanced linearization,~\cite{moajovJFM12} demonstrated that white-in-time forcing with variance proportional to the turbulent energy spectrum can be used to reproduce the DNS-based energy spectrum of velocity fluctuations.~\citet{hwaeck20} determined the wave-number dependence of the variance of stochastic forcing, which is uncorrelated in $t$ and $y$, that minimizes the difference between the Reynolds shear stresses resulting from the mean and the linearized eddy-viscosity enhanced NS equations. Several recent efforts were aimed at matching individual entries of the spectral density matrix $\Sout$ at given frequencies~\citep{bensiparndanles16,benyegsiplec17,towlozyan20} or at capturing the spectral power, $\trace(\Sout)$,~\citep{morsemhencos19}. Finally, compared to the standard resolvent analysis~\citep{moajovtroshamckPOF14}, an optimization-based approach that utilizes a componentwise approach~\citep{rosmck19} offers considerable improvement in matching spectra and co-spectra in turbulent channel flow~\citep{mcmrosmck20}. This further exemplifies the power and versatility of the componentwise input-output viewpoint of fluid flows that was \mbox{introduced in~\citet{jovbamJFM05}.}

	%==========
	% Figure 12  %
	%==========
	
%\newpage
% Summary Points
	\vspace*{-2ex}
\begin{summary}[SUMMARY POINTS]
	\vspace*{-2ex}
\begin{enumerate}

\item The following quote is attributed to Eric Eady: {\em ``It is not the process of linearization that limits insight. It is the nature of the state that we choose to linearize about.''\/} In addition, this review demonstrates that {\em the tools that we use to study the linearized equations are as important as the base flow that we choose to linearize about.}

\item Componentwise input-output analysis {\em uncovers mechanisms\/} for subcritical transition and identifies streamwise streaks, oblique waves, and Orr-Sommerfeld modes as input-output resonances from forcing to velocity components. 

\item Input-output analysis {\em discovers a viscoelastic analogue of the familiar inertial lift-up mechanism\/}. This mechanism arises from stretching of polymer stress fluctuations by a base shear and, even in the absence of inertia, it induces significant amplification that can trigger transition to elastic turbulence in rectilinear flows. 

\item Input-output analysis {\em quantifies impact of forcing and energy-content of velocity components}. It reveals influence of dimensionless groups (e.g., Reynolds and Weissenberg numbers) on amplification of deterministic and stochastic disturbances and identifies relevant spatio-temporal scales as well as the dominant flow structures. 

\item Input-output viewpoint provides a {\em model-based approach to vibrational flow control\/}, where the dynamics are impacted by zero-mean oscillations. Effective strategies for controlling the onset of turbulence and turbulence suppression can be designed by examining the dynamics of fluctuations around the base flow induced by oscillations.

\item Linearized NS equations with stochastic forcing qualitatively predict structural features of turbulent shear flows and provide sufficient flexibility to {\em account for two-point correlations of fully-developed turbulence via low-complexity~models}.

\item Input-output framework provides a {\em data-driven refinement of a physics-based model\/}, guarantees statistical consistency, and captures complex dynamics of turbulent flows in a way that is tractable for analysis, optimization, and control design.

\item Tools and ideas from control theory and convex optimization overcome shortcomings of physics-agnostic machine learning algorithms and enable the development of {\em theory and techniques for physics-aware data-driven turbulence modeling}. 
\end{enumerate}
	\vspace*{-2.1ex}
\end{summary}
 
	\vspace*{-4.5ex}
\begin{issues}[FUTURE ISSUES]
\vspace*{-2ex}
\begin{enumerate}
\item {\bf Complex fluids and complex flows.} Among other emerging applications, input-output analysis is expected to clarify the importance of different physical mechanisms in the presence of surface roughness and free-stream disturbances, and to quantify the impact of modeling uncertainties that arise from chemical reactions and gas surface interactions in hypersonic flows~\citep{canARFM19}.
	
\item {\bf Computational complexity.} For an evolution model with $n$ degrees of freedom, the tools presented in this review require $O(n^3)$ computations. Such computations are routine for canonical flows, but the large-scale nature of spatially-discretized models in complex geometries induces significant computational overhead.

\item {\bf Nonlinear interactions.} Precise characterization of the interplay between high flow sensitivity and nonlinearity in order to capture later stages of disturbance development, identify possible routes for transition, and design effective control strategies for the nonlinear NS equations remains a grand challenge. 

\item {\bf Data-driven techniques.} In spite of the apparent promise of machine and reinforcement learning, a number of challenges have to be addressed, including the development of methods that respect physical constraints, generalize to flow regimes that are not accounted for in the available data, and offer convergence, performance, and robustness guarantees on par with model-based approaches to flow control.

\item {\bf Feedback control.} A host of challenges including estimation using noisy measurements, optimal sensor and actuator placement, efficient computation of optimal and robust controllers, structured and distributed control synthesis, as well as convergence and sample complexity of data-driven reinforcement learning strategies have to be addressed to enable a successful feedback control at high Reynolds numbers.
\end{enumerate}
\vspace*{-3ex}
\end{issues}

%Conclusion
% \vspace*{-6ex}
	\newpage
\section{DISCUSSION}

Herein, we expand on future issues and provide an overview of outstanding challenges.

\vspace*{-3ex}
\subsubsection*{\tc{dred}{Complex fluids and complex flows.}}
	In addition to parallel flows, input-output analysis was utilized to quantify the influence of deterministic~\citep{sipmar13} and stochastic~\citep{ranzarhacjovPRF19b} inputs as well as base flow variations~\citep{brasippramar11} on spatially-evolving boundary layers. In high-speed compressible flows, there is a coupling of inertial and thermal effects, and experiments suggest a significant impact of exogenous disturbances on transition~\citep{FedARFM11}. Traditional receptivity is based on a local spatial analysis~\citep{malAIAA89,berherCTFD91}, and is not applicable to most bodies of aerodynamic interest. For hypersonic vehicles with complex geometry or shock interactions with control surfaces, transition is poorly understood, and empirical testing is typically used to characterize their behavior. Linearization around spatially-evolving base flows in the presence of sharp gradients involves multiple inhomogeneous directions, and even modal stability analysis becomes challenging and computationally demanding~\citep{hildwinicjovcanPRF18,siddwicannicPRF18}. Recently,~\citet{dwisidniccanjovJFM19} employed a global input-output analysis to quantify the amplification of exogenous disturbances and explain the appearance of experimentally observed steady reattachment streaks in a hypersonic flow over a compression ramp. For the laminar shock-boundary layer interaction, this study showed that upstream counter-rotating vortices trigger streaks with a preferential spanwise length scale. Input-output analysis is expected to clarify the importance of different physical mechanisms in the presence of surface roughness and free-stream disturbances, and to quantify the impact of modeling uncertainties that arise from chemical reactions and gas surface interactions on hypersonic flows~\citep{canARFM19}.
	
\vspace*{-3ex}
\subsubsection*{\tc{dred}{Computational complexity.}}
For an evolution model with $n$ degrees of freedom, the tools presented in this review require $O(n^3)$ computations. Such computations are routine for canonical flows, but the large-scale nature of spatially-discretized models in complex geometries induces significant computational overhead. Dominant singular values of the state-transition and frequency response operators can be computed iteratively~\citep{sch07} or via randomized techniques~\citep{halmartro11}. Such computations have been used to conduct nonmodal analysis of complex flows~\citep{jeunicjovPOF16,dwisidniccanjovJFM19}. While in general it is challenging to efficiently solve large-scale Lyapunov equations, efficient iterative algorithms (both in terms of memory and computations) exist for systems with a small number of inputs and outputs and sparse dynamic matrices~\citep{benlipen08}. These are expected to bring the utility of stochastic analysis from canonical channels~\citep{jovbamJFM05} and boundary layers~\citep{ranzarhacjovPRF19b} to flows in complex geometries. 

\vspace*{-3ex}
\subsubsection*{\tc{dred}{Nonlinear interactions.}}
Large amplification of disturbances in conjunction with nonlinear interactions can induce secondary instability of streamwise streaks, their breakdown, and transition~\citep{wal97}. An alternative self-sustaining mechanism shows that turbulence can be triggered by the streamwise-constant NS equations in feedback with a stochastically-forced streamwise-varying linearization~\citep{farioa12,tholiejovfarioagayPOF14}. Nonlinear nonmodal stability analysis identifies initial conditions of a given amplitude that maximize energy at a fixed time~\citep{ker18}. Dissipation inequalities~\citep{ahmvalgaypap19} and a harmonic balance approach~\citep{rigsipcol20} were recently utilized to extend input-output analysis to the nonlinear NS equations, and the theory of integral quadratic constraints~\citep{megran97} was used to study of a phenomenological model of transition~\citep{kalseihem20}. However, it is still an open challenge to precisely characterize the interplay between high flow sensitivity and nonlinearity in order to capture later stages of transition routes and design effective control strategies.

\vspace*{-3ex}
\subsubsection*{\tc{dred}{Data-driven techniques.}}
Machine learning has revolutionized many disciplines, e.g., image recognition and speech processing, and is increasingly used in modeling and decision making based on available data. While the NS equations are often too complex for model-based optimization and control, DNS provides data for reduced-order dynamical modeling~\citep{row05,lum07,sch10,jovschnicPOF14,towschcol18}. Capitalizing on the availability of such data, machine and reinforcement learning have recently been used for flow modeling, optimization, and control~\citep{brunoakou20} and this trend will continue. In spite of the apparent promise, a number of challenges have to be addressed, including the development of methods that respect physical constraints, generalize to flow regimes that were not accounted for in the available data sets, and offer convergence, performance, and robustness guarantees on par with \mbox{model-based approaches to flow control.}

\vspace*{-3ex}
\subsubsection*{\tc{dred}{Feedback control}} offers a more viable approach than sensor-less control for dealing with uncertainties that impact the operation of engineering flows. The scale and complexity of the problem introduce significant challenges for modeling, sensor and actuator placement, and control design. These necessitate the development of model-based and data-driven techniques. In wall-bounded flows at low Reynolds numbers, model-based feedback control has shown significant promise~\citep{josspekim97,bewliu98,hogbewhen03,hogbewhen03b,kimbew07}. Since sensing and actuation are typically restricted to the surface, the flow field needs to be estimated using limited noisy measurements in order to form a control action.~\citet{hoechebewhen05,chehoebewhen06} demonstrated the importance of statistics of disturbances in the design of estimation gains. Alternatively, the data-refined model, given by Equation~\ref{eq.fbk-lnse1}, that matches statistics of turbulent flows can be readily embedded into a Kalman filter estimation framework. Alongside estimation, challenges associated with the optimal sensor and actuator placement~\citep{cherow11,zarmohdhigeojovTAC19}, efficient computation of optimal and robust controllers~\citep{bewlucpra16}, structured and distributed control synthesis~\citep{linfarjovTAC13admm}, as well as convergence and sample complexity of data-driven reinforcement learning strategies~\citep{mohzarsoljovTAC19} have to be addressed to enable a successful feedback control at high Reynolds numbers.

%Disclosure
\vspace*{-4ex}
\section*{DISCLOSURE STATEMENT}

The author is not aware of any affiliations, memberships, funding, or financial holdings that might be perceived as affecting the objectivity of this review. 

% Acknowledgements
\vspace*{-4ex}
\section*{ACKNOWLEDGMENTS}
  
I would like to thank Bassam Bamieh for his mentorship, friendship, and support over the years. Bassam inspired me to work at the interface between fluid mechanics and control and has been an intellectual role model for me from the first day we met. I am grateful to Parviz Moin for welcoming me to the Center for Turbulence Research at Stanford University on numerous occasions; these visits pushed me out of my comfort zone and made me a better researcher. I am indebted to Makan Fardad, Tryphon Georgiou, Satish Kumar, Joe Nichols, and Peter Schmid for our collaborations and for their thoughtful feedback about this review; we enjoy spending time with each other and our scientific interactions are a consequence of our friendship. This work would not have been possible without contributions of my current and former PhD students Anubhav Dwivedi, Gokul Hariharan, Nazish Hoda, Binh Lieu, Fu Lin, Rashad Moarref, Wei Ran, and Armin Zare and without support from the Air Force Office of Scientific Research under Awards FA9550-16-1-0009 and FA9550-18-1-0422, and from the National Science Foundation \mbox{under Award ECCS-1809833.}

\newpage
% References
% \vspace*{-4ex}
%\bibliographystyle{ar-style1}
%\bibliography{../bib/couette,../bib/mj-complete-bib,../bib/mj-unrefereed-bib,../bib/periodic,../bib/covariance,../bib/control-pde,../bib/ref-added-rm,../bib/ref-added-az,../bib/ref-added-wr,../bib/low_rank_bib,../bib/viscoelastic,../bib/hypersonic} 

\end{document}